\documentclass{jpp}

\usepackage{amsmath, amssymb, aas_macros}
\usepackage{epsfig}
\usepackage{color}

\usepackage{graphicx}
\usepackage{natbib}

\usepackage[table]{pstricks}
\usepackage{multido}
\usepackage{pst-all}
\usepackage{pst-node,pst-tree}
\usepackage{lscape}
\usepackage{pst-3dplot}
\usepackage{pstricks-add}
\usepackage{pstricks,pst-slpe}
\usepackage{pgfplotstable}
\usepackage{numprint}
\usepackage{siunitx}
\usepackage{gensymb}
\usepackage{wasysym}

\usetikzlibrary{calc,decorations.markings}

\newcommand\xstrut{\vphantom{\tabular{c}Üg\\Üg\endtabular}}
\newcommand\psBoite[3][white]{\rput(#2){\rnode{#2}{%
\psframebox[fillstyle=gradient,fillcolor=#1,gradbegin=#1,gradend=white]{
\xstrut\makebox [ 2.3cm ] { \tabular { c } #3\endtabular }} } } }

\newcommand{\rot}{\mathbf{\nabla} \times}
\newcommand{\divg}{\mathbf{\nabla}\cdot}

\newcommand{\rlight}{r_{\rm L}}
\newcommand{\Rsph}{R_{\rm sph}}
\newcommand{\Rs}{R_{\rm s}}

\newcommand{\BQ}{B_{\rm qed}}

\newcommand{\ex}{\mathbf{e}_{\rm x}}
\newcommand{\ey}{\mathbf{e}_{\rm y}}
\newcommand{\ez}{\mathbf{e}_{\rm z}}
\newcommand{\er}{\mathbf{e}_{\rm r}}

\newcommand{\ephi}{\mathbf{e}_\varphi}

\newcommand{\masselec}{m_{\rm e}}
\newcommand{\massproton}{m_{\rm p}}

\newcommand{\nuc}{\nu_{\rm B}}
\newcommand{\nup}{\nu_{\rm p}}

\title{Theory of pulsar magnetosphere and wind}

\author[J. P\'etri]%
{J\ls \'E\ls R\ls \^O\ls M\ls E\ns P\ls \'E\ls T\ls R\ls I
  \thanks{Email address for correspondence: jerome.petri@astro.unistra.fr}}

\affiliation{Observatoire astronomique de Strasbourg, Universit\'e de Strasbourg, \\
CNRS, UMR 7550, 11 rue de l'universit\'e, F-67000 Strasbourg, France. \\[\affilskip]}

\pubyear{2016}
\volume{}
\pagerange{}
\date{?; revised ?; accepted ?. - To be entered by editorial office}
\begin{document}

\maketitle

\begin{abstract}
Neutron stars are fascinating astrophysical objects immersed in strong gravitational and electromagnetic fields, at the edge of our current theories. These stars manifest themselves mostly as pulsars, emitting a timely very stable and regular electromagnetic signal. Even though discovered almost fifty years ago, they still remain mysterious compact stellar objects. In this review, we summarize the most fundamental theoretical aspects of neutron star magnetospheres and winds. The main competing models susceptible to explain their radiative properties like multi-wavelength pulse shapes and spectra and the underlying physical processes such as pair creation and radiation mechanisms are scrutinized. A global but still rather qualitative picture emerges slowly thanks to recent advances in numerical simulations on the largest scales. However considerations about pulsar magnetospheres remain speculative. For instance the exact composition of the magnetospheric plasma is not yet known. Is it solely filled with a mixture of $e^\pm$~leptons, or does it contain a non negligible fraction of protons and/or ions? Actually, is it almost entirely filled or mostly empty except for some small anecdotal plasma filled regions? Answers to these questions will strongly direct the description of the magnetosphere to seemingly contradictory results leading sometimes to inconsistencies. Nevertheless, account are given to the latest developments in the theory of pulsar magnetospheres and winds, the existence of a possible electrosphere and physical insight obtained from related observational signatures of multi-wavelength pulsed emission.
\end{abstract}


\section{Introduction}

End product of stellar evolution, neutron stars form a special class of compact objects showing themselves with many different faces \citep{2008PPN....39.1136P, 2013FrPhy...8..679H}. The idea of the existence of neutron stars formed by the gravitational collapse of a star at the end of their life during the explosion of the supernova \citep{1934PNAS...20..254B} was suggested well before their observational evidence that appeared only thirty years later \citep{1968Natur.217..709H}. Studying neutron stars is nowadays without doubt of interest to many areas in theoretical physics and astrophysics. The discovery of pulsars as a sub-class of neutron stars revolutionized astrophysics and revived their theoretical study. Indeed, pulsars can take pride in allowing for many recent advances and progresses in theoretical as well as observational high-energy physics and astrophysics. Just to list some of their direct observational impacts, we mention the confirmation of the existence of neutron stars observed as pulsars \citep{1968Natur.217..709H}, indices on their internal structure, indirect detection of gravitational waves \citep{1975ApJ...195L..51H} and maybe in the future direct detection with the international pulsar timing array \cite{2010CQGra..27h4013H}, detection of the first planetary system \citep{1991Natur.352..311B, 1992Natur.355..145W}, study of quantum processes in a strong magnetic field \citep{2006RPPh...69.2631H, 2015SSRv..191...13L}, motion of matter and photons in strong gravitational fields \citep{2006Sci...314...97K}, survey of the interstellar medium in the Milky Way \citep{2002astro.ph..7156C} by dispersion measure and last but not least survey of the galactic magnetic field in the Milky Way \citep{2006ApJ...642..868H, 2009ApJS..181..557H} by rotation measure. These discoveries highlight the importance of a right understanding of neutron star physics and especially pulsars. We could then take full advantage of our improved knowledge to constrain our theories of gravity and electromagnetism, a quest not reachable on Earth.

In the present paper, we summarize several essential aspects of pulsar physics related to their magnetosphere and wind. Although the general environment of a neutron star is simply described by three ingredients, namely a compact object, rotating and last but not least strongly magnetised, this m\'enage \`a trois brings in already severe complications. These are reported in the overview of Sec.~\ref{sec:MagnetosphereOverview} where the overall electrodynamics is described before plunging deeper into details of the magnetosphere in Sec.~\ref{sec:Magnetosphere}. With the advance of numerical techniques and computer power, the wealth of observations forces us to refine our physical assumptions rendering them more realistic by adding new corrections to the simple magnetospheric view presented in the previous section. Some of these refinements are listed in Sec.~\ref{sec:OtherEffects} and includes general relatively, multipoles, quantum electrodynamics, pair creation and magnetic reconnection. We report then on progresses accomplished via numerical simulations in Sec.~\ref{sec:Simulations}. The dynamics in the magnetosphere is dominated by the electromagnetic field up to a point, the light-cylinder, where particle inertia plays a crucial role. This more remote location is often quoted as the pulsar wind and possesses its intrinsic dynamics distinctly different from the closed magnetosphere, Sec.~\ref{sec:Vent}. The last years have witnessed a dramatic change in the wind paradigm. It became clear that it must be striped around the equatorial plane, Sec.~\ref{sec:VentStrie}, thus leading to a time-dependent view including breakdown of the MHD regime within the stripe. The next decade should bring in more quantitative and qualitative insight into pulsar magnetosphere theories as we bet in the concluding Sec.~\ref{sec:Summary}.


\section{Overview of pulsar magnetospheres}
\label{sec:MagnetosphereOverview}

Soon after the discovery of the first pulsar, it was realized that the central star should be a neutron star. Following simple arguments, a simple but robust image emerged about the main characteristics of this compact object, these being its period of rotation and its surface magnetic field strength. A fast rotating strongly magnetized neutron star in vacuum served as a template for the general understanding of such systems. We remind how scientists came to this conclusion and their implications for current research in the field. We think it useful to point out again the main historical steps because the physics of pulsars, despite fifty years of intensive research, is still in its infancy. 

\subsection{Orders of magnitude}

Although neutron stars have eventually been taken seriously fifty years ago, modelling of its environment is still in its early stages nowadays. Nevertheless it can be indisputably summarized in one sentence: a pulsar hosts a neutron star that is rapidly rotating and strongly magnetized\footnote{The meaning of ``rapid rotation'' and ``strong magnetic field'' needs clarification and will be justified in the following lines.}. Indeed, the hope to explain pulsars with an underlying white dwarf went away very soon after realizing that the observed rotation periods, much less than a second, would disrupt the star because of centrifugal forces induced by stellar rotation at a rate much larger than the break-up velocity limited by the Keplerian frequency. Moreover the collapse of a standard main sequence star to a neutron star with conservation of angular momentum and magnetic flux at the zero's level of approximation can lead to strong magnetic fields as those expected to ignite pulsar electromagnetic activity. Indeed, simple estimates for periods and magnetic fields of pulsars are given by conservation of angular momentum
\begin{equation}
 M_{\rm ns}\,\Omega_{\rm ns}\,R_{\rm ns}^2 = M_*\,\Omega_*\,R_*^2
\end{equation} 
and magnetic flux 
\begin{equation}
 B_{\rm ns}\,R_{\rm ns}^2 = B_*\,R_*^2
\end{equation} 
during the collapse of the progenitor star. $M$, $\Omega$ and $R$ are respectively the mass, the rotation rate and the radius, on one hand of the neutron star with each physical quantity~$X$ indexed by~$X_{\rm ns}$ and, on the other hand the progenitor with index~$X_*$. Assuming mass conservation~$M_{\rm ns}=M_*$ (certainly a too crude assumption), the increase in magnetic field and angular velocity are in the same ratio of
\begin{equation}
 \left(\frac{R_*}{R_{\rm ns}} \right)^2 \approx \numprint{e10}
\end{equation}
for typical main sequence and neutron star radii taken to be about $R_*=\numprint{e6}$~\si{\kilo\meter} and $R_{\rm ns}=\numprint{10}$~\si{\kilo\meter}, respectively. Rotation period of main sequence stars from the Kepler space mission have been observed between 0.2~day and 70~days \citep{2014ApJS..211...24M} and the magnetic field for the Sun is about $\numprint{e-3}$~\si{\tesla}. This leads to $\Omega_{\rm ns} \approx \numprint{0.5}$~\si{\milli\second} and $B_{\rm ns} \approx \numprint{e7}$~\si{\tesla} compatible with actual values of neutron stars. Strongly magnetized stars refers to magnetic field strengths comparable to the quantum critical field given by
\begin{equation}
 \BQ = \frac{\masselec^2\,c^2}{e\,\hbar} \approx \numprint{4.4e9}~\si{\tesla}
\end{equation} 
obtained by equating the electron rest mass~$\masselec\,c^2$ to the cyclotron energy $\hbar\,\omega_B$, $\hbar$ being the reduced Planck constant. A neutron star is also highly compact. Its compactness, defined by the ratio between its Schwarzschild radius~$\Rs=2\,G\,M/c^2$ ($G$ is the gravitational constant) and its actual radius~$R$, is of the order
\begin{equation}
\label{eq:Compacite}
\Xi = \frac{\Rs}{R} \approx \numprint{0.345} \, \left( \frac{M}{\numprint{1.4}~M_\odot} \right) \, \left( \frac{R}{\numprint{12}~\si{\kilo\meter}} \right)^{-1}
\end{equation}
thus close to the most extreme compactness given by a Schwarzschild black hole for which $\Xi=1$. The effects of general relativity will be significant at least close to the stellar surface, in particular around the polar caps where pair creation is supposed to occur. Neutron star mass measurements give an average value of around \numprint{1.5}~$M_\odot$ with a spread of about \numprint{0.5}~$M_\odot$ \citep{2012ARNPS..62..485L}. Most equations of state predict radius about 12~\si{\kilo\meter}. Simultaneous measurements of masses and radii are also of great interest for nuclear physicists to constrain the equation of state of matter above nuclei densities \citep{2016arXiv160302698O}.

\begin{figure}
\centering
\begin{tikzpicture}
\begin{axis}[
  xmode=log,
  ymode=log,
  xlabel=$P$ (s),
  ylabel=$\dot P$(s/s)]
\addplot [color=blue, only marks, mark size=0.8pt] table [y=Pdot, x=P]{pulsar_radio.dat};
\addlegendentry{radio}
\addplot [color=red, only marks, mark=triangle*, mark size=1.5pt] table [y=Pdot, x=P]{pulsar_gamma_non_binaire_ppdot.dat};
\addlegendentry{gamma}
\addplot [color=green, only marks, mark=star, mark size=1.5pt] table [y=Pdot, x=P]{pulsar_gamma_binaire_ppdot.dat};
\addlegendentry{binary}
\end{axis}
\end{tikzpicture}
\caption{$P-\dot P$ diagram of all known pulsars with measured period and period derivatives. Data are from the ATNF Pulsar Catalogue at http://www.atnf.csiro.au/people/pulsar/psrcat/ and Manchester, R. N., Hobbs, G.B., Teoh, A. \& Hobbs, M., AJ, 129, 1993-2006 (2005).}
\label{fig:PPdot}
\end{figure}
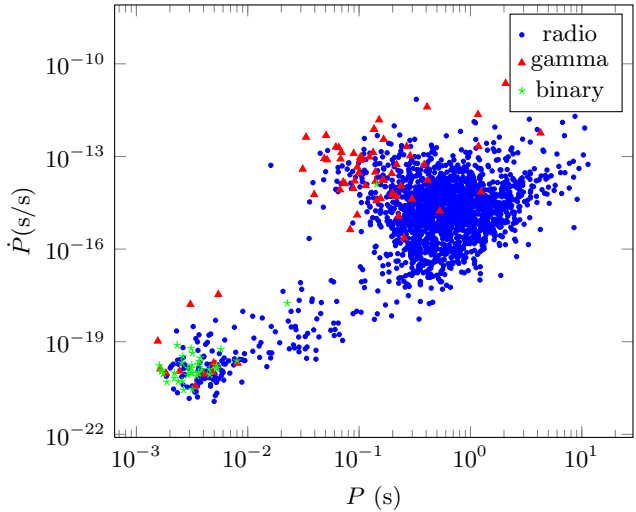

Pulsars are usually compiled in a so called $P-\dot P$ diagram shown in fig.~\ref{fig:PPdot} where $P$ represents the pulsar rotation period and $\dot P$ its period derivative. The latter accounts for the braking of the star through energy and angular momentum losses in vacuum or by a relativistic wind. In fig.~\ref{fig:PPdot}, we separate them into three classes, those seen mainly in radio, those observed in gamma-rays and those being part of a binary system. To date, we know more than \numprint{2000}~pulsars among them about \numprint{100} evolving in binaries. These are only a tiny fraction of the total number of neutron stars in our galaxy estimated to be around~\numprint{e8}-\numprint{e9}.

Although pulsars have been discovered through their radio emission this mechanism remains largely misunderstood. Moreover, only a fraction~of \numprint{e-5} of the rotational kinetic luminosity is converted into radio power. The radio brightness temperature is of the order $T_b \approx \numprint{e25}-\numprint{e28}$~\si{\kelvin}, thus not produced by a usual plasma process but via a coherent emission mechanism that still awaits to be elucidated. To get a more accurate idea of the pulsar machinery, it is compulsory to make a rigorous scrutiny of the physical conditions reigning inside its magnetosphere. The assumption of a rotating magnetic dipole loosing energy per vacuum electromagnetic radiation is unrealistic because we would only expected emission at the rotation frequency~$\Omega$ completely at odds to observations showing a broad band emission spectrum from \si{\mega\hertz} frequencies up to \si{\tera\electronvolt} energies. The rotational braking of the star, the nascent current in the magnetosphere circulating into the wind, the associated particle acceleration and transport of energy from the surface across the light cylinder up to the nebula therefore all require a detailed knowledge about their electrodynamics, especially the longitudinal electric current (along magnetic field lines).

For the sake of simplicity, we essentially distinguish three kind of magnetospheres, or more exactly three fundamental hypotheses to the elaboration of pulsar magnetosphere theory. On one extreme side, we consider a naked star, entirely devoid of plasma in its immediate neighbourhood, the zero-th order formulation so to say. Of course, without plasma there is no high energy emission but if the plasma density remains negligible, the dynamics only weakly depends on the plasma motion and properties. On the other extreme side, we consider a star completely surrounded by a dense plasma, screening the longitudinal electric field~$E_\parallel = \mathbf{E} \cdot \mathbf{B} / B$ that would be imposed in vacuum by the previous assumption. In between these two conflicting starting points, an intermediate model admits the existence of a surrounding partially filled or empty magnetosphere, depending on our pessimistic or optimistic view, called an electrosphere. These three models and possible variations as well as their related observational implications are synthesized in fig.~\ref{fig:ModeleMagnetosphere}. The place of the plasma density cursor is the discriminating parameter. The low density limit leads to non linear plasma wave models \citep{2015Ap&SS.357...52R} whereas the high density limit was developed as a relativistic wind model. Pair production in the magnetosphere is the key process to determine which regime is to be applied and nothing forbid to switch from on regime to the other during plasma transport towards the nebula. \cite{1979SSRv...24..407K} give orders of magnitude for the properties of these waves and winds. We briefly remind the evolution of the ideas concerning pulsar magnetospheres leading to these three alternatives.
\begin{figure}
\begin{center}
\psset{framearc=0.2,shadow=true,fillstyle=solid,shadowcolor=black!55}
\psscalebox{0.9}{
\begin{pspicture}(0,-2)(18,10)
\psBoite[yellow!30]{8,8}{magnetosphere}

\psBoite[cyan!30]{2,6}{vacuum} 
\psBoite[cyan!30]{5,6}{charge\\separated}
\psBoite[cyan!30]{12,6}{almost\\full} 

\psBoite[red!30]{2,4}{Maxwell \\ equation} 
\psBoite[red!30]{5,4}{non-neutral \\ plasma}
\psBoite[red!30]{12,4}{neutral \\ plasma (MHD?)} 

\psBoite[blue!30]{2,2}{cosmic ray\\acceleration} 
\psBoite[blue!30]{5,2}{electrosphere} 
\psBoite[blue!30]{8,2}{polar\\cap} 
\psBoite[blue!30]{11,2}{outer\\gap}
\psBoite[blue!30]{14,2}{slot\\gap}

\psBoite[green!30]{2,0}{monochromatic\\at $\Omega$} 
\psBoite[green!30]{5,0}{radio \& \\ gamma?} 
\psBoite[green!30]{8,0}{radio} 
\psBoite[green!30]{11,0}{gamma} 
\psBoite[green!30]{14,0}{radio \&\\ gamma?} 
\psset{arrowscale=5}
\psline{->}(2,-1.5)(14,-1.5)
\rput(8,-2){Increasing magnetospheric plasma density}
\end{pspicture}
\psset{shadow=false,angleA=-90,angleB=90,linewidth=2pt}
\ncangles{8,8}{2,6}\ncangles{8,8}{5,6}\ncangles{8,8}{12,6}
\ncangles{2,6}{2,4}\ncangles{5,6}{5,4}\ncangles{12,6}{12,4}
\ncangles{12,4}{8,2}\ncangles{12,4}{11,2}\ncangles{12,4}{14,2}
\ncangles{2,4}{2,2}\ncangles{5,4}{5,2}
\ncangles{2,2}{2,0}\ncangles{5,2}{5,0}\ncangles{8,2}{8,0}\ncangles{11,2}{11,0}
\ncangles{14,2}{14,0}
}
\end{center}
\caption{Synthetic view of pulsar magnetosphere models depending on the plasma density in its magnetosphere. The upper cyan boxes indicate the three alternative magnetosphere assumptions. The red boxes describe the regime used to investigate the dynamics. The blue boxes point out the peculiarity of each model. The green boxes summarize the expected emission spectra.}
\label{fig:ModeleMagnetosphere}
\end{figure}

\subsection{Vacuum electromagnetic fields}

The simplest model we can think of is about a vacuum magnetosphere, empty of any plasma or particle. To start with, the internal structure of a neutron star is believed to be in a superconducting and superfluid state. Its electric conductivity is so high that the magnetic field is frozen into the star and could survive for a long time. Moreover because of its rotation, a electromotive field is induced such that the electric field in the corotating frame vanishes,~$\mathbf{E}'=\mathbf{0}$. Transformations of the electromagnetic field from one frame to another require general relativity \citep{1978Ap&SS..58..427K} and not the Lorentz transformations when rotation is considered. The question about electromagnetic fields in rotating frames was raised by \cite{1939PNAS...25..391S} who discussed an illustrative example of two rotating and charged concentric spheres. Following his idea, \cite{1973Ap&SS..24..323W} used the tensorial formalism of general relativity to write Maxwell equations in any rotating coordinate frame. Additional source terms in the inhomogeneous Maxwell equations appear in non inertial frames. From the transformation law between an inertial frame and a rotating frame \citep{1984IJTP...23..441G} we get
\begin{equation}
\label{eq:Eprime}
 \mathbf{E}' = \mathbf E + \mathbf ( \Omega \wedge \mathbf r ) \wedge \mathbf B = \mathbf 0
\end{equation}
where~$\mathbf{r}$ is the position vector and~$\mathbf{\Omega}$ the rotation velocity vector of the star. The interpretation of this relation was not that obvious \citep{1956ApJ...123..508B}. The usual picture of magnetic field line motion has been challenged by \cite{1958AnPhy...3..347N} and should be taken with care. From this equilibrium condition we deduce that the electric and magnetic field are perpendicular in any frame because of the Lorentz invariance of $\mathbf{E} \cdot \mathbf{B}=0$. In other words, magnetic field lines are equipotentials for the electric field. To solve completely the problem of this rotating conductor, we need an assumption about the internal magnetic field. Two simple choices often quoted are an uniform magnetic field inside the star or a point dipole located right at its centre. It is straightforward to show that in both configurations the external magnetic field is dipolar. For a rotator with inclination between rotation axis and either the magnetic moment or the direction of the uniform interior magnetic field depicted by an obliquity~$\chi$, these expressions in spherical polar coordinates~$(r, \vartheta, \varphi)$ in the quasi-static near zone for distance much less than the wavelength $\lambda=2\,\pi\,\rlight$ where $\rlight=c/\Omega$ are given by
\begin{subequations}
  \label{eq:ChampBExtDipolaire}
\begin{align}
  B_r^\texttt{ext} & = \frac{2 \, B \, R^3}{r^3} \, ( \cos\chi \, \cos\vartheta + \sin\chi \, \sin\vartheta \, \cos \psi ) \\
  B_\vartheta^\texttt{ext} & = \frac{B \, R^3}{r^3} \, ( \cos\chi \, \sin\vartheta - \sin\chi \, \cos\vartheta \, \cos\psi ) \\
  B_\varphi^\texttt{ext} & = \frac{B \, R^3}{r^3} \, \sin\chi \, \sin\psi 
\end{align}
\end{subequations}
where $\psi = \varphi-\Omega\,t$ is the instantaneous phase at time~$t$. The external electric field in vacuum is quadrupolar and its components read
\begin{subequations}
\label{eq:ChampEextQuadrupolaire}
\begin{align}
  E_r^\texttt{ext} & = \Omega \, B \, R \, \left[ \frac{R^4}{r^4} \, \cos\chi \, 
    ( 1 - 3 \, \cos^2\theta ) - 3 \, \frac{R^4}{r^4} \, 
    \sin\chi \, \cos\theta \, \sin\theta \, \cos\psi \right] + \frac{Q_*}{4\,\pi\,\varepsilon_0\,r^2} \\
  E_\theta^\texttt{ext} & = \Omega \, B \, R \, \left[ \frac{R^2}{r^2} \, 
    \sin\chi \, \left( \frac{R^2}{r^2} \, \cos 2\, \theta - 1 \right) \, \cos\psi 
    - \frac{R^4}{r^4} \, \cos\chi \, \sin 2\,\theta \right] \\
  E_\varphi^\texttt{ext} & = \Omega \, B \, R \, \frac{R^2}{r^2} \, 
  \left( 1 - \frac{R^2}{r^2} \right) \, \sin\chi \, \cos\theta \, \sin\psi .
\end{align}
\end{subequations}
There is one free parameter depicted by the total charge of the neutron star through the quantity $Q_*$. Indeed, according to Gauss theorem, the asymptotic electric field has only a dominant radial component $E_r = Q_* / 4\,\pi\,\varepsilon_0\,r^2$, that is a monopolar term. 

In order to deduce the electric field inside the star and to fully solve the electromagnetic problem in whole space, we need to distinguish between several magnetizations. Assuming a dipolar magnetic field inside, the electric field inside becomes 
\begin{subequations}
\begin{align}
  E_r^\texttt{int} & = \frac{\Omega \, B \, R^3}{r^2} \, ( \cos\chi \, \sin^2\vartheta - 
    \sin\chi \, \cos\vartheta \, \sin\vartheta \, \cos\psi ) \\
  E_\vartheta^\texttt{int} & = - \frac{\Omega \, B \, R^3}{r^2} \, ( \cos\chi \, \sin 2\,\vartheta +
    2 \, \sin\chi \, \sin^2\vartheta \, \cos\psi ) \\
  E_\varphi^\texttt{int} & = 0 .
\end{align}
\end{subequations}
but if the magnetization is uniform inside the electromagnetic field looks like 
\begin{subequations}
\begin{align}
  B_r^\texttt{int} & = 2 \, B \, ( \cos\chi \, \cos\vartheta + \sin\chi \, \sin\vartheta \, \cos\psi ) \\
  B_\vartheta^\texttt{int} & = 2 \, B \, ( - \cos\chi \, \sin\vartheta + \sin\chi \, \cos\vartheta \, \cos\psi ) \\
  B_\varphi^\texttt{int} & = - 2 \, B \, \sin\chi \, \sin\psi \\
  E_r^\texttt{int} & = 2 \, r \, \Omega \, B \, \sin\vartheta \, (- \cos\chi \, \sin\vartheta + \sin\chi \, \cos\vartheta \, \cos\psi)  \\
  E_\vartheta^\texttt{int} & = - 2 \, r \, \Omega \, B \, \sin\vartheta \, ( \cos\chi \, \cos\vartheta + \sin\chi \, \sin\vartheta \, \cos\psi ) \\
  E_\varphi^\texttt{int} & = 0 \ .
\end{align}
\end{subequations}
These expressions are valid in the near zone where $r\ll\lambda$ because they neglect the displacement current $\varepsilon_0\,\partial_t \mathbf{E}$. Let us have a look on the charge distribution inside the star and at its surface. In this approximation there is no surface current because of the quasi-static assumption. Discontinuities in the magnetic field responsible for this current include corrections of the order $O(\Omega)$. From the perfect conductor condition eq.~(\ref{eq:Eprime}) the density in the absence of electric current, that could be neglected because the advective and displacement terms are of the order $(r/\rlight)^2$, is
\begin{equation}
\label{eq:DensiteCorotation}
\rho_{\rm e} = \varepsilon_0 \, \divg \mathbf{E} = - 2 \, \varepsilon_0 \, \mathbf{\Omega} \cdot \mathbf{B} \ .
\end{equation}
If the magnetization is dipolar, a central point charge exists, given by
\begin{equation}
\label{eq:ChargePonctuelle}
 Q_{\rm c} = \dfrac{8\,\pi}{3} \, \varepsilon_0 \, \Omega \, B \, R^3 \, \cos\chi \ .
\end{equation}
This charge should not be forgotten when computing the full electromagnetic field. The volume charge inside the star is zero between two spherical shells, the remaining distributes on its surface, inducing a discontinuity in the radial component of the electric field which is sustained by a surface electric charge $\sigma_{\rm s} = \varepsilon_0 \, [E_r]$. The central point charge is compensated by the stellar surface charge as summarized in table~\ref{tab:VacuumElectrodynamics1}. In the same vein, for a uniform magnetization, the constant volume charge density is compensated by the surface charge to keep the star electrically neutral whereas the central point charge has disappeared, see table~\ref{tab:VacuumElectrodynamics2}.
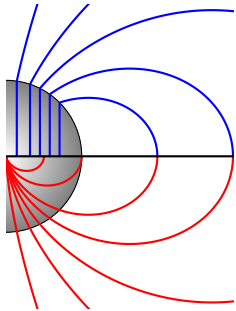
\begin{figure}
\centering
\begin{tikzpicture}
\clip[scale=1] (0,-2) rectangle (3,2);
\begin{scope}[rotate=-90]
\node (B) at (0,0) {};
\filldraw[inner color=white,outer color=gray] (B) circle (1);
\draw [red,thick] plot [samples=300, smooth, domain=0:90, shift={(B)}]
  (xy polar cs:angle=\x, radius= {0.5*sin(\x)*sin(\x)});
\draw [red,thick] plot [samples=300, smooth, domain=0:90, shift={(B)}]
  (xy polar cs:angle=\x, radius= {1*sin(\x)*sin(\x)});
\draw [red,thick] plot [samples=300, smooth, domain=0:90, shift={(B)}]
  (xy polar cs:angle=\x, radius= {2*sin(\x)*sin(\x)});
\draw [red,thick] plot [samples=300, smooth, domain=0:90, shift={(B)}]
  (xy polar cs:angle=\x, radius= {3*sin(\x)*sin(\x)});
\draw [red,thick] plot [samples=300, smooth, domain=0:90, shift={(B)}]
  (xy polar cs:angle=\x, radius= {5*sin(\x)*sin(\x)});
\draw [red,thick] plot [samples=300, smooth, domain=0:90, shift={(B)}]
  (xy polar cs:angle=\x, radius= {10*sin(\x)*sin(\x)});
\draw [red,thick] plot [samples=300, smooth, domain=0:90, shift={(B)}]
  (xy polar cs:angle=\x, radius= {50*sin(\x)*sin(\x)});
\draw [blue,thick] plot [samples=300, smooth, domain=90:180-asin(sqrt(1.0/2.0)), shift={(B)}]
  (xy polar cs:angle=\x, radius= {2*sin(\x)*sin(\x)});
\draw [blue,thick] plot [samples=300, smooth, domain=90:180-asin(sqrt(1.0/3.0)), shift={(B)}]
  (xy polar cs:angle=\x, radius= {3*sin(\x)*sin(\x)});
\draw [blue,thick] plot [samples=300, smooth, domain=90:180-asin(sqrt(1.0/5.0)), shift={(B)}]
  (xy polar cs:angle=\x, radius= {5*sin(\x)*sin(\x)});
\draw [blue,thick] plot [samples=300, smooth, domain=90:180-asin(sqrt(1.0/10.0)), shift={(B)}]
  (xy polar cs:angle=\x, radius= {10*sin(\x)*sin(\x)});
\draw [blue,thick] plot [samples=300, smooth, domain=90:180-asin(sqrt(1.0/50.0)), shift={(B)}]
  (xy polar cs:angle=\x, radius= {50*sin(\x)*sin(\x)});
\end{scope}
\draw [blue,thick] (0.141421,0) -- (0.141421, 0.989949) ;
\draw [blue,thick] (0.316228,0) -- (0.316228, 0.948683) ;
\draw [blue,thick] (0.447214,0) -- (0.447214, 0.894427) ;
\draw [blue,thick] (0.707107,0) -- (0.707107, 0.707107) ;
\draw [blue,thick] (0.57735,0) -- (0.57735, 0.816497) ;
\draw [black,thick] (0,0) -- (5,0) ;
\end{tikzpicture}
\caption{Uniform (upper blue) versus dipolar (lower red) internal magnetic field. Whatever the internal structure, outside the magnetic field is dipolar and the electric field quadrupolar to lowest order in $R/\rlight$.}
\label{fig:UniVsDip}
\end{figure}
Tables~\ref{tab:VacuumElectrodynamics1} and~\ref{tab:VacuumElectrodynamics2} summarizes the electric charge density inside the star and on its surface for both uniform and dipole magnetization and fig.~\ref{fig:UniVsDip} shows the two magnetizations. For completeness, the total volume and surface charges are also computed according to
\begin{subequations}
\begin{align}
Q_{\rm v} & = \iiint \rho_{\rm e} \, r^2 \, \sin \vartheta \, dr \, d\vartheta\, d\varphi \\
Q_{\rm s} & = \oiint \sigma_{\rm s} \, R^2 \, \sin \vartheta \, d\vartheta\, d\varphi
\end{align}
\end{subequations}
leading to the total electric charge of the star by $Q_*=Q_{\rm v}+Q_{\rm s}$.
\begin{table}
\centering
{
\rowcolors{2}{}{green!10}
\begin{tabular}{cc}
\hline
Central point charge~$Q_{\rm c}$ & $\frac{8\,\pi}{3} \, \varepsilon_0 \, \Omega \, B \, R^3 \, \cos\chi$ \\
Volume charge density~$\rho_{\rm e}$ & $-\frac{\varepsilon_0\,\Omega\,B\,R^3}{r^3} \, ( \cos\chi \, ( 1 + 3 \, \cos 2\,\vartheta ) + 6 \, \sin\chi \, \cos \vartheta \, \sin \vartheta \, \cos \psi )$ \\
Surface charge density~$\sigma_{\rm s}$ & $-2\,\varepsilon_0\,\Omega\,B\,R \, ( \cos\chi \, \cos^2 \vartheta + \sin\chi \, \cos \vartheta \, \sin \vartheta \, \cos \psi )$ \\
Total volume charge~$Q_{\rm v}$ & $0$ \\
Total surface charge~$Q_{\rm s}$ & $-Q_c$ \\
Total stellar charge~$Q_*$ & $0$ \\
\hline
\end{tabular}
}
\caption{Properties of vacuum electrodynamics around neutron stars for a dipolar magnetization.}
\label{tab:VacuumElectrodynamics1}
\end{table}

\begin{table}
\centering
{
\rowcolors{2}{}{green!10}
\begin{tabular}{cc}
\hline
Central point charge~$Q_{\rm c}$ & 0 \\
Volume charge density~$\rho_{\rm e}$ & $-4\,\varepsilon_0\,\Omega\,B\,\cos\chi$ \\
Surface charge density~$\sigma_{\rm s}$ & $\varepsilon_0\,\Omega\,B\,R \, ( \cos\chi \, ( 3 \, - 5 \, \cos^2 \vartheta ) - 5 \, \sin\chi \, \cos \vartheta \, \sin \vartheta \, \cos \psi )$ \\
Total volume charge~$Q_{\rm v}$ & $-2\,Q_c$ \\
Total surface charge~$Q_{\rm s}$ & $2\,Q_c$ \\
Total stellar charge~$Q_*$ & $0$ \\
\hline
\end{tabular}
}
\caption{Properties of vacuum electrodynamics around neutron stars for a uniform magnetization.}
\label{tab:VacuumElectrodynamics2}
\end{table}
The neutron star even if surrounded by vacuum could have an atmosphere because of its high surface temperature of $T\approx\numprint{e6}$~\si{\kelvin}. However the thickness of this layer would be very tiny because the height scale for a totally ionized hydrogen gas is
\begin{equation}
 H = \frac{k_{\rm B}\,T}{G\,M\,m_H/R^2} \approx \numprint{4.4}~\si{\milli\meter} \times \left( \frac{M}{1.4~M_\odot} \right) \, \left( \frac{T}{\numprint{e6}~\si{\kelvin}} \right)^{-1} \, \left( \frac{R}{\numprint{e4}~\si{\meter}} \right)^{2}
\end{equation}
At electrostatic equilibrium, the electromotive field displaces charges, initially interior to the pulsar, to its surface where they accumulate to screen this field. Other charges redistribute in such a way that in the rest frame of the star the total electric field vanishes. At the stellar surface appears an electric field of the order
\begin{equation}
  \label{eq:orderGrandeurChampE}
  E = \Omega\,B\,R = \numprint{e13}~\si{\volt/\meter} \ .
\end{equation}
This huge field extracts charges from the surface despite the presence of a potential barrier imposed by the inter and intra-molecular attraction\footnote{We also implicitly neglect the gravitational force. Indeed, the ratio between gravitational and electromagnetic forces is given by eq.~(\ref{eq:ForceGravElec}) for protons and is even a factor~$\frac{\massproton}{\masselec}$ smaller for electrons and therefore completely irrelevant for the physics at work in pulsars.}. This extraction threshold can be neglected without difficulty for a pulsar (at least for the electrons and probably also for the ions). The distinction between particle extraction and no particle extraction leads to different pulsar atmospheric models, a possible explanation for the evolution of pulsar states \citep{1973Natur.241..184E}. The vacuum model could also apply to low density plasmas. By low density \cite{1970Natur.228..348E} meant $n<\numprint{19}$~particles/\si{\meter}$^3$ for instance for the Crab. They proposed a model where particle corotation is only reached at twice the light-cylinder radius, $r=2\,\rlight$. When crossing this surface, particles become highly relativistic and radiate synchrotron photons in regions forming a two armed spiral where $E_\perp>c\,B$.

\subsection{Some historical notes}

The exact analytical solution to the external problem taking into account the boundary condition on the neutron star surface and the displacement current is given by \cite{1955AnAp...18....1D} solution whatever the magnetization, dipolar or uniform. Indeed, as shown in \cite{2015MNRAS.450..714P}, the electromagnetic field in vacuum outside the star is entirely determined by the radial component of the magnetic field at the surface,~$B_{\rm r}$. As this component is the same for both magnetizations, we expect the same solution outside the star. The only difference reflects in the surface charge and current densities, thus accounting for different spindown luminosities and torques exerted on the star.

\begin{figure}
\centering
\input{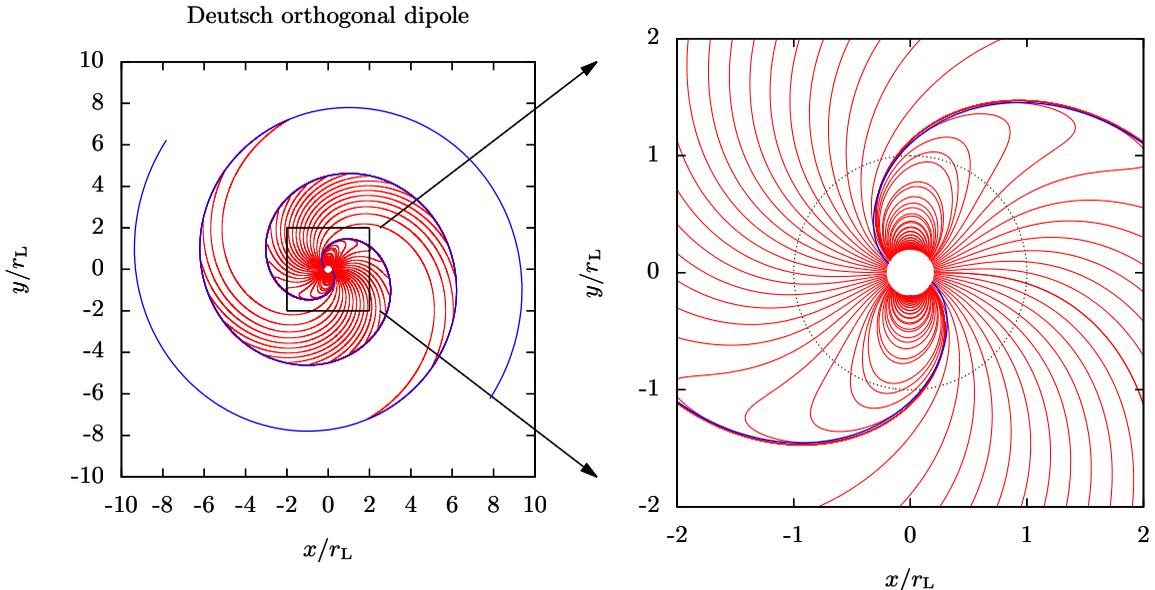}
\caption{Magnetic field lines (red solid lines) of the Deutsch solution for the orthogonal rotator with $R/\rlight=0.2$. The right panel is a zoom into the central region close to the light-cylinder (in dashed black circle of radius unity). The two-armed blue spiral line depicts the large scale wave structure of the electromagnetic field.}
\label{fig:SolutionDeutsch}
\end{figure}

\cite{1955AnAp...18....1D} was the first to compute electromagnetic wave emission emanating from a magnetized star in solid body rotation. He found that for those stars with strong magnetic field and rotating fast, the induced electric field becomes so strong that it is able to accelerate particles of the circumstellar medium to relativistic and even ultra-relativistic speeds. He thought that this phenomenon was the source for cosmic rays, an idea still valid. The rotating magnetized star is therefore at the origin of charge acceleration. At that time, he did not mentioned neutron stars. Moreover, his computations were valid only for a star plunged in vacuum. The star only emits a monochromatic large amplitude electromagnetic wave at a frequency equal to the stellar rotation rate~$\Omega$. The exact analytical solution he found is 
\begin{subequations}
\begin{align}
  \label{eq:DeutschB}
  B_r(\mathbf{r},t) & = 2 \, B \, \left[ \frac{R^3}{r^3} \, \cos\chi \, \cos\vartheta + 
    \frac{R}{r} \, \frac{h^{(1)}_1(k\,r)}{h^{(1)}_1(k\,R)} \, 
    \sin\chi \, \sin \vartheta \, e^{i\,\psi} \right] \\
  B_\vartheta(\mathbf{r},t) & = B \, \left[ \frac{R^3}{r^3} \, \cos\chi \, \sin\vartheta + \left( \frac{R}{r} \, \frac{\frac{d}{dr} \left( r \, h^{(1)}_1(k\,r) \right)}{h^{(1)}_1(k\,R)} + \frac{R^2}{\rlight^2} \, \frac{h^{(1)}_2(k\,r)}{\frac{d}{dr} \left( r \, h^{(1)}_2(k\,r) \right) |_{R}} \right) \, 
    \sin\chi \, \cos \vartheta \, e^{i\,\psi} \right] \\
  B_\varphi(\mathbf{r},t) & = B \, \left[ \frac{R}{r} \, 
    \frac{\frac{d}{dr} ( r \, h^{(1)}_1(k\,r) )}{h^{(1)}_1(k\,R)} \, 
    + \frac{R^2}{\rlight^2}
    \frac{h^{(1)}_2(k\,r)}{\frac{d}{dr} \left( r \, h^{(1)}_2(k\,r) \right) |_{R}}
    \, \cos 2\,\vartheta \right] \, i \, \sin\chi \, \, e^{i\,\psi}
\end{align}
$k=1/\rlight$ is the wavenumber and $h^{(1)}_\ell$ are the spherical Hankel functions of order~$\ell$ satisfying the outgoing wave conditions, see for instance \cite{2005mmp..book.....A}. The induced electric field is then
\begin{align}
  \label{eq:DeutschE}
  E_r(\mathbf{r},t) & = \Omega \, B \, R \, 
  \left[ \left( \frac{2}{3} - \frac{R^2}{r^2} ( 3 \, \cos^2\vartheta - 1 ) \right) \
    \, \frac{R^2}{r^2} \, \cos\chi + 3 \, \sin\chi\, \sin 2\,\vartheta \, e^{i\,\psi}  \,
    \frac{R}{r} \, \frac{ h^{(1)}_2(k\,r)}
    {\frac{d}{dr} \left( r \, h^{(1)}_2(k\,r) \right) |_{R}} \right] \\
  E_\vartheta(\mathbf{r},t) & = \Omega \, B \, R \, 
  \left[ - \frac{R^4}{r^4} \sin 2\,\vartheta \, \cos\chi + \sin\chi\, e^{i\,\psi}  \, \left(
      \frac{R}{r} \, \frac{\frac{d}{dr} \left( r \, h^{(1)}_2(k\,r) \right)}
      {\frac{d}{dr} \left( r \, h^{(1)}_2(k\,r) \right)|_{R}} \, \cos 2\,\vartheta -
      \frac{h^{(1)}_1(k\,r)}{h^{(1)}_1(k\,R)} \right) \right] \\
  E_\varphi(\mathbf{r},t) & = \Omega \, B \, R \, \left[ \frac{R}{r} \, 
    \frac{\frac{d}{dr} \left( r \, h^{(1)}_2(k\,r) \right)}
    {\frac{d}{dr} \left( r \, h^{(1)}_2(k\,r) \right)|_{R}} -
    \frac{h^{(1)}_1(k\,r)}{h^{(1)}_1(k\,R)} \right] \, i \sin\chi \, \cos\vartheta \, e^{i\,\psi}
\end{align}
\end{subequations}
The physical solution is found by taking the real parts of each component, it encompasses a linear combination of the vacuum aligned dipole field and the vacuum orthogonal rotator with respective weights $\cos\chi$ and $\sin\chi$. To complete the solution for arbitrary stellar electrical charge, we add a monopolar electric field contribution due to the stellar surface charge such that
\begin{equation}
 E_r^{\rm mono} = \frac{Q_*-Q_{\rm c}}{4\,\pi\,\varepsilon_0\,r^2}
\end{equation} 
where $Q_*$ is the total electric charge of the star. This term compensates the $\cos\chi/r^2$ decrease of $E_r$ in eq.~(\ref{eq:DeutschE}). Deutsch solution separates space around a magnet into three distinct regions: the near or quasi-static zone where $r\ll \rlight$ and for which the above expressions reduce to the static oblique dipole eq.~(\ref{eq:ChampBExtDipolaire})-(\ref{eq:ChampEextQuadrupolaire}), the transition zone $r\approx \rlight$ and the wave zone $r\gg \rlight$ where the electromagnetic field resembles a transverse electromagnetic plane wave with an elliptical polarization, circular polarization along the rotation axis and linear polarization along the equatorial plane. An example of magnetic field lines in the equatorial plane is shown for the orthogonal rotator as red solid lines in fig.~\ref{fig:SolutionDeutsch}. The radial component of $(\mathbf{B,E})$ decreases like $1/r^2$ whereas the transverse component of $(\mathbf{B,E})$ decreases like $1/r$ typical for radiating fields in three dimensional space. To better catch the geometry of the field lines, let us focus on the perpendicular rotating dipole with $\chi=\pi/2$. In the asymptotic limit when $r\to+\infty$, in the equatorial plane we found a constant ratio 
\begin{equation}
 r \, \frac{B_r}{B_\varphi} = \textrm{cst} \ .
\end{equation}
As explained by \cite{1999PhR...318..227M} there are only two open field lines asymptoting to these Archimedean spirals. Their exact expressions at a fixed time are given in implicit form by
\begin{equation}
 r \, \frac{B_r(\varphi+(r-R)/\rlight)}{B_\varphi(\varphi+(r-R)/\rlight)} = -1
\end{equation}
to be solve for $\varphi$ with respect to $r$. The two solutions are shown as blue solid lines in fig.~\ref{fig:SolutionDeutsch} as a two-armed spiral. Asymptotically, this spiral coincides with the $B_r=0$ loci \citep{1980Ap&SS..67....3K}. \cite{1978Ap&SS..58..427K} gave approximate analytical solutions in the near and wave zone for an uniformly magnetized rotating dipole using a scalar and vector potential description instead of electric and magnetic fields. Subsequently \cite{1980Ap&SS..67....3K} improved the method and gave exact analytical expressions by using rigid-rotation, retardation and radiation operators applied to the static dipole. Then \cite{1981Ap&SS..74..333K} solved the so-called modified Deutsch problem that is, taking into account corotating plasma up to at most the light-cylinder without poloidal current but approximately including inertial effects which were fully treated by \cite{1982Ap&SS..82..441K}. A self-consistent description then required the presence of a disk in the corotation zone \citep{1983Ap&SS..92..113K}. The Deutsch vacuum solution can also be expressed in the corotating frame \citep{1973Ap&SS..24....3F}.

However the presence of plasma modifies that picture because charge acceleration in the magnetosphere leads to an electromagnetic activity detectable on Earth. This activity induces a multi-wavelength emission spectrum as suggested by \cite{1968Natur.218..731G} for neutron stars. The possible association between the Vela supernova remnant and its central pulsar was already discussed by \cite{1968Natur.220..340L}. The first model for an electromagnetically active neutron star was proposed by \cite{1967Natur.216..567P}. Then \cite{1968Natur.219..145P} claimed that a rotating neutron star was the source of energy feeding the Crab nebula with fresh particles and admitted that a strong magnetic field transmits rotational kinetic energy from the star to the nebula via production of high energy particles. From the work of \cite{1955AnAp...18....1D} he deduced the energy radiated by such a star and concluded that a strongly magnetized neutron star located at the centre of the Crab nebula was responsible for the luminosity of its nebula, which was in agreement with observations. This idea was proposed even before the discovery of the first pulsar! He envisaged the existence of a star possessing a purely dipolar magnetic field, its magnetic moment~$\mathbf{\mu}$ making an angle~$\chi$ with respect to the rotation axis. Rotation of magnetic dipole dragged by the star induces emission of a monochromatic electromagnetic wave at the star frequency~$\Omega$. The radiation has a dipolar pattern and its total intensity is given by $L = L^{\rm vac}_\perp \, \sin^2\chi$ where the luminosity of a perpendicular rotator is
\begin{equation}
  \label{eq:RayDipoleMag}
  L^{\rm vac}_\perp = \frac{8\,\pi \, B^2\,\Omega^4\,R^6}{3\,\mu_0\,c^3}
\end{equation}
$B$ is the magnetic field at the equator and $R$ the neutron star radius. A more general prescription for the spin down luminosity, valid in the presence of a plasma, would be to set $ L = f(\chi) \, L^{\rm vac}_\perp$. The function $f$ hides the precise microphysics inside the magnetosphere. We will come back to this point when discussing numerical simulations able to determine~$f$ depending on the plasma regime. In any case, this energy is not extracted from nuclear reactions nor from the collapse of the star. It is drained from the rotational kinetic energy reservoir containing a huge amount of energy estimated to be
\begin{equation}
 E_{\rm rot} = \frac{1}{2} \, I \, \Omega^2  = 2 \, \pi^2 \, I \, P^{-2} \approx \numprint{1.97e39} ~\si{\joule} \, \left( \frac{I}{\numprint{e38}~\si{\kilogram\metre\squared} } \right) \, \left( \frac{P}{1~\si{\second}} \right)^{-2}
\end{equation}
with $I$ the stellar moment of inertia, equal to $\frac{2}{5}\,M\,R^2$ for a homogeneous sphere. The power radiated exhausts this energy~$E_{\rm kin}$ and generates a luminosity following the relation
\begin{equation}
\label{eq:Luminosite}
  L_{\rm rot} = - \frac{dE_{\rm rot}}{dt} = -I\,\Omega\,\dot{\Omega} = 4 \, \pi^2 \, I \, \dot P \, P^{-3} \approx \numprint{3.95e24}~\si{\watt} \, \left( \frac{I}{\numprint{e38}~ \si{\kilogram\metre\squared} } \right) \, \left( \frac{\dot P}{\numprint{e-15}} \right) \, \left( \frac{P}{1~\si{\second}} \right)^{-3}
\end{equation}
with a typical spindown time scale of $\tau = P/2\,\dot P$ known as the characteristic age of the pulsar. A useful information about the brake efficiency is depicted by the braking index defined by
\begin{equation}
 n = \frac{\Omega \, \ddot\Omega}{\dot\Omega^2} \ .
\end{equation}
Without any a priori knowledge of the secular evolution of all pulsar parameters such as magnetic field~$B$, electric equivalent radius~$R_{\rm el}$, moment of inertia~$I$, inclination angle~$\chi$, the braking index according to vacuum magnetodipole losses is
\begin{equation}
\label{eq:IndiceFreinage}
 n = 3 + \frac{\Omega}{\dot\Omega} \, \left[ 2 \, \frac{\dot B}{B} + 2 \, \dot\chi \, \cot \chi + 6 \, \frac{\dot R_{\rm el}}{R_{\rm el}} - \frac{\dot I}{I} \right] \ .
\end{equation}
The electric equivalent radius~$R_{\rm el}$ is a fictive boundary of the star accounted for replenishing the corotating magnetosphere with plasma that from an electrical point of view is indistinguishable from the star. Such concept of radius was introduced by \cite{1997MNRAS.288.1049M} to account for spindown properties of the Crab pulsar.

Energy losses are accompanied by a torque exerted on the neutron star that brakes its rotation according to eq.~(\ref{eq:Luminosite}), thus applying a torque along the rotation axis~$\ez$ but also a torque in the perpendicular plane tending to align the magnetic moment with the rotation axis: the anomalous torque. In the vacuum solution, this happens following the integral of motion $\Omega(t)\,\cos\chi(t) = \textrm{cst}$ \citep{1970ApL.....5...21M, 1970ApJ...159L..81D} deduced from the spindown torque $\dot \Omega \propto \Omega^3 \, \sin^2 \chi$ and therefore a braking index (keeping other parameters constant in time) evolving in time according to
\begin{equation}
 n = 3 + 2 \, \cot^2 \chi(t) .
\end{equation}
For a filled magnetosphere, loss by a charged wind from the poles induces an increase of obliquity with a decrease of rotation rate because of the integral of motion $\Omega(t) \,\sin\chi(t) = \textrm{cst}$, see \cite{2015SSRv..191..207B}. Assuming a spindown like $\dot \Omega \propto \Omega^3 \, \cos^2 \chi$ the braking index now becomes
\begin{equation}
 n = 3 + 2 \, \tan^2 \chi(t)
\end{equation}
which also stays above $n=3$, conflicting with measurements of braking index for eight pulsars summarized in \cite{2015PhRvD..91f3007H}. However, the spindown torque obtained by \cite{2015SSRv..191..207B} seems to be based on an unphysical solution. \cite{1987ApJ...322..822M} demonstrated that the torque in realistic magnetospheres is always aligning because, independently of any details, open magnetic field lines always bent backward with respect to rotation. Moreover, as already pointed out by \cite{1972Ap&SS..19..249S}, the vacuum results should not straightforwardly transpose to the more realistic plasma filled magnetosphere. Indeed, the plasma filled magnetosphere evolution of the inclination angle offers another interpretation of braking index larger than~3 \citep{2016ApJ...823...34E}. In the same vain, \cite{2014MNRAS.441.1879P} accounted for plasma filled magnetospheres in the force-free and MHD limit contributing to the total torque and therefore to the subsequent obliquity evolution.

Applied to the Crab nebula, formula~(\ref{eq:Luminosite}) indicates that the pulsar furnishes a power of the order~\numprint{e31}~\si{\watt}, a value remarkably close to what the surrounding nebula radiates. Thus, it is the rotational braking of the pulsar that feeds the nebula with particles and energy. Such a braking needs a gigantic magnetic field estimated by equating the power lost by the neutron star eq.~(\ref{eq:Luminosite}) with the magnetodipole emission of an oblique rotator eq.~(\ref{eq:RayDipoleMag}) to obtain
\begin{equation}
 B = \sqrt{-\frac{3\,\mu_0\,c^3\,I\,\dot\Omega}{8\,\pi \,f(\chi) \, \Omega^3\,R^6}} = \sqrt{\frac{3\,\mu_0\,c^3\,I\,\dot P \, P}{32\,\pi^3 \,f(\chi) \,R^6}} \approx \frac{\numprint{1.01e8}~\si{\tesla}}{\sqrt{f(\chi)}} \, \left( \frac{I}{\numprint{e38}~\si{\kilogram\metre\squared} } \right) \, \left( \frac{\dot P}{10^{-15}} \right)^{-1/2} \, \left( \frac{P}{1~\si{\second}} \right)^{-1/2}.
\end{equation}
For the Crab this gives about~\numprint{e8}~\si{\tesla}. \cite{1969ApJ...157.1395O} were the first to envisage such magnetic field strengths. \cite{1969PhRvL..22..728G} have also investigated the acceleration of particles to very high energy pushed by such large amplitude low frequency electromagnetic waves. This intensity of the field was confirmed by the synchrotron spectra of the pulsar. However, this model does not explain the origin of the pulsed radio emission because it does not describe how to produce and accelerate particles, the magnetosphere being empty. Noticing that radiation needs particle acceleration it became quickly clear that the magnetosphere could not remain empty. Several scenarios have therefore been proposed. \cite{1968Natur.218..731G} explained radio emission by a conglomerate of electrons in corotation with the star. This idea of formation of a bunch of electrons responsible for the coherent emission has then been invoked many times in recent models.

\cite{1969ApJ...157..869G} examined in details the aligned rotator. They noticed that an empty magnetosphere cannot last for a reasonable time because of strong electric fields induced by rotation of the magnetic moment, pulling particles out from the surface and dragging them in corotation with the star up to the light cylinder. Farther away a wind is formed, made of charged particles. The polar caps represent therefore a first choice region to explain radio emission. It is strictly speaking not a model for real pulsars because no pulsation is predicted for an aligned configuration assuming axisymmetry. Never mind, \cite{1969PASAu...1..227G} stipulated that the physics of an oblique rotator should not be very different from that of an aligned rotator. The very popular hollow cone model was born \citep{1969ApL.....3..225R, 1975ApJ...196...51R}. Although an aligned rotator requires less effort because of axisymmetry, \cite{1971NPhS..233..149M} recognized that an oblique rotator could deviate significantly from the aligned case leading to secular evolution of the pulsar geometry by for instance precession.

\cite{1970Natur.227..465S, 1971ApJ...164..529S} introduced the first real model for pulsars by injection of particles at the polar caps. These primary particles emit gamma-ray photons through curvature radiation, photons that in turn disintegrate into secondary electron/positron pairs. A cascade develops and the charged flow is controlled by this space charge. The coherence of the emission is provided by bunches of electrons and positrons circulating in opposite direction. Later on even photohadronic pair production in the pulsar magnetosphere were considered by \cite{1979ApJ...228..536J}.

\cite{1975ApJ...196...51R} improved the model of \cite{1970Natur.227..465S} by introducing the discharge and drifting subpulses phenomena. These models require polar caps as sources of relativistic particles. The sign of the charge available on these caps depends on the scalar product~$\mathbf \Omega \cdot \mathbf B$ deduced from eq.~(\ref{eq:DensiteCorotation}), thus having sometimes electrons sometimes ions present on the surface, in other words two classes of pulsars. Such segregation was never observed, no such distinction should be expected.

\cite{1972ARA&A..10..427R} gave an early review on pulsars known at that time. Simplifying analytical treatment without sacrificing essential physics is always a good idea. Indeed \cite{1973Ap&SS..24..289M, 1982MNRAS.199..211D} and \cite{1983MNRAS.204.1125D} made attempts to model pulsar electrodynamics in 2D cylindrical coordinates that is invariant under translation along the $z$-axis, to get better physical insight without dealing with the full 3D complexity but keeping the important non axisymmetric property. Such approach pioneered by \cite{1976MNRAS.175..257M} and took over by \cite{1979AuJPh..32..681B} to investigate particle inertia effects was however never pursued later.

Particle acceleration in a two-fluid plasma was discussed for an aligned rotator by \cite{1974ApJ...193..217S} and \cite{1975ApJ...201..431H} and extended to an oblique geometry by \cite{1975ApJ...201..719H}.

On an experimental side, only a handful of laboratory experiments have been performed to study neutron star magnetosphere among them the Terella by Birkeland beginning of the 20th century \citep{birkeland1908norwegian} to study polar aurora in gas-discharge experiments and more recently the one by \cite{1979AuJPh..32...71E}.

\subsection{General picture}

Although all models are based on fundamental ideas to explain radio emission, the theory is inconsistent and does not solve the question of the global circuit for the electric current and charge loading. How do charges circulate within this magnetosphere? Moreover, the magnetic field in the nebula remains to intense to be only a relic of the explosion and the presence of relativistic particles indubitably reveals that the source must come from the central pulsar.

As we saw, rotation of the neutron star combined to the strong magnetic field produce avalanches of electron/positron pairs. Vacuum solutions are not stable. The magnetosphere is necessarily filled with at least leptons maybe also protons and/or ions. To first approximation, plasma effects should screen the longitudinal electric field, that is the component of $\mathbf E$ along magnetic field lines should vanish, $E_\parallel =0$, meanwhile cancelling any acceleration of particles. If this were not the case, charges would be immediately accelerated towards appropriate regions to cancel this electric field component. Screening implies an abundance of electron/positron pairs not restricted by any microphysics but only by the requirement to cancel the $E_\parallel$ component. However, exact electric field screening in the polar caps has been challenged by \cite{1998MNRAS.295L..53S, 2002MNRAS.336..233S} who solved Poisson equation in the gap. The acceleration time is very short with respect to the period, about $\tau_B = 1/\omega_B \approx$~\numprint{e-20}~\si{\second}. A contradiction appears already at this point. Indeed, we required to have plasma flowing along field lines to produce multi-wavelength radiation but if these are not accelerated how should they radiate? In a strong magnetic field, particles are restricted to stay in their fundamental Landau level. Indeed the energy levels are quantized in the plane perpendicular to magnetic field lines according to
\begin{equation}
 E_n = \sqrt{ \left( 1 + \left(2\,n+2\,s+1 \right)\,\frac{B}{B_{\rm c}} \right) \, \masselec^2\,c^4 + p_\parallel^2\,c^2 }
\end{equation}
where $n$ represents the quantum number characterising the excitation degree of the level and $s=\pm1/2$ symbolise the electron spin \citep{1978PhRvD..18.1053D}. All energy levels are degenerated with an arbitrary choice of the spin except for the fundamental level~$n=0$ for which $s=-1/2$. Actually, degeneracy is lifted through higher order interactions between particles and the radiation field \citep{1982A&A...115...90H, 1991ApJ...380..541P}. However particles are free to move along magnetic field lines and need a parallel component of the electric field $E_\parallel\neq0$ in order to accelerate and radiate. 

\begin{figure}
 \centering
\begin{tikzpicture}
\begin{scope}
\clip[scale=1] (-2,-3.5) rectangle (2,3);
\shade[inner color=white, outer color=yellow] (-4,-4) rectangle (4,4);
\fill[green] (0,0) -- (2,0.74) -- (2,-2.5) -- cycle;
\fill[green] (0,0) -- (-2,2.5) -- (-2,-0.73) -- cycle;
\draw [black] plot [samples=300, smooth, domain=0:360]
  (xy polar cs:angle=\x, radius= {3*sin(\x-60)*sin(\x-60)});
\draw [black] plot [samples=300, smooth, domain=0:360]
  (xy polar cs:angle=\x, radius= {7*sin(\x-60)*sin(\x-60)});
\draw [black] plot [samples=300, smooth, domain=0:360]
  (xy polar cs:angle=\x, radius= {20*sin(\x-60)*sin(\x-60)});
\shade[inner color=yellow, outer color=orange, rotate=60] (-0.4,-0.4) rectangle (0.4,0.4);
\shade[inner color=red, outer color=blue] plot [samples=300, smooth, domain=0:360]
  (xy polar cs:angle=\x, radius= {2.3*sin(\x-60)*sin(\x-60)});
\shade[outer color=gray, inner color=white] plot [samples=300, smooth, domain=0:360]
  (xy polar cs:angle=\x, radius= {2.1*sin(\x-60)*sin(\x-60)});
\draw [black] plot [samples=300, smooth, domain=0:360]
  (xy polar cs:angle=\x, radius= {1.5*sin(\x-60)*sin(\x-60)});
\draw[thick, green, ->] (0,-2) -- (0,2) node [above] {$\Omega$}; 
\draw[rotate=-30, thick, red, ->] (0,-2) -- (0,2) node [above right] {$\mu$}; 
\draw[rotate=-70, thick] (0,-4) -- (0,4) ; 
\draw[rotate=39, thick] (0,-4) -- (0,4) ; 
\draw[thick, black] (2,-4) -- (2,4) ; 
\draw (1.7,-2.7) node [below] {$\rlight$} ;
\draw[thick, black] (-2,-4) -- (-2,4) ; 
\filldraw[inner color=white,outer color=black] (0,0) circle (0.2);
\end{scope}
\draw[black,<-,thick] (0.125,0.25) -- (4,2) node [right,text=orange] {polar cap} ;
\draw[black,<-,thick] (1.875,-0.25) -- (4,0) node [right,text=blue] {slot gap} ;
\draw[black,<-,thick] (1.875,0.325) -- (4,1) node [right,text=green] {outer gap} ;
\draw[black,<-,thick] (1.,-0.5) -- (4,-1) node [right,text=gray] {closed magnetosphere} ;
\draw[black,<-,thick] (0.0,-2.5) -- (4,-2) node [right,text=yellow] {base of the wind (open field lines)} ;
\end{tikzpicture}
 \caption{Schematic view of the magnetosphere within the light-cylinder. Sizes of the gaps are not to scale.}
 \label{fig:Magnetosphere}
\end{figure}
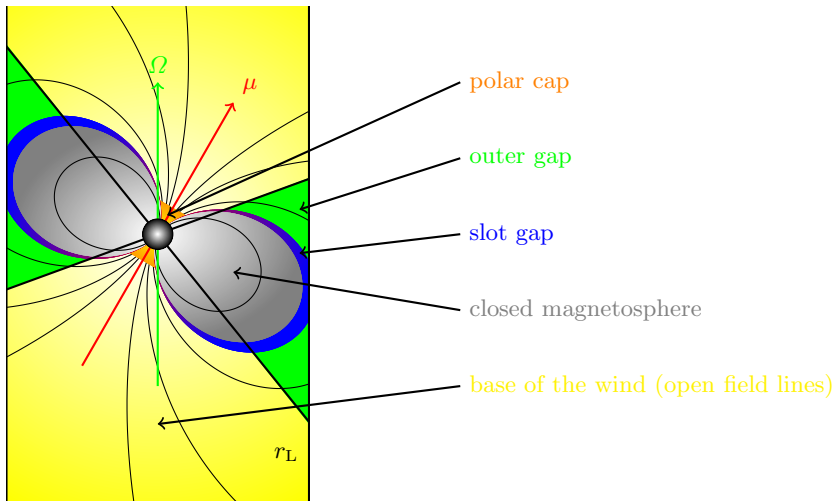

Clearly $E_\parallel=0$ should not hold everywhere in the magnetosphere. We will come back to that point later when discussing possible gaps in the magnetosphere. As emphasized by \cite{1997MNRAS.287..262S}, the determination of the accelerating electric field in the vacuum gaps should be treated as a global problem including the current circuit flowing in the magnetosphere as he did earlier in \cite{1991ApJ...378..239S}. Particle acceleration cannot be studied locally with special boundary conditions but consistently with the large scale plasma configuration. Nevertheless, let us summarized the essential features of a pulsar magnetosphere so far
\begin{itemize}
 \item a plasma corotating with the star in the ideal MHD or even force-free approximation: solid body rotation dictated by the star and free motion along field lines.
\item a light cylinder: corotation stops outside this cylinder of radius~$\rlight=c/\Omega$. Particle inertia becomes important, magnetic field lines are significantly deformed and swept back by this mass load.
\item a magnetic topology with open and closed field lines. Closed field lines are imprisoned inside the light cylinder, plasma is corotating, no motion along field lines is permitted. Open field lines cross the surface of the light cylinder and their foot are anchored in the polar caps. Particles escape freely to infinity along these field lines.
\item a light surface: surface where the intensity of the electric field become equal to that of the magnetic field, $E=c\,B$. The electric drift approximation is violated, particles suffer acceleration, the ideal MHD or force-free approximation breaks down. The light surface and the light cylinder do not coincide, the first surface could be rejected to infinity for sufficiently strong longitudinal currents.
\item polar caps: regions around the magnetic poles where open field lines are attached to, deviation from force-free is expected to produce radio emission \citep{1971ApJ...164..529S, 1975ApJ...196...51R}.
\item slot gaps: small elongated excision volumes along the last closed field line within the magnetosphere, essentially empty of charges allowing pair creation \citep{1979ApJ...231..854A, 1983ApJ...266..215A}, emergence of high energy radiation and acceleration of particles \citep{2003ApJ...598.1201D}.
\item for the aligned rotator, in the equatorial plane, transition from closed to open field lines goes through a so-called Y-type neutral point (for short Y-point) at a radius $R_{\rm Y}$. It is generally assumed that $R_{\rm Y}=\rlight$ but more generally it should satisfy $R\leqslant R_{\rm Y}\leqslant\rlight$. For instance \cite{1971ApJ...169L...7S} used the prescription $R_{\rm Y} = R^{1-\eta} \, \rlight^\eta$ with $\eta\in[0,1]$.
\item outer gaps: large almost empty volumes in the magnetosphere, between the null surface (where $\rho=0$) and the last closed field line with copious pair creation via $\gamma + \gamma \rightarrow e^+ + e^-$ \citep{1986ApJ...300..500C, 1986ApJ...300..522C, 1995ApJ...438..314R}. Synchrotron emission from these gaps were studied by \cite{2001ApJ...546..401C}.
\item annular gaps: region between the critical field line and the null surface \citep{2004ApJ...606L..49Q}.
\end{itemize}
The scheme of fig.~\ref{fig:Magnetosphere} is an illustration of some important quantities introduced above. Possible finite temperature of the plasma is not accounted for but thermally supported hot magnetospheres were suggested by \cite{1974MNRAS.166..409H}. Also the situation outside the light-cylinder is quite different from the regime inside it. Indeed the pattern of charges and current distribution present outside the light-cylinder are superluminal even if the particles themselves remain subluminal. Such motions generate radiation qualified as Schott radiation by \cite{1985MNRAS.215..701D} and to be distinguished from Cerenkov radiation. A analogy with Cerenkov emission was nevertheless put forward by \cite{1981ApJ...251..674A}. This flow outside the light-cylinder will be discussed in the pulsar wind theory sec.~\ref{sec:Vent}.

In a series of papers by \cite{1976MNRAS.175..645A, 1976ApJ...203..226A, 1976ApJ...206..822A, 1976ApJ...204..889A, 1976MNRAS.177..661A} it was claimed that the transition between the corotating magnetosphere and the wind should go through a shock discontinuity and not via a continuous MHD flow. Singular surfaces in the magnetosphere were also found by \cite{1976MNRAS.177..415B}.

\section{Theory of pulsar magnetospheres} 
\label{sec:Magnetosphere}

Establishing a consistent model of pulsar physics requires an accurate and quantitative description of the magnetospheric structure, the dynamics and radiative outputs, that is, the magnetic field topology, the current flowing inside and outside the light-cylinder and particle acceleration mechanisms. Such a study in the general case is very difficult to conduct. Simple situations are instead treated but keeping the problem interesting from a physical point of view. The hypotheses usually accepted are the following
\begin{itemize}
\item the magnetosphere is filled with a pair plasma screening the electric field such that~$\mathbf{E}\cdot\mathbf{B}=0$ everywhere. This means that all charged particles adapt their motion to maintain a vanishing acceleration along field lines, thus $E_\parallel=0$. Spatially localized slight deviations from this rigorous $E_\parallel=0$ fulfilment are expected to ignite electromagnetic activity in the magnetosphere. Subleties in achieving $E_\parallel\neq0$ lead to different plasma regimes involving a plethora of gap and cap models.
\item particles follow an electric drift motion superposed to a translation along field lines.
\item the regime is stationary and at least for earlier models assumed axisymmetric (aligned rotator).
\item primary particles emanate from the surface of the star, there is no pair creation.
\item the plasma is quasi-neutral, which means that the space charge is overwhelmed by a background much more dense neutral plasma.
\item sometimes the opposite is claimed that is a plasma entirely charge separated, in other word a truly non neutral plasma.
\item gravity and pressure (temperature) forces are neglected compared to electromagnetic forces.
\end{itemize}
Let us explain in more details important implications of all these assumptions.

\subsection{Filled magnetospheric model}

An aligned rotator in vacuum does not radiate because dipolar magnetic emission cancels for zero obliquity, $\chi=0$. But if plasma cohabits within the magnetosphere, the current generated by the plasma motion induces a braking of the star through torques exerted on the stellar crust. This idea was formulated by \cite{1969ApJ...157..869G}. But where does this plasma come from? At first sight, the gravitational field is sufficiently intense to retain particles at its surface but nevertheless this hypothesis is wrong. Indeed, the strong magnetic field combined to the rotation of the star generates a potential drop at the stellar surface hardly sustainable for the charges in the crust. The electric field component aligned with the magnetic field, of the order of $E_\parallel \approx$\numprint{e10}~\si{\volt/\meter} is able to pull them out. The discontinuity of $E_{\rm r}$ when crossing the surface/vacuum interface provokes a surface charge density constrained to spread over vacuum because of $E_\parallel \neq 0$. Comparing the coulombian force to the gravitational attraction for a proton, we estimate the ratio
\begin{equation}
\label{eq:ForceGravElec}
 \frac{f_{\rm elmag}}{f_{\rm grav}} = \frac{e\,E_\parallel\,R^2}{G\,M\,\massproton} \approx \numprint{e9}
\end{equation}
and a value $\masselec/\massproton\approx\numprint{2000}$ larger for electrons. The gravitational force is completely negligible. The vacuum around the star is unstable and must replenish with charges.

\cite{1969ApJ...157..869G} supposed that the electromagnetic environment of the magnetosphere is described by a plasma corotating with the star up to the light cylinder and magnetic field lines with no toroidal component. A fundamental difference exists between closed and open field lines. On the latter, a current circulation is launched from the polar caps, regions with cone opening angle given by simple geometrical arguments in equation~(\ref{eq:RayonCalottePolaire}). Unfortunately, estimation of the energy loss by magnetodipole radiation furnishes the same order of magnitude as the Deutsch-Pacini model. It is therefore impossible to assess which of both models, empty or fully filled magnetosphere, is really pertinent for pulsars.

The light cylinder is an imaginary cylindrical surface whose axis is parallel to the rotation axis of the star and possesses a radius corresponding to a distance to the centre of the neutron star for which the corotation speed reaches the speed of light. The radius of the light cylinder is thus defined by 
\begin{equation}
  \label{eq:RayonCylindreLumiere}
  \rlight = \frac{c}{\Omega} \ .
\end{equation}
Physically, corotation is insured by the drift motion of particles in the electromagnetic field at the electric drift velocity given by
\begin{equation}
 \mathbf{v}_{\rm Edrift} = \frac{\mathbf{E} \wedge \mathbf{B}}{B^2}
\end{equation}
It does not depend on the nature of the particles (mass, charge) but uniquely on the structure of the electromagnetic field $(\mathbf E, \mathbf B)$. Therefore this electric drift cannot induce any current except if there is a deviation from charge neutrality whereby a convective current exists. This drift does not forbid motion along magnetic field lines. Indeed, for a perfectly conducting plasma, in the ideal MHD regime, the drift speed according to eq.~(\ref{eq:Eprime}) becomes
\begin{equation}
\mathbf{v}_{\rm Edrift} = \mathbf{\Omega} \wedge \mathbf{r} - \frac{\mathbf{B} \cdot ( \mathbf{\Omega} \wedge \mathbf{r})}{B^2} \, \mathbf{B}
\end{equation}
which clearly indicates contribution from corotation recognizable in the first term on the right hand side, plus a sliding along field lines recognizable in the second term on the right hand side. In order to avoid exceeding the speed of light, field lines have to bent to induce a toroidal component~$B_\varphi\neq0$. This points out the dichotomy between closed and open field lines. The fundamental problem of pulsar electrodynamics was to find a reasonable expression for this parallel current. Numerical simulations have been able to answer satisfactorily this question as we discuss in Sec.~\ref{sec:Simulations}. \cite{1984MNRAS.209..285A} considered interesting alternative models carefully by studying the force-free surfaces. They tempted also to include vacuum gaps between the neutron star and the force-free regions as well as particle exchange via charged polar beams. \cite{2012Ap&SS.342...79L} noticed that the current required to flow within the magnetosphere does not necessarily match the pair production rate and its flow within the polar caps. The mismatch could induce modulation of radio emission.

\subsection{Ideal MHD and force-free limit}

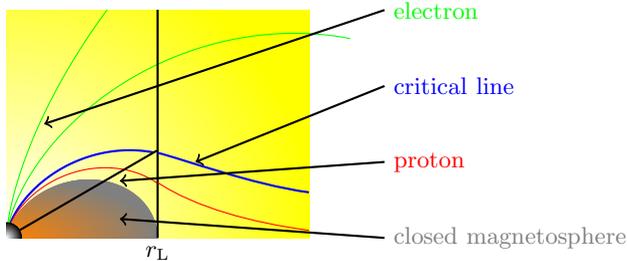
\begin{figure}
 \centering
\begin{tikzpicture}
\begin{scope}
\clip[scale=1] (0,0) rectangle (5,3);
\shade[inner color=white, outer color=yellow] (-4,-4) rectangle (4,4);
\draw [green] plot [samples=300, smooth, domain=0:60]
  (xy polar cs:angle=\x, radius= {1*sin(\x-90)*sin(\x-90)});
\draw [blue, thick] plot [samples=300, smooth, domain=30:90]
  (xy polar cs:angle=\x, radius= {3*sin(\x-90)*sin(\x-90)});
\draw [green] plot [samples=300, smooth, domain=30:90]
  (xy polar cs:angle=\x, radius= {7*sin(\x-90)*sin(\x-90)});
\draw [green] plot [samples=300, smooth, domain=60:90]
  (xy polar cs:angle=\x, radius= {20*sin(\x-90)*sin(\x-90)});
\draw [red] plot [samples=300, smooth, domain=20:90]
  (xy polar cs:angle=\x, radius= {2.4*sin(\x-90)*sin(\x-90)});
\shade[inner color=orange, outer color=gray] plot [samples=300, smooth, domain=0:360]
  (xy polar cs:angle=\x, radius= {2.*sin(\x-90)*sin(\x-90)});
\draw[rotate=-60, thick,black] (0,0) -- (0,2.3) ; 
\draw[thick, black] (2,0) -- (2,4) ; 
\filldraw[inner color=white,outer color=black] (0,0) circle (0.2);
\draw [blue,thick] (2,1.12)  .. controls (2.3,1.05)  and (3.25,0.7) ..  (4,0.6)  ; 
\draw [red] (2,0.73)  .. controls (2.5,0.4)  and (3.25,0.2) ..  (4,0.1)  ; 
\end{scope}
\draw (2,-0.0) node [below] {$\rlight$} ;
\draw[black,<-,thick] (0.5,1.5) -- (5,3) node [right,text=green] {electron} ;
\draw[black,<-,thick] (2.5,1) -- (5,2) node [right,text=blue] {critical line} ;
\draw[black,<-,thick] (1.5,0.75) -- (5,1) node [right,text=red] {proton} ;
\draw[black,<-,thick] (1.5,0.25) -- (5,0) node [right,text=gray] {closed magnetosphere} ;
\end{tikzpicture}
 \caption{First model of a pulsar magnetosphere as proposed by \cite{1969ApJ...157..869G}. The neutron star is symbolized by a circle on bottom left. The open field lines let a charged wind escape from the poles. The closed field lines are filled with the corotation density and do not support any electric current.}
 \label{fig:ModeleGJ}
\end{figure}

In this most studied approximation, the magnetosphere is sufficiently populated with plasma in order for the conductivity in the medium to become infinite or in other words that all component of the electric field parallel to the magnetic field to be immediately screened, $E_\parallel=0$. Moreover, the electromagnetic field dominates the dynamics of the magnetosphere to several orders of magnitude with respect to pressure, gravity and inertia. The Lorentz force on a plasma element, treated as a one component fluid, is therefore null. Its vanishing leads to the so-called force-free approximation
\begin{equation}
 \label{eq:force_free}
 \rho_{\rm e} \, \mathbf E + \mathbf j \wedge \mathbf B = \mathbf 0 
\end{equation}
where $\rho_{\rm e}$ is the charge density and $\mathbf j$ the current density. The electric field is orthogonal to the magnetic field $\mathbf E \cdot \mathbf B=0$. Implicitly magnetic energy density $\frac{B^2}{2\,\mu_0}$ dominates against any other kind of energy and notably the one related to particles inertia. This corresponds to the vanishing mass limit. Moreover no dissipation is associated with this regime, ideal MHD applies to the flow of velocity field $\mathbf{v}$ and 
\begin{equation}
 \label{eq:MHD_ideale}
 \mathbf E + \mathbf v \wedge \mathbf B = \mathbf 0 \ .
\end{equation}
From eq.~(\ref{eq:force_free}) and eq.~(\ref{eq:MHD_ideale}) we deduce that the current density is made of a convective term related to charge separation~$\rho_{\rm e} \, \mathbf v$ and to a field aligned current~$j_\parallel$ thus
\begin{equation}
 \mathbf j = \rho_{\rm e} \, \mathbf v + j_\parallel \, \mathbf{B} / B .
\end{equation}
\cite{1969ApJ...157..869G} postulated simply that the magnetosphere was entirely filled up to the light cylinder, fig.~\ref{fig:ModeleGJ}. The magnetic field~$\mathbf{B}$ aligned with the magnetic moment~$\mathbf{\mu}$ and rotating at the angular speed~$\mathbf{\Omega}$, generates an electromotive field from which forces are sufficient to overcome gravity, creating a magnetosphere filled with plasma ejected from the surface of the star. In the aligned case evoked here, magnetic dipolar emission as suggested by Pacini disappears, there is no more braking through magnetodipole radiation but through acceleration of charges in the magnetosphere as explained later. If we assume that the reservoir of particles is infinite, the magnetosphere will be entirely saturated with ions and electrons up to the light cylinder with a density of charge insuring corotation of the whole system according to Maxwell-Gauss equation
\begin{equation}
  \label{DensiteCharge}
  \rho_{\rm e} = \varepsilon_0\, \divg\, \mathbf{E} = \rho_{\rm cor} + \mathbf{j} \cdot (\mathbf{\Omega} \wedge\mathbf{r})/c^2 \ .
\end{equation}
The corotation charge density is given by $\rho_{\rm cor} = -2\,\varepsilon_0\, \mathbf{\Omega}\cdot\mathbf{B}$ with the associated particle density number $n_{\rm cor} = \rho_{\rm cor}/e$. $\rho_{\rm cor}$ is the density required to screen the longitudinal electric field. If the current density is purely corotating then $\mathbf{j} = \rho_{\rm e} \, \mathbf{\Omega} \wedge\mathbf{r}$ and the density simplifies into
\begin{equation}
  \label{eq:DensiteCorot}
  \rho_{\rm e} = \frac{\rho_{\rm cor}}{1 - (\mathbf{\Omega} \wedge\mathbf{r})^2/c^2} \ .
\end{equation}
In that case, the density diverges at the light-cylinder unless $\mathbf{\Omega} \cdot \mathbf{B}=0$ there. The MHD or force-free approximation requires a particle density number much larger that the minimum required by the corotation, that is $n\gg n_{\rm cor}$ to insure almost perfect charge neutrality. Thus if pair creation is ineffective, such high densities could not be reached and the neutral fluid regime should be replaced by a non neutral plasma behaviour. The denominator of eq.~(\ref{eq:DensiteCorot}) brings in a relativistic correction in $(1-r^2\,\sin^2\vartheta/\rlight^2)$ due to the magnetospheric currents modifying the structure of the magnetic field, phenomenon very perceptible in the vicinity of the light cylinder. Indeed, corotation of the magnetospheric charge with the pulsar generates an electric current $\mathbf{j} = \rho_{\rm e} \, \mathbf{v}$ modifying through its effects the initial configuration of the magnetic field. This self-consistent current leads to more important effects when approaching the light cylinder. It is responsible for certain relativistic effects in particular the determination of the corotation density. The magnetic perturbations induced by these corotating currents have a tendency to repel field lines in a direction opposite to the pulsar (plasma diamagnetic effect). These far away field lines inflate to infinity until they open up. Mass loading causes field lines to sweep back thus generating a longitudinal current $j_\parallel$ that is difficult to estimate solely on ground of first principles.

The magnetosphere then splits into two regions, one with closed field lines and the other with open field lines. Both kind of field lines are in solid body rotation. Brought back to the level of the stellar surface, open field lines focus into a small zone in the vicinity of the magnetic poles, the polar caps which have a radius not larger than
\begin{equation}
  \label{eq:RayonCalottePolaire}
  r_{\rm cp} = R \, \vartheta_{\rm cp} = R \, \arcsin \sqrt{\frac{R}{\rlight}} \approx R \, \sqrt{\frac{R}{\rlight}}
\end{equation}
assuming vacuum dipolar field lines whose polar equation is $r=\lambda\,\sin^2\vartheta$. These estimates do not include distortion due to either retardation effects around the light-cylinder and already present in the Deutsch solution or magnetospheric currents. The region enclosed inside the light cylinder is entirely filled with plasma at the corotation density~$\rho_{\rm e}$. If the intrinsic magnetic field is dipolar, positively charged regions are separated from negatively charged regions by a conical interface with opening angle defined by the condition~$\mathbf{\Omega} \cdot \mathbf{B}=0$ according to eq.~(\ref{eq:DensiteCorot}). Some magnetic field lines enclose simultaneously charges of both signs, which rise the question of the existence of a process able to explain how this transport can be produced. Open magnetic field lines going beyond the light cylinder let particles definitely escape from the pulsar contributing to the total electric current. They are divided into electron supported flow and proton supported flow, delimited by the critical field line which is at the same electric potential as the interstellar medium. The star loses then charges from the polar caps through the formation of a charged wind which is a situation that cannot last for ever. The power released by these escaping particles is comparable to the power radiated by the magnetic dipole of \cite{1968Natur.219..145P}. The characteristic braking time scale and pulsar age will then remain the same. Quantitative results will be discussed in the paragraph about numerical simulations in sec.~\ref{sec:Simulations}. Note that in some versions of \cite{1969ApJ...157..869G} model, electron-positron pairs are formed during the period of magnetosphere filling. Positive charges are then made of positrons.

Although being able to explain the origin of particles, this model suffers from internal inconsistency problems bound to the endless discharge of the pulsar and to the thorny issue of the current closure. Moreover, \cite{2001MNRAS.322..209S} have demonstrated through numerical simulations that this model of magnetosphere entirely filled with corotating plasma is unstable. They observed a collapse to a new charge distribution similar to the one obtained by \cite{1985A&A...144...72K}, see later the section~\ref{sec:Electrosphere} discussing about the electrosphere. 
We remind that the electric field produced in vacuum by a rotating star is known since the work by \cite{1947PhRv...72..632D} and for the oblique rotator filled with plasma since~\cite{1965JGR....70.4951H} thus well before \cite{1969ApJ...157..869G}.
Whether the force-free solution can strictly apply outside the light-cylinder or not was questioned by \cite{1978MNRAS.183..771B} who showed that a small parallel electric field must exist in order to allow for a finite speed of particle along field lines.

\subsection{The pulsar equation}


The current flowing along magnetic field lines in the magnetosphere constitutes an inescapable unknown to the dynamics of pulsars. In order to determine it self-consistently, the problem has to be solved from the surface of the star up to infinity. This task is very arduous but real progresses have been made the last decade thanks to numerical simulations. But before, let us remind the main approaches before this new era of informatics. \cite{1973ApJ...180..207M} was the first to compute the exact structure of a 2D axisymmetric magnetosphere in the absence of field aligned current, $j_\parallel=0$. The fundamental equation for these corotating field lines was given through the magnetic flux function~$\psi$ related to the magnetic field by~$\mathbf{B} = \frac{1}{r} \, \mathbf{\nabla} \psi \wedge \mathbf{e}_\varphi$. The magnetic flux function~$\psi$, in presence of a longitudinal current satisfies a relation established independently by \cite{1973ApJ...182..951S} and by \cite{1973ApJ...180L.133M} see also \cite{1973ApJ...183..967J}, it is written as
\begin{equation}
  \label{eq:EquationPulsar}
  \frac{\partial^2 \psi}{\partial r^2} + \frac{\partial^2 \psi}{\partial z^2}
-\frac{1}{r}\frac{\rlight^2+r^2}{\rlight^2-r^2} \,\frac{\partial \psi}{\partial r}
= -\frac{A(\psi)\,A'(\psi)}{\rlight^2-r^2} .
\end{equation}
It is often named the pulsar equation. \cite{1974ApJ...187..359E} gave another derivation of the pulsar equation and made some useful comments about the underlying hypothesis. The function~$A(\psi)$ is a priori arbitrary, but it must verify some alfvenic regularity conditions at the light cylinder. It is related to the poloidal current~$I$ by~$\mu_0\,I=2\,\pi\,A$. The singularity at~$r=\rlight$ imposes a strong constraint on the function~$A$ that must satisfy the regularity condition~$2\,\rlight\,\frac{\partial \psi}{\partial r}=A\,A'$. In the absence of longitudinal currents $A=0$ and far away from the light cylinder, $r\ll\rlight$, the multipolar expansion of the field in vacuum is retrieved. Note also that this equation is singular on the light cylinder~$r=\rlight$. It can be shown that these field lines are perpendicular to the light cylinder. This leads to a very important physical conclusion: the Poynting vector does not possess a component normal to the light cylinder which means that the electromagnetic energy flux through the light cylinder vanishes. In the absence of longitudinal current, the plasma filling the magnetosphere screens the dipolar field, no magnetodipole emission is allowed. The energy loss of the pulsar cannot come from the action of a current circulating in the interior of its magnetosphere without crossing the light cylinder. Moreover, the solution exterior to the light cylinder has no influence on the interior solution. Solving equation~(\ref{eq:EquationPulsar}) for the flux function~$\psi$, \cite{1973ApJ...180..207M} obtained the shape of the magnetic field interior to the light cylinder. 
The plasma has a tendency to deform field lines in the direction of an increase of total magnetic flux extending to the light cylinder, magnetic energy in the vicinity of the cylinder is also increased. \cite{1979MNRAS.188..385M} and \cite{1979MNRAS.188..799M} extended this model by adding small gaps between ions and electrons along the null surface. Perturbations of this null surface by for instance charge depletion in the charge separated plasma is unstable against vacuum gap formation in its vicinity. Following arguments detailed by \cite{1973NPhS..246....6H}, replenishment is forbidden. Later \cite{1981MNRAS.194...95H} studied the properties of vacuum gaps with finite temperature plasmas. A current flows out of the null surface where replenishment is impossible but acceleration of particles to very high energies is expected in the huge potential drop limited by pair production \citep{1976ApJ...203..209C}. \cite{1974MNRAS.167..457O, 1975MNRAS.170...81O} suppressed the hypothesis of corotation introduced by \cite{1973ApJ...180..207M} and computed the magnetic field configuration in such a situation. \cite{1973ApJ...182..951S} introduced particles inertia but assuming that it remains small and without giving exact solutions. Inertial effects were also the topic of \cite{1979MNRAS.189..709S} who presented first results in \cite{1980MNRAS.192..409S}. The problem of an oblique rotator has not been studied. Let us cite the work by \cite{1999MNRAS.309..388M} who determined the pulsar magnetosphere in the case of a perpendicular rotator. The current formed by the particle flow but also the displacement current act to distort field lines. 

Progress were accomplished by \cite{1999ApJ...511..351C} who managed to treat numerically the regular singularity along the light cylinder. According to their results, it seems that only one such function~$A$ exists for which the solution crossing the light cylinder possess no discontinuity. However other solutions have been found by \cite{2006MNRAS.368.1055T} if the singular point is located inside the light-cylinder, translating the Y-point~$R_{\rm Y}$ as proposed by \cite{1971ApJ...169L...7S}. Equating the force balance between the Y-point and the centrifugal force \cite{1972ApJ...173L..33R} found a braking index of $n=7/3$ for the Crab in agreement with observation at that time and also in agreement with the period-pulse width relation. Closed field lines do not necessarily stop at the light-cylinder but maybe already well within it, at the so called force-balance radius where gravitational and centrifugal forces compensate each other \citep{1973ApJ...181..161R}. As pulsars spin down, the Y-point moves outwards at a rate depending on reconnection efficiency, the two extreme cases being on one hand no shift thus $R_{\rm Y}=cst$ and on the other hand very efficient readjustment of the magnetic topology leading to $R_{\rm Y}=\rlight$. This could have interesting implications for the death line of pulsars \citep{2006ApJ...643.1139C}. Where ever the location of the Y-point, the circuit must be closed by a return current. The path taken by this return current may be along the last open field lines, the so-called separatrix but not necessarily. Indeed the dynamics of this Y-point, even only described locally, is still delicate and controversial. In addition, it is not clear how it influences the global structure of the magnetosphere \citep{2003ApJ...598..446U}. The solution found by \cite{1999ApJ...511..351C} is not unique. The delicate point concerns the current sheet in the equatorial plane that was introduced so to say by hand in order to provide the current closure of the electric circuit. In an attempt to remove this arbitrariness of the current sheet, \cite{2014ApJ...781...46C} constructed another force-free solution for the axisymmetric rotator that takes off the separatrix. The current sheet only exists outside the light-cylinder, starting at the Y-point. Dissipation occurs only in this current sheet which obeys to a different dynamics compared to the standard pressure supported discontinuity. Particle acceleration in the radiation reaction limit can effectively dissipate the Poynting flux within this current sheet \citep{2016JPlPh..82c6303C}. A disk wind and jet geometry completely removing the current sheet in all space is also not excluded \citep{2006ApJ...652.1494L, 1990ApJ...350..732S}. As these authors emphasized, such kind of solutions are also not unique. The current sheet problem instigated \cite{2003PThPh.109..619O} to look deeper into the force-free and MHD solutions of an axisymmetric rotator. They showed that the drift approximation is violated at several light-cylinder radii. See also \cite{2014PASJ...66...25T} for another method to construct current sheet free magnetospheres. In any case the pulsar generates an electric current originating from the polar caps. By action of the Laplace force on the stellar crust, it brakes its rotation. The separatrix is also a privileged place to produce the observed pulsed emission \citep{2007ApJ...667L..69G}.

\subsection{Oblique rotator}

A general method to deal with force-free electrodynamics was developed by \cite{1997PhRvE..56.2181U} without assuming axisymmetry through introduction of Euler potentials. However oblique rotators are much more complex to study because the magnetic field does not reduce to a flux function~$\psi$ as it was the case for the axisymmetric problem. Only numerical simulations solving the time-dependent Maxwell equations can give realistic solutions for the structure of the magnetospheric currents and fields. These results represent major advances toward a self-consistent modelling of pulsar magnetospheres. This novel approach was made possible thanks to progress in numerical methods for simulations of relativistic and magnetized flows. We come back to this point in Sec.~\ref{sec:Simulations}. It is too rarely quoted that the net charge of the pulsar even surrounded by a corotating magnetosphere deviates from zero. The exact value of this charge depends on the obliquity~$\chi$ and vanishes only for a perpendicular rotator \citep{1975ApJ...201..783C}. Charges are distributed within the magnetosphere and within the star itself, their relative filling depending on general-relativistic effects. A Hamiltonian approach revealed also useful to grab general interesting properties of oblique pulsar magnetospheres \citep{1972MNRAS.158...13E, 1976MNRAS.174..125E}. Small obliquity magnetospheres can be treated as perturbation of the aligned case \citep{1982MNRAS.198..405M}.

\subsection{Energy losses}

Knowing the global electrodynamics of pulsar magnetosphere, the entire current circuit is accessible. Therefore the electromagnetic torque exerted in its interior and on its surface
\begin{equation}
 \mathbf{K} = \int \mathbf{r} \wedge ( \mathbf{j} \wedge \mathbf{B} ) \, dV + \int \mathbf{r} \wedge ( \mathbf{i}_{\rm s} \wedge \mathbf{B} ) \, dS 
\end{equation}
can be computed. \cite{2014PhyU...57..799B} asserted that for an orthogonal rotator with $\chi=90$\degree, the toroidal magnetic field component is much less than the poloidal one in such a way that
\begin{equation}
 \left. {B_\varphi} \right|_{\rlight} \approx \sqrt{\frac{R}{\rlight}} \, \left. {B_{\rm p}} \right|_{\rlight} 
\end{equation}
and thus a spindown rate much smaller for the orthogonal case compared to the aligned case. However, simulations show that the spindown is the same in both geometries within a factor two, therefore the current in the magnetosphere must be much higher in the orthogonal rotator to compensate for the decrease in Poynting flux. \cite{2014PhyU...57..799B} also claimed that in such a magnetosphere $\Omega(t)\,\sin\chi(t)=\textrm{cst}$. The obliquity has a tendency to increase with time on a timescale $\tau_\chi\approx P/2\dot P$ conflicting with the vacuum expectations.

\subsection{Quantitative magnetospheric structure}

In some special cases, exact analytical solutions have been found with and without longitudinal currents. They are summarized in chapter two of \cite{2010mfca.book.....B}. They represent interesting models to understand and quantify the back reaction of the current onto the magnetosphere. Let us briefly remind some general comments. In the force-free case, solutions only exist in regions where $E<c\,B$ otherwise the force-free condition would be violated. Moreover, if $j_\parallel=0$, it can be shown that the magnetic field must be perpendicular to the light cylinder and therefore no Poynting flux crosses this surface. So we get the important results that no spindown is allowed in the force-free regime if there is no longitudinal current. In addition, the poloidal current tends to concentrate magnetic field lines towards the equator. If $j_\parallel \gg j_{\rm cor}$ then the light surface is rejected to infinity otherwise with $j_\parallel \ll j_{\rm cor}$ there exists a natural boundary on the force-free region given by $E=c\,B$ and close to the light-cylinder. \cite{2013ApJ...764..129P} gave recently also a new exact analytical solution for the axisymmetric magnetosphere. Special focus along the magnetic axis was also performed by \cite{2012MNRAS.427..514P}.

Let us stress that the space charge really available in a pulsar can drastically deviate from the charge necessary to screen the electric field, several processes are list below
\begin{itemize}
 \item particles inertia \citep{1974ApJ...192..713M}.
 \item curvature of field lines \citep{1981ApJ...248.1099A}.
 \item general relativistic effect \citep{1990SvAL...16..286B, 1992MNRAS.255...61M}.
 \item inefficient particles extraction \citep{1975ApJ...196...51R}.
\end{itemize}

\section{Other effects on the magnetosphere}
\label{sec:OtherEffects}

So far, most of the pulsar magnetosphere investigations tried to solve Maxwell equations in flat space-time assuming the lowest order magnetic field structure: a rotating dipole. There are several caveats to these assumptions. First it is clear that close to the neutron star, especially at the polar caps, strong gravity effects would distort the electromagnetic fields due to space-time curvature and frame dragging. Second, it is not excluded that higher multipolar components exist in the magnetosphere, producing polar cap shapes very different from the dipole. These could in principle be observed in the pulsed radio emission through the pulse profile and phase resolved polarization signature and maybe also in high energy light curves. Third, young pulsars and even more so for magnetars, the magnetic field strength approaches or exceeds the critical value~$\BQ$. Quantum electrodynamics corrections should then be applied to the magnetosphere according to for instance the effective Euler-Heisenberg Lagrangian \citep{1936ZPhy...98..714H}. A direct indisputable consequence of QED is pair creation in the vicinity of the surface, a crucial effect to fill the magnetosphere with charged particles. We briefly comment on these issues in this section.

\subsection{General relativity}

Soon after the discovery of the first pulsar, there was no doubt that it harbours a strongly magnetised and rotating neutron star. The electromagnetic field generated in vacuum for such a rotator was known since the work by \cite{1955AnAp...18....1D} although applied to non compact stars. But we know that neutron star are very compact because of a compactness parameter given by eq.~(\ref{eq:Compacite}), i.e. they resemble almost to black holes.

Few but exact analytical solutions exist for the structure of the magnetic field in strong gravitational fields. For instance expressions for a static magnetic dipole field were given by \cite{1964ZETF...47..1030G} and by \cite{1974PhRvD..10.3166P}. Multipoles have been given by \cite{1970Ap&SS...9..146A}. As for flat space-time \citep{1973ApJ...180L.133M}, useful expressions exist for the force-free monopole in Schwarzschild metric, see \cite{2011PhRvD..83l4035L} although frame dragging was discarded.

\cite{1974AnPhy..87..244C} showed in the case of an aligned rotator that the electric field induced by the dragging of inertial frames can be as important as the field induced by the rotation itself. These results were generalized for an oblique rotator a few years later by \cite{1980Ap&SS..70..295C} thanks to a formalism developed previously by \cite{1974PhLA...47..261C, 1974PhRvD..10.1070C, 1975PhLA...54....5C}. This demonstrated clearly that a quantitative analysis of the acceleration processes and radiation in the vicinity of the neutron star can only be done by a treatment of Maxwell equations in the presence of a strong gravitational field. In this way \cite{1976GReGr...7..459P} looked for an approximate solution to Maxwell equations in curved space-time let it be Schwarzschild or Kerr, through a linearized approach using the Newman-Penrose formalism \citep{1962JMP.....3..566N}. He computed the emission of electromagnetic waves in vacuum for a rotating dipole in general relativity with an expression for the Poynting flux $\dot{E}$ depending on~$R/\rlight$. He found the following expression for a Schwarzschild metric
\begin{equation}
 \dot{E}_{\rm gr} \approx \frac{\dot{E}_{\perp}}{1+(R/\rlight)^2} \, \left( 1 - \frac{3}{2} \, \frac{\Rs}{R} + \frac{\Rs}{R} \times \textrm{correction in} (\frac{R}{\rlight}) \right)
\end{equation}
thus close to Deutsch, $\dot{E} \approx \dot{E}_{\perp}\,(1-(R/\rlight)^2)$ expectation for $\Rs=0$. $\dot{E}_{\rm gr}$ is the general-relativistic spindown luminosity and $\dot{E}_{\perp}$ the flat space time spindown luminosity. \cite{1992MNRAS.255...61M} investigated the influence of space-time curvature and frame dragging of inertial frames on the electric field at the polar caps of a pulsar. \cite{1995ApJ...449..224S} studied in details the electric field in Schwarzschild metric for an aligned rotator in vacuum and plunged in a plasma. The result is an important increase in the electric field at the surface of the star, implying a larger charge density and therefore an acceleration of charges more efficient with possible consequences on the high energy emission of pulsars \citep{1994ApJ...425..767G}. The aligned rotator was revived by \cite{2000PThPh.104.1117K} for investigations of particle acceleration in vacuum. \cite{2001MNRAS.322..723R}, \cite{2002MNRAS.331..376Z}, \cite{2004MNRAS.352.1161R} and  \cite{2013MNRAS.433..986P} computed the effects of general relativity on the electromagnetic field around a slowly rotating neutron star. \cite{1997ApJ...485..735M} and \cite{2003ApJ...584..427S} were concerned about particle acceleration around polar caps in curved space-time. \cite{2004MNRAS.348.1388K} looked for approximate analytical solutions to the oblique rotator problem in vacuum in general relativity. They furnished an approximate numerical solution, expanded to first order. A treatment of the magnetosphere with help of Grad-Shafranov equation has been exploited by \cite{2005MNRAS.358..998K}. \cite{2010MNRAS.408..490M} studied the influence of neutron star oscillations in general relativity on the corotation density in the magnetosphere for a aligned rotator. \cite{2016MNRAS.455.3779P} made an extensive study of force-free pulsar magnetospheres in general relativity. General relativity seems to play a decisive role for efficient pair creation at the surface \citep{2015ApJ...815L..19P, 2016arXiv160405670B}.

Note that black hole magnetospheres can be treated similarly to neutron stars, except for the presence of an event horizon for the former. The problem of this horizon is solved by a change of spatio-temporal coordinates to the Kerr-Schild metric for instance. This coordinate transform permitted the numerical study of the monopolar solution of the black hole magnetosphere presented by \cite{2004MNRAS.350.1431K}. He used a 3+1 formulation of electrodynamics in general relativity. It is useful for pulsars and black holes, and nicely summarized by \cite{2011MNRAS.418L..94K}. Space-time is decomposed in a ``absolute'' time and a three dimensional ``curved space'' to come back to more traditional hyperbolic systems for Maxwell equations in flat space-time plat. \cite{2007ChJAA...7..743Y} used the same formalism for the force-free regime for the magnetosphere of an axisymmetric black hole. \cite{2008ApJ...684.1359M} extended the general-relativistic field to a special spacetime geometry called NUT space (Newman-Unti-Tamburino).

\subsection{Multipoles}

Most of pulsar emission models assume a dipolar magnetic field anchored right at the centre of the star. This hypothesis is certainly correct far from the star, around the light cylinder and beyond, since the high order multipoles~$\ell$ decrease with radius faster than low order ones, like $r^{-(\ell+1)}$. But nothing forbids the existence of significant multipolar components in the vicinity of the star. Multipoles are easily induced by a rotating decentred dipole. The consequences of an off-centred dipole on neutron star proper motion and torque was the main topic in \cite{1979ApJS...41...75R}. Following the same line, \cite{1972Ap&SS..16..130C} showed how to compute force-free multipole components close to the surface with an extension to include general-relativistic effect \citep{1973ApJ...186..267C}. \cite{1979ApJS...41...75R} developed a general formalism for computing the multipolar electromagnetic moments of a neutron star therewith explaining the high velocity of pulsars by asymmetric radiation when its progenitor exploded, an early idea by \cite{1975ApJ...201..447H}. \cite{1991ApJ...373L..69K} studied the influence of multipoles on the estimate of millisecond pulsars magnetic field and rotational braking via their braking index. \cite{2002MNRAS.334..743A} discussed the role of multipoles on the radiation processes and pair creation in the magnetosphere and \cite{2003ARep...47..613K} evoked the influence on the current emanating from the polar caps.
\cite{2010MNRAS.409.1077B} showed an alteration of radiative dipolar magnetic losses because of the presence of multipolar components. Obviously, the polar caps geometry is strongly tributary to multipolar components \citep{1996A&A...310..135Z} with important consequences on radio emission but also on pair creation in such fields \citep{1980MNRAS.192..847J, 2011ApJ...726L..10H}. Magnetic multipoles also have an impact on accretion processes to spin up neutron stars to millisecond periods. The derived spin-up line in the $P-\dot P$ diagram could constrain multipole moments \citep{1993ApJ...408..160A}.

Very recently, \cite{2015A&A...573A..51B} and \cite{2015MNRAS.450..714P} gave exact analytical expressions for any multipolar electromagnetic field in vacuum. It represents a generalisation of the Deutsch field solutions in terms of spherical Hankel functions. \cite{2015MNRAS.453.3540A} investigated the influence of an aspherical shape of the neutron star onto its rotational motion and showed that even a very small ellipticity leads to a precession of period compatible with timing residuals. They took into account the presence of a plasma in the magnetosphere. Aspherical shapes can also give rise to multipolar fields.

Observational support for the presence of multipoles are given already for main sequence stars. \cite{1974MNRAS.169..471S} looked at decentred dipole in stars with a displacement along the magnetic axis. Off centred dipole is already present in AP Stars to solve the asymmetry problem between the north and south hemisphere \citep{1970ApJ...159.1001L}. An off-centred dipole is also the preferred way to explain Zeeman line profile as explained in \cite{1974ApJ...187..271B}. In the context of high-energy processes around compact objects, the radio emission of PSR~J2144-3933 is explained with a novel model about pair creation in the magnetosphere \citep{2000ApJ...531L.135Z} or simply by the presence of intense multipolar components of the surface magnetic field in all radio pulsars \citep{2001ApJ...550..383G, 2002A&A...388..235G}. 

Magnetospheric topologies that deviate slightly or significantly from a pure dipole represent attractive explanations for many electromagnetic phenomena occurring in the neighbourhood of neutron stars. Twisted magnetospheres are especially investigated to understand flares in magnetars \citep{2009ApJ...703.1044B, 2011A&A...533A.125V, 2015MNRAS.447.2821P, 2016arXiv160502253A} and also to account for the mode switching and related spindown changes in intermittent pulsars \citep{2016arXiv160607989H}.

\subsection{Quantum electrodynamics}

The ultra strong magnetic field inferred from the global energetics of pulsars approaches or even exceeds the quantum critical value of $\BQ$. Quantum electrodynamics is therefore required to correctly describe the physics in such field. Pair creation is the most important effect, feeding the neutron star surrounding with fresh and ultra relativistic electron/positron pairs. But this process works on a local scale useful to understand microphysics phenomena. The question arises of the impact of quantum electrodynamics on the global energetic evolution of pulsar spin-down luminosity. This topic was touched by several authors and it seems that even for magnetar field strengths, the corrections from QED remain weak \citep{1997JPhA...30.6475H}. QED effects can be combined to general relativity in the 3+1 formalism as shown by \cite{2015MNRAS.451.3581P}. Applications for a static oblique dipole are given by \cite{2016MNRAS.456.4455P}. Recent numerical simulations of GRQED and GRFFQED rotating dipole confirm the absence of significant corrections to the spindown \citep{2016arXiv160705935P}.

On a smaller scale, the strong magnetic field anchored into the neutron star induces vacuum birefringence and modifies the way electromagnetic waves propagate in vacuum \citep{2001MNRAS.327.1081H, 2003PhRvL..91g1101L, 2003MNRAS.338..233H, 2015CMMPh..55.1857G}. Especially, two normal modes that have mutual orthogonal polarisation travel at different speed in the magnetosphere \citep{2006RPPh...69.2631H, 2014PhRvD..90b3011D}. \cite{2016APh....73....8A} estimated that the delay observed by a detector at Earth would be of the order $\Delta t\approx$\numprint{e-8}\si{\second}, but unfortunately too weak for current instrumentations. \cite{2005PhRvD..71f3002D} showed that gravity can be combined with QED to study light propagation in a realistic neutron star environment. See also \cite{2016JCAP...05..042F} for a broader discussion about force-free theories including a general Lagrangian not necessarily issued from quantum electrodynamics. Bending of light ray due to QED effects was also mentioned by several groups like \cite{1982Natur.295..215S, 1984Ap&SS.102..327S} and \cite{2003A&A...399L..39D}. Non linear electrodynamics induces a supplementary redshift compared to gravitation rendering the compactness $M/R$ difficult to estimate \citep{2004ApJ...608..925M}.

\subsection{Pair creation}

Along field lines with strong intensity, electrons and positrons copiously radiate synchrotron photon on a very short cooling time scale of about $\tau_{\rm syn} \approx c/\omega_B^2 \, r_{\rm e} \approx$\numprint{e-15}\si{\second} thus much smaller that the pulsar period. To a good approximation, we can say that in vacuum, leptons reach quasi instantaneously ultra relativistic speed as soon as they a created around the poles. Besides synchrotron emission, curvature radiation furnishes also numerously photons disintegrating in this magnetic field via the channel
\begin{equation}
 \gamma + B \rightarrow e^- + e^+
\end{equation}
according to a probability per unit length given by \cite{1966RvMP...38..626E} from which we deduced a mean free path of
\begin{equation}
 l \approx \frac{4.4}{\alpha_{\rm sf}} \, \lambda_{\rm e} \, \frac{\BQ}{B_\perp} \, \textrm{exp} \left( \frac{4}{3\,\chi} \right).
\end{equation}
with
\begin{equation}
\chi = \frac{\epsilon_\gamma}{2 \, \masselec\,c^2} \, \frac{B_\perp}{\BQ}
\end{equation}
valid in the limit $\chi\ll1$.
Curvature emission is not the only source of electron/positron pairs in polar caps. Indeed, it was realized in the middle 1990-s \citep{1993Ap&SS.200..251K, 1995ApJ...445..736S, 1996ApJ...468..338L, 1996A&A...310..135Z} that inverse Compton scattering of the thermal radiation from the star surface provides gamma-rays decaying in pairs at energies of primary electrons much smaller than a few TeV required by the classical models. The process has been extensively studied in the literature.

Secondary plasma generations are created following two different models
\begin{itemize}
\item for \cite{1975ApJ...196...51R}, particles can not freely escape from the surface thus producing a charge density different from corotation $\rho \neq \rho_{\rm cor}$. The longitudinal electric field builds up approximatively like 
\begin{equation}
\frac{dE_\parallel}{dh} = \frac{\rho-\rho_{\rm cor}}{\varepsilon_0}
\end{equation}
in the corotating frame where $E_\parallel$ is the parallel electric field and $h$ the altitude measured from the surface. Particles do not circulate freely and the longitudinal electric field becomes with the Ruderman-Sutherland field $E_{\rm RS}$ and $H$ the size of the gap
\begin{equation}
 E_\parallel = E_{\rm RS} \, \frac{H-h}{H}
\end{equation}
\item to the contrary for \cite{1977ApJ...217..227F, 1978ApJ...222..297S, 1979ApJ...231..854A}, they circulate freely leading to the boundary conditions on the gaps as $E_\parallel (h=0) = E_\parallel (H=0) =0$ and corotation charge density $\rho = \rho_{\rm cor}$.
\end{itemize}
Primary particles are believed to reach $E \approx$\numprint{e7}~MeV whereas secondary pairs only painfully reach $E\approx$\numprint{e2}-\numprint{e4}~MeV with a particle distribution function close to $N(E)\propto E^{-2}$ and a multiplicity $\kappa \approx$\numprint{e3}-\numprint{e4} \citep{2010mfca.book.....B}. The global picture of the polar outflow is a primary beam of charged particles with high Lorentz factor~$\gamma_{\rm b} \approx \numprint{e7}$ producing cascades of $e^\pm$ pairs at a multiplicity $\kappa \approx$\numprint{e2}-\numprint{e4} with lower Lorentz factor~$\gamma_{\rm s} \approx \numprint{e2}-\numprint{e3}$ \citep{2002ApJ...581..451A}. These pairs are produced by resonant or non resonant inverse Compton scattering, depending on the neutron star surface temperature, and curvature radiation \citep{2001ApJ...554..624H, 2001ApJ...560..871H}. Radiation emanating from this pair creation process was investigated by \cite{2015IJTP...54..645L} using the Vlasov-Maxwell equations.

\subsection{Magnetic reconnection}

Removing the ideal MHD or force-free regime by adding resistivity or other dissipative effects leads to violation of the flux freezing condition. In addition, a rearrangement of the magnetic topology is expected via magnetic reconnection. This process can drastically perturb the initial magnetic configuration and induces non stationary states in the magnetosphere. Already in the early 70s \cite{1971NPhS..232..144S} claimed that an explosive inflation of the magnetospheric plasma could explain the observed glitch phenomena in the Crab pulsar. Tearing instability were very popular in the 80s and applied for an electron-positron plasma in pulsar magnetospheres by \cite{1987Ap&SS.134..181S} as an onset for magnetic reconnection events. \cite{2005A&A...442..579C, 2007A&A...466..301C} invoked reconnection in the vicinity of the light-cylinder with possible implications for the braking index \citep{2007A&A...472..219C}. The slowing down of the neutron star inflates the size of the light-cylinder. Open field lines in the vicinity of the separatrix must reconnect and close switching to the closed and corotating part of the plasma. This can be seen as a shift in the Y-point. According to \cite{1976ApJ...203..226A} the force-free condition or more generally the ideal MHD regime is violated whenever the criterion $(L\,c/\omega_p)^2\ll 1$ is nor more satisfied where $L$ is a typical gradient scale of macroscopic quantities, and $\omega_p$ the plasma proper frame frequency. Therefore corotation cannot be maintain outside a critical radius~$r_{\rm c}<\rlight$ comprised inside the light-cylinder and depending on local plasma conditions. The exact expression for the particle Lorentz factor near the light-cylinder found by \cite{1976ApJ...203..226A} has been criticized by \cite{1977Ap&SS..51..239B}. He forgot to include inertial drift in his integral of motion \citep{1980Ap&SS..72..251B, 1980AuJPh..33..771B}. Recently \cite{2014MNRAS.443.2197B} showed that bulk flow acceleration to very high Lorentz factor like $\Gamma \approx (1-r^2/\rlight^2)^{-1}$ can occur close to the light-cylinder provided magnetic field lines are swept forward in the hope to explain very high energy pulsed emission of the Crab. An alternative production of these VHE would be by parallel electric field acceleration in the outer gap close to the light-cylinder \citep{2012MNRAS.424.2079B} but this mechanism has been criticized by \cite{2014MNRAS.442L..43H}.

\subsection{Magnetospheric oscillations}

Already in 1965, \cite{1965Natur.205..787C} suggested that neutron star oscillations could generate electromagnetic activity in a neutron star magnetosphere and account for the X-ray observations in the Crab nebula known at that time. Investigations of oscillations above the polar caps have been conducted by many authors to generate for instance radio wave \citep{1978Ap&SS..53..377R}, two-stream instabilities maintaining oscillations converted into radiation \citep{1993AstL...19...14L} or to explain the drifting subpulses \citep{2004ApJ...609..340C}. Oscillations could triggered magnetospheric activity and related radio emission of normal pulsars and magnetars \citep{2015ApJ...799..152L}. Stellar oscillations impact on the maximum Lorentz factor of particles accelerated in the polar cap \citep{2012A&A...540A.126Z}. Oscillations in magnetars are able to shift the radio emission generation threshold to less restrictive regions in the $P-\dot P$ diagram \citep{2012MNRAS.419.2147M}. Oscillations are relevant for magnetar quasi-periodic oscillations \citep{2009MNRAS.395..443A}. \cite{2014PTEP.2014b3E01K} included some resistivity prescription and perturbed the magnetosphere through torsional shear oscillations hoping to explain X-ray and gamma-ray flares in magnetars. The presence of a plasma oscillating in the magnetosphere modifies the energy loss depending on the oscillation frequency compared to vacuum \citep{2000MNRAS.316..734T}.

Usually the matter content of pulsar magnetospheres is assumed to be made of light particles, leptons that do not contribute significantly to the total mass of the pulsar. Nevertheless, several authors considers the effect of mass loaded magnetosphere, in particular to shift the Y-point inside the light cylinder towards the star \citep{1977SvA....21..432P}. This author also predicted possible episodes of mass ejection to be related to the glitches. \cite{2015ApJ...805..106T} argued that depending on the plasma and magnetic energy content within the magnetosphere, the pulsar may expel matter sporadically outside the light-cylinder. 

An important question raised recently by several authors concerns the connection between the neutron star interior to its exterior, i.e. its magnetosphere. Usually the former community assumes vacuum outside whereas the latter community assumes specific boundary conditions on the surface. Obviously, this approach is neither satisfactory for the first nor for the second community. Thus any realistic solution to pulsar magnetosphere should join smoothly the interior field to the exterior field. Such rather new investigations including both domains are possible as demonstrated by \cite{PhysRevD.89.084045} who computed the Poynting flux depending on equations of state and compactness. Matching ideal MHD simulations with the force-free schemes was used by \cite{2013PhRvD..88j4031P} to compute neutron star magnetospheres. \cite{2014MNRAS.437....2G} and \cite{2015MNRAS.447.2821P} undertook similar studies. \cite{2015ApJ...799...23B} even attempted to compute the full solution inside and outside with realistic equations of state and account of all fundamental interactions including general relativity and results from quantum mechanics.

\subsection{Non-corotating and highly rotating magnetosphere}

In the simplest description of the plasma motion, the electromagnetic field achieves a configuration imposing perfect corotation of particles at most up to the light-cylinder. We stress that this fact is an assumption and not a result of the model. Therefore the question ``Does pulsar magnetosphere really corotate with the underlying neutron star?'' is meaningful. Such questioning was the subject of several papers among them \cite{2012ApJ...745..169M, 2014MNRAS.437..262M, 2016JPlPh..82b6302M} who were preoccupied by the neglect of the inductive electric field in MHD or force-free magnetospheres. Differential rotation of for instance the open field lines due to potential drop above the polar cap changes the value of the electric current density and impacts on the braking index \citep{2007Ap&SS.308..575T, 2007MNRAS.379..605T}. 

Dissipation regions were charges are able to cross field lines are compulsory to close the current and transfer angular momentum outside the light cylinder. \cite{1988MNRAS.232..277F} and \cite{1988MNRAS.232..303F} looked for such solutions.

\cite{1955AnAp...18....1D} electromagnetic field expressions, although being an exact analytical solution to Maxwell equations, fails to give an accurate picture of relativistically rotating neutron stars when $R/\rlight\rightarrow1$ because it assumes non relativistic rotation. \cite{1992ApJ...401L..27B, 1994A&A...283.1018B} gave an answer to this problem and showed an analogy with synchrotron radiation. \cite{1995Ap&SS.234...57D} proposed an alternative derivation of this relativistic rotating dipole. The increase in spindown power is counterbalanced by gravitational effects when the mass of the dipole is added \citep{2013NewA...25...38H}.

\section{Numerical simulations}
\label{sec:Simulations}

Searching for an analytical solution to the problem of the magnetospheric structure is very cumbersome or even impossible in a realistic situation. Another complementary approach allowing deeper and more quantitative insight consists to perform numerical simulations of the temporal evolution of the magnetosphere. We then hope to observe relaxation to a stationary equilibrium state. The level of complexity of these simulations relies on the approximation used to described the behaviour of plasmas interacting with the stellar electromagnetic field, radiative corrections and self-consistent treatment of particle injection through pair formation. Starting with the crudest physical description known as the force-free approximation, useful to investigate neutron star but also black hole magnetospheres on the largest scales, several other plasma regimes have been or should be explored in the future. The so far most extensively studied are
\begin{itemize}
\item force-free (magnetodynamics): charge and current carriers have no or negligible mass. They respond instantaneously to the external electromagnetic field to furnish the required charge and current densities imposed by the evolution of the fields. The matter stress-energy tensor vanishes. No energy dissipation occurs.
\item resistive magnetodynamics: in order to allow for dissipation and transfer of energy from the field to the particles, some resistive terms are added to the force-free current. The resistivity prescription is not unique and loosely constrained. Motion of the plasma is not solved.
\item magnetohydrodynamics (MHD): particle inertia is taken into account and the full stress-energy tensor, matter and field, is solved. Simulations are performed in the ideal limit or in the resistive regime.
\item multi-fluids: the electron/positron plasma does not strictly follow the MHD system because both particle species have the same mass. The usual MHD ordering according to the masses is therefore impossible. Multi-fluid schemes evolve each species independently, the coupling going through electromagnetic interactions via Lorentz forces. Binary collisions between particles irrespective of their species is treated following Monte Carlo techniques.
\item fully kinetic treatment: convenient to account for individual particle acceleration towards distribution functions that are out of thermal equilibrium. Needs to solve the full Vlasov-Maxwell equations and thus very expensive computationally.
\item radiation reaction limit: particles in pulsar magnetospheres radiate copiously up to the point where any acceleration is compensated by radiation reaction. In this special case, particle motion can be solve analytically to give an expression for the velocity (equal to the speed of light) only in terms of the external electromagnetic field. It represents an interesting alternative to the full Vlasov-Maxwell approach in the strong radiation reaction limit.
\end{itemize}
Let us pinpoint the merit of all these approximations.

\subsection{Force-free electrodynamics (FFE)}

Force-free electrodynamics leads to some degeneracy in its physical interpretation. Indeed, under such hypothesis, two interpretations are plausible
  \begin{itemize}
  \item either the plasma is non neutral therefore completely charge separated. This corresponds to a weak density of particles in the magnetosphere with $(n_+,n_-)\approx n_{\rm cor}$. Complete charge separation has been criticized by \cite{1973A&A....27..413S}.
  \item either ideal MHD applies. This implies a quasi-neutral plasma therefore a large particle density number in this same magnetosphere with $n = n_+ + n_- \gg n_{\rm cor}$ with $n_+ \approx n_-$ but a small difference to let room for a possible small electric charge density such that there  $|n_+-n_-| \ll n$.
\end{itemize}
Which of this view prevails in a realistic magnetosphere? It depends on the injection rate of charged particles, a direct consequence of efficient pair formation in the vacuum gaps, a still unsolved problem. Nevertheless nebulas seem to prefer the second option of a dense plasma, we will explain why in the section about pulsar winds. However, major problems arise from unconstrained global features, namely
\begin{itemize}
  \item the total charge of the star and its surrounding magnetosphere remains unconstrained and worse not necessarily null. However, \cite{1976Natur.259...25J, 1976ApJ...206..831J} gave an argument to constrain the electric charge of the star to such a value to stop leakage towards the nebula, assuming that only electrons leave the star.
  \item in the same vein the total electric current does not necessarily vanish.
  \item as a corollary the total charge of the system ``star+magnetosphere'' is neither necessarily conserved nor constrained (so back to first point).
\end{itemize}

This did not prohibit \cite{2006ApJ...648L..51S} to realise the first three dimensional simulation of an oblique rotator. The electric current is only a function of the electromagnetic field, charges must adjust themselves their position and velocity to be able to furnish the required charge and current density fulfilling the force-free condition eq.~(\ref{eq:force_free}) such that \citep{1999astro.ph..2288G}
\begin{subequations}
\begin{align}
 \label{eq:J_Ideal}
  \mathbf j & = \rho_{\rm e} \, \frac{\mathbf{E}\wedge \mathbf{B}}{B^2} + \frac{\mathbf{B} \cdot \rot \mathbf{B} / \mu_0 - \varepsilon_0 \, \mathbf{E} \cdot \rot \mathbf{E}}{B^2} \, \mathbf{B} \\
 \rho_{\rm e} & = \varepsilon_0 \, \divg \mathbf E \ .
\end{align}
\end{subequations}
Independently \cite{2009A&A...496..495K} implemented a similar algorithm but treating the boundary conditions more satisfactorily with help of perfectly matched layers (PML), a technique described by \cite{1994JCoPh.114..185B, 1996JCoPh.127..363B}. \cite{2012MNRAS.420.2793K} then extended the simulation box to tenth of $\rlight$. Note that in all these simulations, no account of the current parallel to the magnetic field is taken, only the electric drift current, the term proportional to $\mathbf E \wedge \mathbf B$ in expression~(\ref{eq:J_Ideal}) is really computed, a limitation due to the finite difference scheme they used. The force-free aligned rotator has been reinvestigated by several authors like \cite{2005PhRvL..94b1101G}, \cite{2006MNRAS.368L..30M} and \cite{2006MNRAS.368.1055T}. A analytical study of the influence of the relativistic space charge limited outflow was undertaken by \cite{2005ApJ...630..454M}. Lastly, the computation of light curves associated to these simulations performed by \cite{2012ApJ...754L...1K} offers an efficient test to check the conjectured hypothesis. \cite{2004MNRAS.349..213G} attempted to look analytically about non dissipative force-free magnetospheres using Fourier transform techniques in 2D.

Let us emphasize some drawbacks of the first ever 3D simulations
\begin{itemize}
 \item the ratio $R/\rlight=0.2$ is too large, it corresponds to an unrealistic pulsar of period as low as 1~ms.
 \item the cartesian geometry does not permit a satisfactory treatment of boundary conditions at the stellar surface.
 \item the outer bound of the numerical box leads to inconvenient reflections polluting the interior of the domain for long time runs.
 \item these simulations use $\mathbf E \cdot \mathbf B$ cleaning techniques which in effect introduce a parallel electric current that shorts out this $\mathbf E \cdot \mathbf B$. This method achieves the same purpose as the parallel electric current term in equation~(\ref{eq:J_Ideal}).
 \item a current sheet forms, separating field lines attached to the north pole from those attached to the south pole. It represents a singular surface difficult to catch numerically and physically not realistic. The ideal MHD or force-free approximation fails, dissipation should play an important role in this current sheet.
\end{itemize}

\begin{figure}
  \begin{center}
 \input{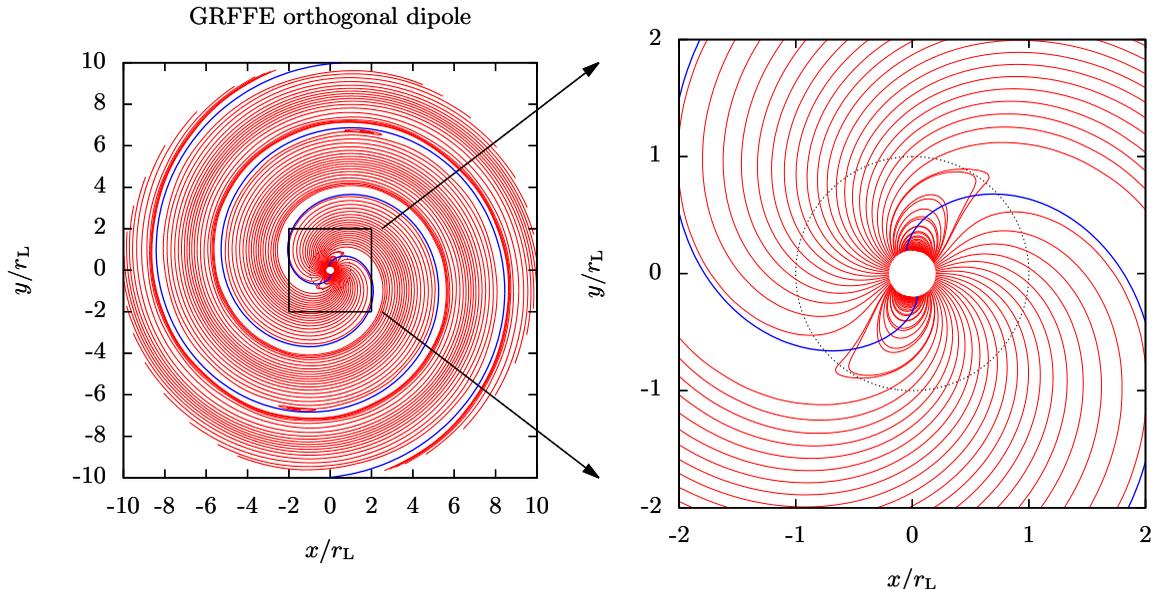} 
\caption{Magnetospheric structure of the perpendicular rotator for a general relativistic dipole magnetic field with $R/\rlight=0.2$ and $R/\Rs=2$. The distances are normalized to the light cylinder radius. A spiral arm form where field lines change polarity. This special geometry is at the heart of the striped wind model, Sec.~\ref{sec:Vent}.}
    \label{fig:MagnetosPerp}
  \end{center}
\end{figure}

\cite{2012MNRAS.424..605P} has partially eliminated some of these drawbacks by formulating a new algorithm to solve Maxwell equations with help of pseudo-spectral methods. The main idea is to expand the unknown fields into vector spherical harmonics. Application examples of this technique in electromagnetism are available in \cite{1978AmJPh..46..849L} and \cite{1985EJPh....6..287B}. At the same time, \cite{2012MNRAS.423.1416P} developed a similar technique but only in axisymmetric geometry that was recently reinvestigated by \cite{2016MNRAS.455.4267C}. The superiority of this novel method is indisputable, from both the point of view of numerical precision, boundary condition treatment and computational resources. The perpendicular rotator is shown in fig.~\ref{fig:MagnetosPerp}, the structure of the magnetic field lines in the equatorial plane are visible in red solid lines. To ease the comparison with the vacuum rotator we overlap the two-armed spiral in blue solid line in order to localize the discontinuity. Adjustment is done by eye and it is necessary to add a small phase shift with respect to vacuum to correctly reproduce the sheet. From the force-free simulations, the power radiated by Poynting flux for an oblique rotator can be deduced and fitted with a simple relation
\begin{equation}
  \label{eq:L_dot}
  L_{\rm sp} \approx \frac{3}{2} \, L_\perp^{\rm vac} \, ( 1 + \sin^2\chi )
\end{equation}
in agreement with \cite{2006ApJ...648L..51S}. The presence of a magnetospheric plasma multiplies by three these losses compared to vacuum. The aligned rotator also radiates at a rate of $L_{\rm sp} \approx \frac{3}{2} \, L_\perp^{\rm vac}$. This contrasts radically with the solution for an aligned rotator in vacuum which does not radiate. Moreover \cite{2012MNRAS.424..605P} demonstrated that the total charge of the star+magnetosphere system does not vanish except in the particular case of a perpendicular rotator. This is reminiscent of the point charge located at the stellar centre eq.~(\ref{eq:ChargePonctuelle}). It is questionable how such a charge could subsist without cancellation by attraction of particles of the opposite sign from the surrounding. To conclude about those simulations, the luminosity of a plasma filled magnetosphere is of the same order of magnitude as the dipole in vacuum. It is therefore delicate to make a definite distinction observationally between these two models simply by inspection of the power radiated.

\subsection{Resistive force-free electrodynamics}

The current sheet appearing in the above mentioned simulations is an artefact of the force-free approximation. It would also appear in ideal MHD simulations. In this region, a non negligible resistivity should soften the discontinuity. The force-free electrodynamics in its simplest form does not allow for dissipation in the flow because it corresponds to an infinite conductivity. Although it should not be such a drawback for the global magnetospheric structure, it is really a pain to elucidate locally the location of emission regions where particle acceleration occurs and radiation is produced. The force-free approximation can not account for particle acceleration nor for pulsed emission in the magnetosphere. This impossibility goes back to the condition $\mathbf E \cdot \mathbf B = 0$ which is to restrictive. We should allow for a $E_\parallel$ and/or for regions where $E>c\,B$. Such prescriptions have been proposed for the current in the magnetosphere, alleviating the prescription in eq.(\ref{eq:J_Ideal}). This requires a parallel electric field that by the force-free assumption does not exist. To circumvent this disadvantage, some less restrictive magnetodynamics regimes have been developed, a kind of resistive magnetodynamics. Whereas the prescription in the force-free limit leads to a definite an precise expression for the current density, it is less clear how to impose this current when the flow becomes dissipative or resistive. There is no unique prescription to generalize Ohm's law in this regime. Some degree of freedom is available for the exact expression of the current $\mathbf{j}$. Several examples of a kind of generalized resistive Ohm's law for a relativistic pair plasma have been suggested by \cite{2012ApJ...746...60L}, \cite{2012ApJ...749....2K} and \cite{2003MNRAS.346..540L}. \cite{2008JCAP...11..002G} proposed an approximation called strong field electrodynamics and giving the current
\begin{equation}
 \mathbf j = \frac{\rho_{\rm e} \, \mathbf E \wedge \mathbf B + \sqrt{ \rho_{\rm e}^2 + \gamma^2 \, \sigma^2 \, E_0^2/c^2} \, (E_0 \mathbf E / c + c\,B_0 \mathbf B)}{B^2+E_0^2/c^2} \ .
\end{equation}
The $\sigma$ parameter cannot be interpreted as a conductivity because for $\sigma=0$ the vacuum case is not retrieved. Actually this expression is valid for a plasma entirely charge separated and subject to radiation reaction in the ultra-relativistic regime, see below. Indeed, for $\sigma=0$ we found 
\begin{equation}
 \mathbf j = \frac{\rho_{\rm e} \, \mathbf E \wedge \mathbf B + |\rho_{\rm e}| \, (E_0 \mathbf E / c + c\,B_0 \mathbf B)}{B^2+E_0^2/c^2} \ .
\end{equation}
Another approximation consists of writing Ohm's law in the fluid rest frame where the electric and magnetic field are aligned and then Lorentz transform to the lab frame. In such a way \cite{2012ApJ...746...60L} found an expression function of the fluid velocity along the field lines, $\beta_\parallel\,c$ which remains undetermined, such that
\begin{equation}
 \mathbf j = \frac{\rho_{\rm e} \, \mathbf E \wedge \mathbf B + ( -\beta_\parallel \, \rho_{\rm e} + \sqrt{ \gamma^2 \, ( 1-\beta_\parallel^2) } \, \sigma \, E_0 / c) \, (E_0 \, \mathbf E / c + c \, B_0 \, \mathbf B)}{B^2+E_0^2/c^2} \ .
\end{equation}
The minimal hypothesis they choose was to set $\beta_\parallel=0$ for lack of better knowledge about the longitudinal speed. The current then simplifies into
\begin{equation}
\label{eq:CourantOhmMinimal}
 \mathbf j = \frac{\rho_{\rm e} \, \mathbf E \wedge \mathbf B + \gamma \, \sigma \, E_0 / c \, ( E_0 \, \mathbf E / c + c\,B_0 \, \mathbf B)}{B^2+E_0^2/c^2}
\end{equation}
which is Ohm's law for a relativistic quasi neutral plasma 
\begin{equation}
 \mathbf j = \gamma \, \sigma \, ( \mathbf E + \mathbf v \wedge \mathbf B - (\mathbf E \cdot \mathbf v ) \, \mathbf v) + \rho_{\rm e} \, \mathbf v
\end{equation}
with a drift speed given by 
\begin{equation}
 \mathbf v_{\rm drift,res} = \frac{\mathbf E \wedge \mathbf B}{B^2+E_0^2/c^2}
\end{equation}
and an associated Lorentz factor
\begin{equation}
 \gamma^2 = \frac{E_0^2+c^2\,B^2}{E_0^2+c^2\,B_0^2}
\end{equation}
The current is then exactly the one obtained from the minimal hypothesis with $\beta_\parallel=0$.
The origin of the conductivity was not explicitly stated in these works but turbulence in a relativistic plasma could account for sharp variation of the effective conductivity within the magnetosphere. According to \cite{1974SvA....18..211K} the conductivity increases with distance to the star, which is opposite to the FIDO model used by \cite{2014ApJ...793...97K}. The latter work used an Ohm law given by
\begin{equation}
 \mathbf{j} = \rho_{\rm e} \, \frac{\mathbf E \wedge \mathbf B}{B^2+E_0^2/c^2} + \sigma \, \mathbf{E}_\parallel
\end{equation}
which has been reexplored using spectral methods by \cite{2016arXiv160602707C}. Earlier attempts to design a generalized Ohm law are given by \cite{1977AuJPh..30..471B, 1977PhLA...60..309B}. Switching between vacuum and high conducting magnetosphere furnishes an explanation for the braking index variation during on and off states \citep{2012ApJ...746L..24L}.

\subsection{Ideal and resistive MHD}

Since the determination of this resistivity is debated, it seems more judicious to relax the resistive magnetodynamics condition and explore the MHD realm, including particle inertia, and even pressure, as  for the aligned rotator which is also a more realistic approach. We can also take advantage of the remarks made by \cite{2006MNRAS.367...19K}. The most satisfactory method would certainly require a multi-fluid or better a kinetic approach. The MHD approach to pulsar magnetosphere was performed by \cite{2013MNRAS.435L...1T}. \cite{2009MNRAS.398..271K} adopted this multi-fluid track performing two-fluid cold plasma simulations but so far only for the aligned rotator. Some analytical description of a two-fluid axisymmetric pulsar magnetosphere is given by \cite{2015MNRAS.446.2243P}. Resistive methods are usually constrained by stiff source terms depending on the conductivity parameter. Such difficulties are overcome by introducing implicit-explicit Rung-Kutta (IMEX) schemes as implemented by \cite{2013MNRAS.431.1853P}. MHD type of waves exist in pulsar magnetospheres but need to take the charge density into account \citep{2011A&A...535L...5U}. Simulations of monopoles and axisymmetric  dipoles performed by \cite{2006MNRAS.368.1717B} revealed that depending on the resistivity, the location of the Y point can shift well inside the light-cylinder modifying the spindown rate from the standard acceptance that $R_{\rm Y}=\rlight$.

\subsection{Kinetic methods}

The full description of the plasma would require solutions of the Vlasov-Maxwell equations. It offers the most detailed view of the magnetospheric plasma configuration and allows deep diagnostics of particle acceleration regions. Unfortunately these equations are numerically very difficult to solve because distribution functions are defined in six dimensions, three space coordinates and three momenta coordinates. A less stringent technique employs particle in cell (PIC) methods. They were successfully applied by \cite{2007MNRAS.376.1460W} to elucidate the link between the active magnetosphere and the pulsar wind by including a possible pair creation mechanism with radiation reaction forces. These works were took up later with a better resolution, a higher number of particles by \cite{2011MNRAS.418..612W}. They set up an electrostatic approximation, neglecting the feedback of the current onto the magnetic field, supposed dipolar and immutable. However this is justified only when the magnetospheric current density is weak. \cite{2012PASJ...64...43Y} described very similar studies. \cite{2010PASJ...62..131U} focused on a detailed study of the Y~point in the aligned rotator, i.e. the cusp point of the last field line just grazing the light cylinder. These PIC methods are indisputably much more promising to model the pulsar magnetosphere.

Starting with a full 3D PIC code, \cite{2014ApJ...785L..33P} studied axisymmetric configurations. Then \cite{2015ApJ...801L..19P} presented 3D PIC simulations of the pulsar magnetosphere with conclusions very similar to the MHD magnetosphere performed by \cite{2013MNRAS.435L...1T}. \cite{2014ApJ...795L..22C} used PIC simulations and included pair creation from the polar cap up to the light cylinder to look at the filling properties of the magnetosphere. They found solutions very similar to the force-free limit for large pair injection but relaxation to an electrosphere for too low injection rates. \cite{2015MNRAS.449.2759B} looked at the transfer of energy between the Poynting flux and the particle. He found that roughly 20\% goes into particle acceleration and up to 50\% if the electric field is not sufficiently screened by the presence of a plasma. Similarly, \cite{2015MNRAS.448..606C} performed 2D axisymmetric PIC simulations of the aligned rotator to look at particle acceleration in the equatorial plane containing the current sheet (that otherwise in 3D would be called the striped wind part). Depending on the particle injection rate, they found that up to 30\% of magnetic energy is dissipated within several light-cylinder radii.

Very recently, first attempts have been made by \cite{2016MNRAS.457.2401C} to include radiation reaction self-consistently into a fully 3D PIC code in order to observe particle acceleration and jointly to extract light curves. PIC simulations do not put particles arbitrary into the magnetosphere as does force-free but rely on a more microphysics explanation of pair creation. The crucial point is to adjust the efficiency to realistic values that are unfortunately largely unconstrained.

\subsection{GRFFE}

The trend to move to more quantitatively accurate magnetospheres via numerical simulations requires more physical inputs to catch the full complexity of pulsar electrodynamics. General relativistic effects should be accounted for to get precision better than 20\%. As the quality and quantity of multi-wavelength observations increased drastically the last decades, those refinements become compulsory. The 3+1 formalism has been extensively used to computed general relativistic force-free solution for the neutron star magnetosphere. Vacuum solution of Deutsch kind but in general relativity are discussed in \cite{2013MNRAS.433..986P}. The numerical simulations based on a pseudo-spectral code are described in \cite{2014MNRAS.439.1071P} and extended to a discontinuous Galerkin approach in \cite{2015MNRAS.447.3170P, 2016MNRAS.455.3779P}. The conclusions drawn from SRFFE simulations remain valid and the physics is not changed. However frame dragging seems to be required to enhance the pair production in the polar caps \citep{2015ApJ...815L..19P} to get sufficiently high plasma densities. \cite{2015Ap&SS.356..301R} proposed general-relativistic corrections to the charge density along open field lines in the slow rotation approximation and including a possible deformation of the star.

\subsection{GRFFQED}

QED effects are compulsory on a microscopic scale to trigger pair cascades in the strong magnetic field of a neutron star. Single or multiple photons interactions and disintegration into leptons are the main channels to feed the magnetosphere with a plasma. The question arises of the effect of these strong fields onto the macroscopic scale of the order the light-cylinder radius. Currently, investigations have been performed to account for lowest order corrections induced by QED to the total spindown luminosity and electromagnetic field structure around neutron stars. Because the corrections remain weak, less than the fine structure constant for field strengths $B\le$\numprint{e10}\si{\tesla}, preliminary results for vacuum rotators show that QED effects are irrelevant as far as the global dynamics is concerned \citep{2016arXiv160705935P}. Plasma effects in the force-free regime are also investigated but no drastic changes are found compared to vacuum.

\subsection{Radiation reaction limit}

So far, fluid simulations treated radiation in a post processing fashion, after computing the magnetosphere structure in a force-free, MHD or resistive approximation. There is no back reaction of emission onto particle dynamics. Because pulsar magnetospheres contain ultra-relativistic particles radiating copiously in all wavelengths, radiative corrections to particle trajectories can be easily treated in the radiation reaction limit assuming a stationary balance between acceleration and emission.
Indeed, in the electromagnetic field prevailing in the pulsar magnetosphere, the plasma suffers strong radiation reaction, invalidating the condition $\mathbf E \cdot \mathbf B=0$. Particles are braked and feel a kind of frictional force directed oppositely to their velocity such that $\mathbf f_{\rm rad} = - K \, \mathbf v$. By definition the constant $K$ is positive and can be derived explicitly as follows. In a stationary regime, a particle of charge $q$ is pulled by Lorentz and radiation reaction forces such that $\mathbf f_{\rm Lorentz} + \mathbf f_{\rm rad} = \mathbf 0$ or
\begin{equation}
\label{eq:Aristote}
 q \, ( \mathbf E + \mathbf v \wedge \mathbf B ) = K \, \mathbf v
\end{equation}
Following the reasoning of \cite{1999stma.book.....M} it is possible to derive the speed of any particle in a prescribed electromagnetic field in the limit where their speed is equal to c, which is a good approximation in pulsar magnetospheres. We notice that 
\begin{equation}
\label{eq:KvB}
 q \, \mathbf E \cdot \mathbf B = K \, \mathbf v \cdot \mathbf B \neq 0 . 
\end{equation} 
The constant $K$ is solution of
\begin{equation}
 K^4 \, v^2 - q^2 \, ( E^2 - v^2 \, B^2 ) \, K^2 - q^4 \, (\mathbf E \cdot \mathbf B)^2 = 0
\end{equation}
assuming that the speed of the particles are near to speed of light, we solve for $K$ to obtain
\begin{equation}
 K^2 \approx \frac{q^2}{2\,c^2} \, \left[ E^2 - c^2 \, B^2 \pm \sqrt{(E^2 - c^2 \, B^2)^2 + 4 \, c^2 \, (\mathbf E \cdot \mathbf B)^2 }\right] \ .
\end{equation}
The solution with negative sign has to be rejected because $K^2<0$. $K$ is solution of the following Lorentz invariant system 
\begin{subequations}
 \begin{align}
  E^2 - c^2 \, B^2 & = c^2 \, K^2 / q^2 - c^2 \, B_0^2 \\
  \mathbf E \cdot \mathbf B & = c \, K \, B_0 / |q|
 \end{align}
\end{subequations}
$c\,K/|q|$ represents the intensity of the electric field in the frame where electric and magnetic field are aligned. The constant $c\,K/|q|>0$ can be linked to the previous discussion about resistive force-free electrodynamics. In that section it was depicted by the letter $E_0$.

In the special case where $\mathbf E \cdot \mathbf B = 0$ we get for $E < c\,B$ the condition $K^2=0$ and for $E > c\,B$ the condition $K^2=\frac{q^2}{c^2} \, ( E^2 - c^2 \, B^2)$. In the case of a general weak electric field, $E \ll c\,B$ then $K^2 = q^2 \, (\mathbf E \cdot \mathbf B/c\,B)^2 $. Solving for the speed starting from eq.~(\ref{eq:Aristote}) and using eq.~(\ref{eq:KvB}) we obtain
\begin{equation}
 ( K^2 + q^2 \, B^2 ) \, \mathbf v = q^2 \, \mathbf E \wedge \mathbf B + q \, K \, \mathbf E +
q^3 \, \frac{\mathbf E \cdot \mathbf B}{K} \, \mathbf B \ .
\end{equation}
The velocity can be decomposed into a drift motion superposed to a motion along $\mathbf E$ and $\mathbf B$ such that 
\begin{subequations}
 \begin{align}
  \mathbf v_{\rm drift,rr} & = \frac{\mathbf E \wedge \mathbf B}{E_0^2/c^2 + B^2} \\
  \mathbf v_{\rm EB,rr} & = \textrm{sign}(q) \, \frac{E_0 \, \mathbf E/c^2 + B_0 \, \mathbf B}{E_0^2/c^2 + B^2} \ .
 \end{align}
\end{subequations}
For a vanishing magnetic field, the particle whatever its charge moves at the speed of light along the electric field with velocity $\mathbf v = \textrm{sign}(q) \, c \, \mathbf E / E$ as expected from an almost instantaneous acceleration on time scale much shorter than any other dynamical time (zero inertia limit). These results about the speed of particles in the radiation reaction limit have already been given by \cite{1985MitAG..63..174H} and \cite{1989A&A...225..479F} who also solve numerically the equation of motion including radiation reaction. \cite{1990A&A...238..462F} discussed the validity of the Lorentz-Dirac equation in pulsar magnetospheres. A detailed study by \cite{1986ApJS...61..465L} integrating numerically the Lorentz Dirac equation for electrons and protons showed the particle orbits and maximum attainable energy for a perpendicular rotator. The radiation reaction limit, much simpler to implement as the full equation of motion, has been applied to solve the pulsar magnetosphere topology by \cite{2013arXiv1303.4094G}, \cite{2012arXiv1205.3367G} and \cite{2016arXiv160600162C}. They claim that two types of pulsars should exist: those that are very dissipative and those that are not. \cite{1974A&A....36..267F} studied radiation reaction in pulsar magnetosphere also in the context of cosmic ray acceleration.

From this general expression of the particle velocity in an electromagnetic field, we can prescribe an electric current density including the motion of electrons and positrons. Considering a reference particle density number~$n_0$ and introducing the pair multiplicity parameter by~$\kappa$, the charge  density $\rho_{\rm e}$ is deduced from Maxwell-Gauss equation and furnishes the reference particle number density. Let us use $n_0$ as a free parameter such that
\begin{subequations}
 \begin{align}
 \rho_{\rm e} & = \varepsilon_0 \, \divg \mathbf E \\
 e \, n & = |\rho_{\rm e}| \\
 \rho_{\rm e} & = e \, ( n_+ - n_- ) \\
 \mathbf j & = e \, ( n_+ \, \mathbf v_+ - n_- \, \mathbf v_- )
 \end{align}
\end{subequations}
$n$ corresponds to the particle density required for the minimal hypothesis of a totally charge separated plasma. In order to estimate the particle density number, we start from a fully charge separated plasma and add neutral pairs $e^\pm$ with a multiplicity~$\kappa$. We must distinguish between two kinds of regions. If the space charge is positive we choose a background electron density null and add pairs. Primary positrons are at a number of $e\,n=\rho_{\rm e}$ such that 
\begin{subequations}
 \begin{align}
 n_- & = \kappa \, n_0 \\
 n_+ & = \kappa \, n_0 + \frac{\rho_{\rm e}}{e} = \kappa \, n_0 + n \ .
 \end{align}
\end{subequations}
If the space charge is negative we choose a background density of positrons null and add pairs. Primary electrons are then at a number of $e\,n=|\rho_{\rm e}|$ such that 
\begin{subequations}
 \begin{align}
 n_- & = \kappa \, n_0 + \frac{|\rho_{\rm e}|}{e} = \kappa \, n_0 + n \\
 n_+ & = \kappa \, n_0 \ .
 \end{align}
\end{subequations}
In all cases, we notice that the total density of pairs is the same and given by $n_+ + n_- = n + 2\,\kappa\,n_0$. Noting that the speed $\mathbf v_{EB}$ of electrons is opposite to that of positrons, because $K>0$ by assumption and $V_{EB} \propto \textrm{sign}(q)$, the current becomes
\begin{equation}
 \mathbf j = \rho_{\rm e} \, \mathbf v_{\rm drift,rr} + (|\rho_{\rm e}|+2\,\kappa\,n_0\,e) \, \mathbf v_{EB} \ .
\end{equation}
For a mono-fluid description, we introduce the fluid velocity by 
\begin{equation}
 \mathbf v = \frac{n_+ \, \mathbf v_+ + n_- \, \mathbf v_-}{n_++n_-} = \mathbf v_{\rm drift,rr} + \frac{\rho_{\rm e}}{|\rho_{\rm e}|+2\,\kappa\,n_0\,e} \, \mathbf v_{EB} \ .
\end{equation}
For a quasi-neutral plasma, the pair multiplicity is very high $\kappa\gg1$ which means that $|n_+ - n_-|\ll n_+ + n_-$. Therefore $|\rho_{\rm e}| \ll |\rho_{\rm e}| + 2\,\kappa\,n_0\,e$ and to first approximation the fluid velocity is simply equal to the electric drift motion $\mathbf v = \mathbf v_D$. Injecting this expression in Ohm's relativistic law, we get the current from the minimal hypothesis of eq.~(\ref{eq:CourantOhmMinimal}).

\subsection{Observational signature of magnetospheric structure}

Remotely diagnosing magnetospheric activity in pulsar physics requires predictions or better a posteriori adjustments of dynamical and geometric parameters such as particle injection rate, obliquity of pulsar and inclination of line of sight. Several works in the last decade tried to matched recent gamma-ray light-curves obtained from Fermi/LAT \citep{2013ApJS..208...17A} assuming different plasma regimes. For instance \cite{2010ApJ...715.1270B} tested the simple vacuum dipole and compared predicted light-curves with observations. In a second trial \cite{2010ApJ...715.1282B} used the force-free magnetosphere obtained from previously mentioned simulations with seemingly better fits. Actually the latter model should not radiate because force-free is dissipationless. The location of emission sites is left at the discretion of the physicists. The same and other authors used results from resistive or more generally speaking dissipative prescriptions to compute characteristic pulse profiles \citep{2012ApJ...754L...1K, 2014ApJ...793...97K, 2015ApJ...804...84B}. \cite{2015ApJ...804...84B} tried to get observational signatures of a dissipative magnetosphere through the computation of gamma-ray light-curves. They used the Force-free Inside Dissipative Outside (FIDO) model described by \cite{2014ApJ...793...97K} to best fit the data. Why no dissipation should apply inside the light-cylinder remains mysterious on a physical ground. Full PIC simulations also start to predict light curves although results are still preliminary \citep{2016MNRAS.457.2401C}. They do not make assumptions about emission sites, they are self-consistently determined by the simulations themselves. Sometimes, breakdown of force-free regime is stated in the vicinity of the light-cylinder, allowing efficient particle acceleration and associated intense X-ray and gamma-ray emission \citep{1994MNRAS.271..621M}. Angular momentum is carried away by the relativistic wind and current closure must occur outside the light-cylinder \citep{1994MNRAS.269..191S}. Moreover, possible synchro-Compton emission in the vicinity of the light-cylinder was already reported by \cite{1975Ap&SS..33..111F}.

To summarizes in a very condensed way the results obtained so far from numerical simulations of relativistic plasmas in force-free, MHD or kinetic regimes we show the fitted spindown luminosities in table~\ref{tab:FluxPoyntingModel}. Note that for all the above mentioned simulation results the Y-point is locate at the light-cylinder and therefore the spindown rate implicitly assumes that $R_{\rm Y}=\rlight$. However according to the axisymmetric FFE magnetosphere constructed by \cite{2006MNRAS.368.1055T}, this slowdown is drastically enhanced when the Y-point is shifted well inside the light-cylinder. As claimed by \cite{2010MNRAS.408L..41T}, the mode changing and nulling of some pulsars could be interpreted by a movable Y-point.
\begin{table}
\centering
\begin{center}
{
\rowcolors{2}{}{green!10}
\begin{tabular}{ccc}
\hline
\rowcolor{yellow} 
Plasma regime & $f(\chi)$ & Ref. \\
\hline
Vacuum & $ \approx (1-a^2) \, \sin^2 \chi$ & \citep{1955AnAp...18....1D} \\
QED vacuum & $\approx (1.0-a^2+O(\alpha_{\rm sf})) \, \sin^2 \chi$ & \cite{2016arXiv160705935P} \\
GR vacuum & $\approx (1.0+1.1\,a) \, \sin^2 \chi$ & \citep{2014MNRAS.439.1071P, 2016MNRAS.455.3779P} \\
GRQED vacuum & $\approx (1.0+1.1\,a+O(\alpha_{\rm sf})) \, \sin^2 \chi$ & \cite{2016arXiv160705935P} \\
FFE & $\approx \frac{3}{2} \, ( 1.0 + 1.2 \, \sin^2 \chi ) $ & \citep{2006ApJ...648L..51S, 2012MNRAS.424..605P} \\
FFQED & $\approx \frac{3}{2} \, ( 1.0 + 1.2 \, \sin^2 \chi )$ & \cite{2016arXiv160705935P} \\
GRFFE & $\approx \frac{3}{2} \, ( 1.1 + 1.6\,\sin^2 \chi ) $ & \citep{2016MNRAS.455.3779P} \\
GRFFQED & $\approx \frac{3}{2} \, ( 1.1 + 1.6 \, \sin^2 \chi )$ & \cite{2016arXiv160705935P} \\
MHD & $\approx \frac{3}{2} \, ( 1.0 + 1.2 \,\sin^2 \chi ) $ & \citep{2013MNRAS.435L...1T} \\
\hline
\end{tabular}
}
\end{center}
\caption{Spindown luminosity expectations from simulations assuming different plasma regimes. The results are $f(\chi) = f_0 + f_1 \, \sin^2 \chi$ with $a=R/\rlight$ and for $R=2\,\Rs$ in GR. For force-free simulations, the coefficients $f_0, f_1$ depend slightly on~$a$, they are not included here but given for $a=0.1$.}
\label{tab:FluxPoyntingModel}
\end{table}

\section{Electrosphere models}
\label{sec:Electrosphere}

All previous models assumed a magnetosphere entirely filled with a relativistic plasma made essentially of electron/positron pairs at a high multiplicity factor~$\kappa\gg1$ (but still not enough to fully explain observations). This implies a quasi-neutral state of the plasma. However this configuration is plausibly unstable depending on the rate of particle injection from the polar caps as observed in recent numerical simulations. The simplest idea consists therefore to construct a nearly corotative electrosphere, that is a magnetosphere partially filled with a non-neutral plasma in which charged particles, from one species or another (electrons, positrons, protons or ions), are present and rotate at a speed close but not equal to that of the star. If this non neutral plasma enters in solid body rotation with the star, then from a purely electrical point of view, nothing will distinguish this charge separated space region from the star. The neutron star can then equivalently be seen as a larger sphere of radius~$R_{\rm el}$ introduced in the braking index eq.~(\ref{eq:IndiceFreinage}). The impossibility to exceed the speed of light and the hypothesis of synchronous solid body rotation shows that this electrosphere cannot extend farther than the radius of the light cylinder. Its extension could be even less if plasma is in over-rotation as found in simulations from the middle 80s and beginning 2000. Curiously, electrospheres are neither well known nor seriously studied by authors interested in pulsar physics. We remind useful characteristics of this atypical model hoping to rise again its attractiveness. The properties of the neutron star electrosphere has been extensively studied in \cite{PetriThese2002} PhD thesis.

\subsection{Non neutral plasma behaviour}

The electrosphere model possesses a very different behaviour from that of a quasi-neutral plasma filled magnetosphere used in force-free or MHD theory. In an electrosphere, the plasma is non neutral and shows properties often opposed to those of a neutral plasma. Table~\ref{tab:PlasmaNeutrevsNonNeutre} summarizes the divergent features of the two kind of plasmas. 
\begin{table}
 \centering
{
\rowcolors{2}{}{green!10}
 \begin{tabular}{ll}
\hline
\rowcolor{yellow} 
 non neutral plasma & neutral plasma \\
\hline
one sign charge & neutral on large scales with boundary effects \\
easily trapped & unstable, diffusion, hardly trapped \\
long time scale & short time scale \\
finite volume & diffuse in space \\
sharp interface with vacuum & smooth transition \\
fixed charge density & charge density not constrained \\
freezing around the Debye length weak & recombination for a small Debye length weak \\
\hline
\end{tabular}
}
 \caption{The differences in behaviour between neutral and non neutral plasmas.}
 \label{tab:PlasmaNeutrevsNonNeutre}
\end{table}
Among them, we are particularly interested in particle confinement in electromagnetic traps with variable geometry. Depending on the topology of the magnetic and electric fields, let them be absent, constant, monopolar, dipolar or quadrupolar, the volume of the charge separated regions will show various shapes. Table~\ref{tab:PiegeagePlasma} furnishes a list of traps often used by plasma physicists. A pulsar resembles may be to a rotating Terrella.
\begin{table}
 \centering
{
\rowcolors{2}{}{green!10}
 \begin{tabular}{llll}
\hline
\rowcolor{yellow} 
 TRap geometry & Name & Plasma configuration & Remarks \\
\hline
B=0; E=monopole & Charged sphere & Keplerian disk & Classical atom \\
B=0; E=oscillating quadrupole & Paul & Time dependent & \\
B=const; E=quadrupole & Penning & Rigidly rotating sphere & \\
B=const; E=electrodes & Malmberg & Rigidly rotating ellipse & \\
B=dipole; E=monopole & Charged Terrella & Rigidly rotating disk & \\
B=dipole; E=quadrupole & Rotating Terrella & Domes and disk & Pulsar? \\
\hline
\end{tabular}
}
 \caption{Different trapping systems for a non neutral plasma. Each configuration of the electromagnetic field generates a specific shape of the space charge distribution.}
 \label{tab:PiegeagePlasma}
\end{table}
Non neutral plasmas are well studied in laboratory experiments because they are easy to confine for a long time \citep{1999RvMP...71...87D}. There are some analogies between charge separated plasmas and hydrodynamics as pointed out by \cite{1978MNRAS.182..735W} who discussed it in the context of non neutral pulsar magnetospheres.

The process of formation of this electrosphere is the following\footnote{It may be unrealistic because the magnetosphere builds up during the collapse of the progenitor and the formation of a neutron star. Nevertheless it helps to find a way to construct such solutions.}. The strongly magnetized and rotating neutron star generates surface and volume charge distributions dictated by the law of electrostatic equilibrium of a perfect conductor in its rest frame. The electric field drags particles out of the surface towards stable equilibrium positions, the so called force-free surfaces (FFS). Particles spread in the immediate stellar surrounding, filling a space charge region forming an extended atmosphere called {\it electrosphere}. The extension of this atmosphere is not dictated by thermal pressure as it would for the traditional concept of an atmosphere but rather by the electromagnetic forces acting on the charge separated gas. As for the filled magnetosphere, the electrospheric current disturbs the magnetosphere when it approaches the light cylinder. However, if over-rotation is important as we show below, this feedback could lead to perceptible magnetic perturbations already well within the light-cylinder. Moreover, owing to the strong magnetic field, all these particles quickly de-energize to their fundamental Landau level through synchro-photon emission, forbidding any motion perpendicular to magnetic field lines. They are therefore constrained to move along these field lines progressively filling the electrosphere. But then how to fill it? Will charges of opposite sign occupy one same region of space to reach a quasi neutral state or will they form what we call a charge separated electrosphere where positive and negative zones are exclusively populated by particles of one sign? Let us have a look on different models tempting to give an answer to this question, sometimes in an arbitrary manner.

\begin{figure}
\begin{center}
 \includegraphics[width=0.75\textwidth]{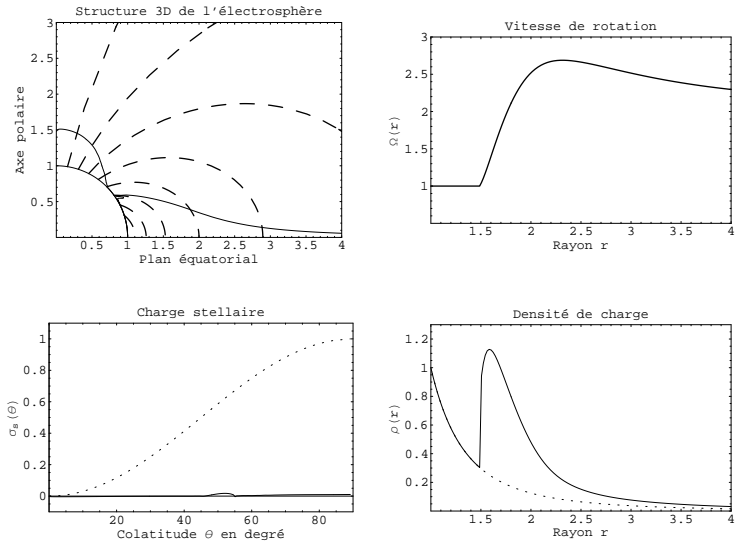} 
\caption{\label{fig:Electrosphere} Electrosphere model of an aligned rotator obtained from semi-analytical solutions \citep{2002A&A...384..414P}.} 
\end{center}
\end{figure}

Given the predominance of electromagnetic forces compared to gravitational forces and any other phenomenon related to particle inertia, it is justified to neglect their mass. Only the Lorentz force exerts a significant action. In electrostatic equilibrium this force vanishes at all places where matter subsists. In this way, in populated regions the law $\mathbf{E} + \mathbf{v}_{\rm cor} \wedge \mathbf{B} = \mathbf{0}$ is valid and electric and magnetic field are again perpendicular $\mathbf{E}\cdot\mathbf{B}=0$ as inside the star or in the force-free limit. For the sake of simplicity, we ignore relativistic effects, an approximation that is justified for an electrosphere remaining at a reasonable distance of the light cylinder, $r\ll \rlight$. Some generalisations are obviously conceivable. Building on the method invented by \cite{1985A&A...144...72K}, \cite{2002A&A...384..414P} have shown the existence of such solutions for an aligned rotator, with an extension confined well inside the light-cylinder. The solution possesses an equatorial disk in differential rotation and two domes of charge opposite to that of the disk, fig.~\ref{fig:Electrosphere}. This differential rotation imposes a velocity larger than the stellar rotation, a new but also very important aspect with deep consequences for the stability and long term evolution of such plasmas. A pulsar maybe represents an astrophysical application of particle trapping in a rotating Terrella.

\subsection{Expectations from the model}

The corotation of the electrosphere with the star stops at the light cylinder, for~$r>\rlight$ or even at shorter distances if over-rotation happens. Several developments have been proposed to replace the notion of corotation, some of them are presented here. An important theorem due to \cite{1974ApJ...190..391P} derived from not too restrictive assumptions shows that a magnetosphere finite in extent with large vacuum gaps and in force-free equilibrium cannot be in corotation with the star everywhere. As a consequence, differential rotation is an intrinsic property of electrospheres or more generally speaking magnetospheres with vacuum gaps \citep{1979ApJ...227..579M}. Differential rotation of plasmas around neutron stars was expected since the early days of pulsar theory as accounted by the relativistic hydrodynamical study of \cite{1974ApJ...192..703H} who showed the finite extension of the corotation part up to a critical magnetic surface. The trapping regions are defined by the force-free condition $\mathbf{E} \cdot \mathbf{B} = 0$ corresponding to an extremum in the electrostatic potential (maximum for positive charges and minimum for negative charges). Such sites represent therefore equilibrium places for one species, positive if the potential is minimal or negative if the potential is maximal. These regions are called force-free surfaces (FFS) and have been extensively studied by many authors as a base for the geometry of this electrosphere \citep{1989ApJS...71..583T, 1994ApJ...431..718T}. \cite{1978ApJ...222..675J} emphasized also the notion of corotational drift surface (CDS) in the arbitrary inclination angle case. \cite{1976ApJ...206..831J} extensively studied the electrostatic problem of neutron star atmospheres in a series of papers \citep{1978ApJ...222..675J, 1979ApJ...227..266J, 1980ApJ...238.1081J, 1980ApJ...237..198J, 1981ApJ...251..665J, 1981ApJ...247..650J}. \cite{1976SvA....20...23R} computed analytical models of the regions around neutron stars filled with electrons and protons/positrons. For the aligned case, he found a proton belt and an electron cap. For a small misalignment he got an electron filament current along the poles. He already noted the importance of the total electric charge of the star. He refined his model in several subsequent papers about the electron polar cap shape \citep{1977Ap&SS..51...59R, 1985Ap&SS.117....5R}, electron and particle acceleration \citep{1979Ap&SS..66..401R, 1987SvA....31..640R, 1987Ap&SS.132..353R} and the influence on pair creation in the magnetosphere \citep{1981Ap&SS..75..423R, 1982Ap&SS..88..173R, 1984Ap&SS.107..381R} and the global structure of the magnetosphere \citep{1988Ap&SS.143..269R, 1989Ap&SS.158..297R}. According to this literature, the 70s and 80s were the favourable periods to work on the topic discussed in this section. Or maybe it has to be traced back to the perspicacity of these authors only.

\subsubsection{Charged wind}

The improperly called open field lines\footnote{In an electrosphere, no large current is able to truly open field lines because no current exists.}, that is those who do not close inside the light cylinder, let particles escape from the poles. A {\it charged wind} made either of positrons/protons/ions either of electrons, following the sign of the charge, leaves the magnetosphere from the magnetic poles. The electromagnetic field in the zone of the charged wind finds its source in the current and the distribution of charges induced by these escaping particles. This current, responsible for the loss of charges around both polar caps, should discharge the neutron star. This discharge can not last for ever, so we must think of either a complete electrostatic equilibrium state or either that the current loop closes somewhere in the system. In order to circumvent this difficulty, this current loop should close inside the light cylinder, which is difficult because of the constraint imposed to the particles to stay on drifting orbits along field lines. This difficulty is known as the problem of current closure. This is why \cite{1983ZhETF..85..401B, 1993ppm..book.....B} suggest the notion of active magnetosphere. A similar argument leading to transfield flows was suggested by \cite{1986Ap&SS.119..105S, 1988MNRAS.233..405S}.

\subsubsection{Active magnetosphere}

Given that the current above the poles streams away from the pulsar, a return current must necessarily exist and be pointed towards the pulsar. The current closure is insured thanks to the violation of the electric drift approximation close to the light surface because $E>c\,B$. Indeed, when approaching this light surface, the drift velocity becomes equal to the speed of light, particle energy increases quickly forcing acceleration and consequently violating the ideal MHD approximation. Electrons and positrons are allowed to cross field lines along this light surface, electrons going in one direction and positrons in opposite direction. The current closes through the magnetic surface lying on the frontier of the dead zone, boundary surface separating open from closed field lines.

\subsubsection{Electrostatic equilibrium}

The current closure can be avoided if the system evolves towards an electrosphere in equilibrium with an extension less than the light cylinder radius~$\rlight$. The distribution of atmospheric charges would be in an equilibrium state under the action of the electromagnetic field, equilibrium that can be qualified as electrostatic. In a second approximation, we can envisage the breakdown of the frozen in theorem through the development of instabilities allowing the passage of a resistive and turbulent current. The turbulence is an effect felt by each particle in addition to the macroscopic electric field responsible for the drift motion such as microscopic electric fluctuations similar to micro-fields responsible for coulombian collisions, but much more intense. In these conditions particles can exceptionally deviate from the trajectories indicated by field lines. This aspect is related to non neutral plasma instabilities developing in the electrosphere. This suggestion is also an alternative for the current closure problem. Deviation from a pure equilibrium is required to ignite an electromagnetic activity in the pulsar and to hope to observe emission. Depending on the charge load in this electrosphere, if it almost entirely fills the light-cylinder, we could speak about a partially filled magnetosphere but with huge gaps.

\subsection{Magnetosphere partially filled}

Empty region in the magnetosphere where introduced by~\cite{1973NPhS..246....6H, 1975MNRAS.171..619H} to solve some contradictions appearing in the model of~\cite{1969ApJ...157..869G}. Indeed, in the latter, the suppression of equatorial charges can not be compensated by ions emanating from the star because there exist no mean to accelerate them from the poles without conveying negative charges at the surface. From this, \cite{1973NPhS..246....6H} conclude that the electrosphere should split at the interface between positive and negative space charges in order to let room for empty regions denoted traditionally by gaps. Moreover, if an electron-ion or an electron-positron pair migrates to a negatively charged zone, the positive particle would immediately be attracted by the positive charge region\,\footnote{This shows the extreme stability of a non neutral plasma and whose properties are drastically different from those of a traditional neutral plasma.} by crossing the gap. This motion seems at first sight paradoxical but is the result of the electromotive field. The system settles down to a new stable equilibrium state after the perturbation decayed. The same thing would happen if an electron-ion or electron-positron pair would be located in the positively charged zone. Several theoreticians have contributed to the study of the properties of this charge separated magnetosphere including vacuum gaps. But none of these authors have presented a self-consistent electrostatic model of the charge repartition in the electrosphere apart those resorting to numerical techniques.

\subsection{PIC and fluid simulations}

In the middle of the 1980, \cite{1985A&A...144...72K} presented a stationary solution for the self-consistent electrosphere, stable and finite in extent, avoiding all the complications caused by the limit of the light cylinder radius and the current closure problem. More refined calculations have then been presented by \cite{1985MNRAS.213P..43K}. The purely numerical approach employed revealed a structure physically more realistic than those from \cite{1969ApJ...157..869G} with huge gaps between charge separated regions of opposite sign. Using a boundary element method described in \cite{1989Ap&SS.161..145S}, \cite{1989Ap&SS.161..187S} confirmed the electrospheric structure followed several years later by \cite{1993A&A...268..705Z}.

The basic idea to construct this model was to pull charges out from the surface of the pulsar to spread them in vacuum following magnetic field lines until reaching an equilibrium state in which electric field and magnetic field are perpendicular, the force-free surfaces. To achieve their goal, they used a N-body code method in which charges were symbolised by rings to account for the symmetry of the configuration, an aligned rotator. These rings were obliged to follow field lines from which they were emitted until they immobilize in the potential wells. In estimating the electric field, care must be taken from the contribution of the star itself, its central point charge (for a dipolar magnetization), as well as from the rings. As soon as charges were at the right places, in an equilibrium position, they generate a potential at the surface of the pulsar and a novel distribution of charges by electrostatic influence. These new charges must also be sent into the electrosphere until complete exhaustion of charges located at the stellar crust. Simulations performed for different values of the total charge of the system showed that stable solutions in electrostatic equilibrium really exist. Unfortunately these simulations did not gave any clues to the exact structure of the electrosphere. Indeed, the nature of these simulations did not permit to compute the plasma density nor the precise shape of the frontier separating electrosphere and vacuum, the discretization of charges leading to only a crude representation. \cite{2009ApJ...690...13M} took over this technique. They developed a 3D electromagnetic PIC code in order to construct electrospheres in the general case of an oblique rotator with better resolution of the plasma configuration and a larger number of particles (or ring in axisymmetry). Following the same line, starting from a magnetosphere solution \`a la force-free of Goldreich-Julian type, \cite{2001MNRAS.322..209S} showed that it is unstable and collapse to an electrosphere. On a more fundamental side, \cite{1989PhRvA..40.3769Z, 1991Ap&SS.176..105Z} analysed trapped particle trajectories inside the magnetosphere and wave field. They found bounded orbits outside the light-cylinder and speculated about radiation from those particles.

Particle techniques are useful but fluid approaches are less noisy and offer a complementary view. Let us briefly mention some early attempts. \cite{1974PhRvL..32.1019K} performed self-consistent relativistic two-fluid simulations of the aligned pulsar magnetosphere and found closed field lines even beyond the light-cylinder which seems to contradict theoretical expectations. It is not clear if this is due to their dissipative scheme or massive particle effects \citep{1978MNRAS.182..157W} but the distinction between neutral and charge separated plasma is essential. The only source of charge being the star, \cite{1975ApJ...202..762K} two-fluid simulations showed closed field lines everywhere and particle crossing magnetic surface due to strong electric fields induced by charge separation. The volume and surface charge distribution within the star has been given by \cite{1976PhRvL..36..686P} who also pointed out the importance of the central point charge. Early numerical techniques are described by \cite{1976CoPhC..12....9P}.

\subsection{Stability}

The electrosphere found in simulations clearly shows a differential rotation of the equatorial disk. This feature was not observed in force-free simulations. This new degree of freedom stores kinetic energy that is released via instabilities arising due to the plasma differential rotation. This rotation can strongly impact on the structure and dynamics of the magnetosphere. A linear analysis performed by \cite{2012A&A...541A.117U} revealed growth rates of the order of the rotation period leading to a plasma diffusion within the magnetosphere on very short time scales. Non neutral plasma instabilities contribute also strongly to modify the traditional view of the magnetosphere. The diocotron and magnetron instabilities allow efficient diffusion of charges through field lines and breaks the frozen in approximation of the magnetic field. According to the work of \cite{2002A&A...387..520P} and \cite{2003A&A...411..203P, 2007A&A...464..135P} the diocotron instability seems to efficiently diffuse charges. Its growth rate is comparable to the rotation velocity of the star thus acting on a very short time scale. Inclusion of relativistic effects as reported by \cite{2007A&A...469..843P} or for the magnetron instability detailed in \cite{2008A&A...478...31P} leave these conclusions unchanged. 2D electrostatic PIC simulations of \cite{2009A&A...503....1P} have definitively shown the importance of these effects on pulsar electrodynamics. MHD type instabilities of non-neutral plasmas can lead to short time variability in the magnetosphere possibly related to radio emission fluctuations \citep{2014A&A...563A..29U}. Moreover, the evolution of the non-neutral plasma, especially in the disk, has to satisfy some conservation laws \citep{2005A&A...434..405A} stipulating that an isolated disk, i.e. without particle injection, will remain confined in the vicinity of the neutron star.

\begin{figure}
\centering
\begin{tikzpicture}[photon/.style={decorate,decoration={snake,post length=1mm}}]
\begin{scope}
\clip[scale=1] (0,0) rectangle (5,3);
\draw[] (0,0) -- (4.5,0) ;
\filldraw [green,thick] (0.2,0.4)  .. controls (1,1.)  and (1.5,2)  ..  (0.85,3.1) -- cycle  ; 
\filldraw [red,thick] (0,0.)  .. controls (0.3,1.2)  and (2.,0.2) ..  (3,0) -- cycle  ; 
\filldraw[inner color=white,outer color=black] (0,0) circle (0.5);
\draw[blue,thick,->] (2.3,1) arc (40:60:3) ;
\draw[blue,thick,<-] (3.1,0.1) arc (30:50:3) ;
\end{scope}
\draw[thick, black] (4,-0.2) -- (4,4) ; 
\draw[thick, black] (0,-0.2) -- (0,4) ; 
\draw (4,-0.1) node [below] {$\rlight$} ;
\draw (2.5,1.) node {$e^\pm$} ;
\draw[black,<-,thick] (0.7,1.6) -- (5,3) node [right,text=green] {dome} ;
\draw[black,<-,thick] (3,1) -- (5,2) node [right,text=blue] {gaps with pair creation} ;
\draw[black,<-,thick] (1.2,0.2) -- (5,1) node [right,text=red] {disk in differential rotation} ;
\draw[->,thick,red,photon] (1,-0.2) -- (3,-0.2) node [right] {$e^+$} ;
\draw[<-,thick] (2,-0.2) -- (5,0.3) node [right,text=red] {equatorial current} ;
\draw[->,thick,green,photon] (0.2,1) -- (0.2,3) node [right] {$e^-$} ;
\draw[<-,thick] (0.2,2) -- (5,4) node [right,text=green] {polar wind} ;
\end{tikzpicture}
\caption{An electrospheric model for pulsars. Adapted from \cite{PetriThese2002}. The activity of this dead electrosphere could be revived by an equatorial current transporting charges across field lines due to non neutral plasma instabilities and a polar wind made of charges of opposite sign to compensate for the equator loss of charges.}
\label{fig:ModeleElectrosphere}
\end{figure}
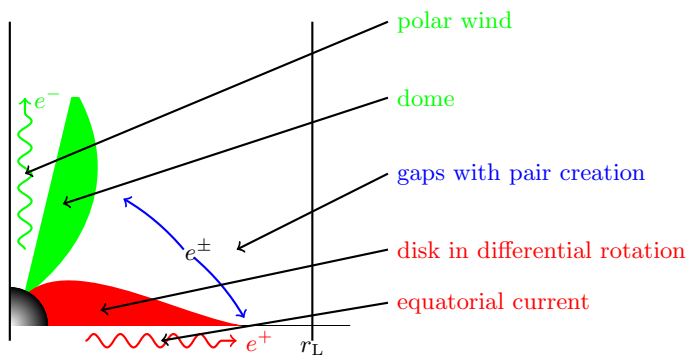

To conclude the pulsar magnetosphere/electrosphere story, table~\ref{tab:PulsarModele} summarizes the basic models of a pulsar and table~\ref{tab:Pulsarparameter} estimates the essential parameters for the characteristics quantities of a pulsar magnetosphere. Fig.~\ref{fig:ModeleElectrosphere} summarized schematically the revival of an electrosphere as an active pulsar with leptonic outflows along the rotation axis and equatorial plane. Early particle simulations of \cite{2007MNRAS.376.1460W} tend to prove the possibility of formation of such charged winds.

The plasma inside the light-cylinder is at the base of the wind we know describe.

\begin{table}
 \begin{center}
{
\rowcolors{2}{}{green!10}
\begin{tabular}{lll}
\hline
\rowcolor{yellow} 
Model &  & Reference \\
\hline
Oblique rotator in vacuum & & \cite{1955AnAp...18....1D} \\
Neutron star & & \cite{1967Natur.216..567P} \\
Bunch of corotating particles & & \cite{1968Natur.218..731G} \\
Aligned rotator and plasma source & & \cite{1969ApJ...157..869G} \\
Aligned rotator and pair creation & & \cite{1970Natur.227..465S} \\
Polar cap, cavities, discharge & & \cite{1975ApJ...196...51R} \\
Outer gaps & & \cite{1986ApJ...300..500C} \\
Slot gaps & & \cite{1983ApJ...266..215A} \\
Trapping of charges & & \cite{1985MNRAS.213P..43K} \\
Keplerian disk& & \cite{1981ApJ...251..654M} \\
\hline
 \end{tabular}
 }
 \end{center}
\caption{\label{tab:PulsarModele}The essential models describing the magnetosphere activity of a pulsar.}
\end{table} 

\begin{table}
 \begin{center}
{
\rowcolors{2}{}{green!10}
\begin{tabular}{llll}
\hline
\rowcolor{yellow}
Quantity & Estimate & Second & Millisecond \\
\hline
Mass $(M_\odot$) & $M$ & 1.4 & 1.4 \\
Radius (km) & $R$ & 12 & 12 \\
Moment of inertia (kg\,m$^2$) & $I=\frac{2}{5}\,M\,R^2$ & $1.6\times10^{38}$ & $1.6\times10^{38}$ \\
Period (s) & $P$ & 1 & $10^{-3}$ \\
Rotation velocity (rad/s) & $\Omega=\frac{2\,\pi}{P}$ & 6.283 & 6\,283 \\
Braking ($s/s$) & $\dot{P}$ & $10^{-15}$ & $10^{-18}$ \\
Luminosity (W) & $L=4\,\pi^2\,I\,\dot{P}\,P^{-3}$ & $6.3\times10^{24}$ & $6.3\times10^{30}$ \\
Magnetic field at surface (T) & $B=\sqrt{\frac{3\,\mu_0 \, c^3}{32\,\pi^3}} \, \frac{\sqrt{I\,P\,\dot{P}}}{R^3}$ & $7.4\times10^{7}$ & $7.4\times10^{4}$ \\
Magnetic field at $\rlight$ (T) & $B_L=B\,\frac{R^3}{\rlight^3}$ & $1.6\times10^{-3}$ & $1.6\times10^{3}$ \\
Magnetic moment (A\,m$^2$) & $\mu=4\,\pi\,\frac{B\,R^3}{\mu_0}$ & $1.7\times10^{27}$ & $1.7\times10^{24}$ \\
Electric field (V/m) & $E=\Omega\,B\,R$ & $7.5\times10^{12}$ & $7.5\times10^{12}$ \\
Gavitational/electric force & $\frac{G\,M\,\masselec}{R^2\,e\,E} $ & $9.7\times10^{-12}$ & $9.7\times10^{-12}$ \\
Light cylinder radius (km) & $\rlight=\frac{c}{\Omega}$ & 47\,700 & 47.7 \\
Polar cap radius (m) & $R_{\rm cp}=R\,\sqrt{\frac{R}{\rlight}}$ & 190 & 6\,017 \\
Potential drop across a polar cap (V) & $\Delta \phi_{\rm cp} = \frac{\Omega\,B\,R^3}{\rlight}$ & $2.2\times10^{13}$ & $2.2\times10^{16}$ \\
Potential drop from pole to equator (V) & $\Delta \phi =\Omega\,B\,R^2$ & $9.0\times10^{16}$ & $9.0\times10^{16}$ \\
Particle number density at $R$ (m$^{-3}$) & $n = 2\,\varepsilon_0\,\frac{\Omega\,B}{e}$ & $6.9\times10^{16}$ & $6.9\times10^{16}$ \\
Particle number density at $\rlight$ (m$^{-3}$) & & $1.1\times10^{6}$ & $1.1\times10^{15}$ \\
Particle flux ($s^{-1}$) & $\mathcal{F} = \frac{4\,\pi\,\varepsilon_0}{e} \, \Omega^2 \, B \, R^3$ & $7.5\times10^{29}$ & $7.5\times10^{32}$ \\
Plasma frequency at $R$ (Hz) & $\nup = \frac{1}{2\,\pi} \, \sqrt{\frac{n\,e^2}{\varepsilon_0\,\masselec}}$ & $2.3\times10^{9}$ & $2.3\times10^{9}$ \\
Plasma frequency at $\rlight$é(Hz) & & $9.4\times10^3$ & $2.9\times10^{8}$ \\
Cyclotron frequency at $R$~(Hz) & $\nuc = \frac{e\,B}{2\,\pi\,\masselec}$ & $2.8\times10^{18}$ & $2.8\times10^{15}$ \\
Cyclotron frequency at $\rlight$ (Hz) & & $4.5\times10^{7}$ & $4.5\times10^{13}$ \\
Characteristic age (years) & $\tau = \frac{P}{2\,\dot{P}}$ & $1.6\times10^7$ & $1.6\times10^7$ \\
Gravitational potential energy (J) & $E_{\rm g} = \frac{3}{5}\,\frac{G\,M^2}{R}$ & \numprint{2.6e46} & \numprint{2.6e46} \\
Rotational kinetic energy (J) & $E_{\rm k} = \frac{1}{2}\,I\,\Omega^2$ & \numprint{3.2e39} & \numprint{3.2e45} \\
Magnetic energy (J) & $E_{\rm B} = \frac{4\,\pi}{3}\,\frac{B^2\,R^3}{2\,\mu_0}$ & \numprint{1.62e34} & \numprint{1.62e28} \\
Thermal energy (J) & $E_{\rm th} = \frac{3}{2}\,N\,k\,T$ & \numprint{3.4e40} & \numprint{3.4e40} \\
\hline
 \end{tabular}
 }
 \end{center}
\caption[Pulsar fundamental parameters]{\label{tab:Pulsarparameter} The fundamental parameters of a normal and a millisecond pulsar.}
\end{table}

\section{Pulsars winds}
\label{sec:Vent}

It is often assumed that pulsars lose their rotational kinetic energy through the formation of an ultra-relativistic and magnetized wind, made essentially of leptonic~$e^\pm$ pairs, and not just magnetodipole losses in vacuum which would contradict broad band pulsed emission. This energy, drawn from the rotational kinetic energy of the central star, is extracted via the Lorentz force exerted at the stellar crust $\iint \mathbf{i}_{\rm s} \wedge \mathbf{B} \, dS$ and carried away in an electromagnetic wave: the Poynting flux where $\mathbf{i}_{\rm s}$ is the surface charge current and $dS$ the surface element. If surface charges are present, the electric force also contributes to the spindown in the form $\iint \sigma_{\rm e} \, [\mathbf{E}] \, dS$ where $\sigma_{\rm e}$ is the surface electric charge and $[\mathbf{E}]$ the jump in electric field across the same surface. Schematically, from an electrical point of view, the system generates a potential drop, the magnetized star delivering a potential difference equal to that between the centre and the rim of a polar cap, electric wires are replaced by open magnetic field lines and the resistive charge by the nebula acting as a calorimeter. The wind expands from the external parts of the pulsar magnetosphere, through the vicinity of the light cylinder, up to the neighbouring nebula and feeding it with freshly made ultra-relativistic particles. Evolving in a magnetic field, theses particles emit synchrotron and inverse Compton radiation, detectable as for instance in the famous Crab nebula. \footnote{See \cite{2008ARA&A..46..127H} for a review about the Crab and \cite{2009ASSL..357..421K} for a summary about pulsar wind and nebula (PWN) theory. Moreover, the catalog of PWNs can be found in Roberts, M.S.E., 2004, `The Pulsar Wind Nebula Catalog (March 2005 version)', McGill University, Montr\'eal, Quebec, Canada (available on the World-Wide-Web at "http://www.physics.mcgill.ca/~pulsar/pwncat.html").}
As a general picture, magnetized ultra-relativistic winds are thought to find their source in a compact object, neutron star or black hole. The flow, dominated by the Poynting flux, helps in the modelling of some quasars and gamma-ray burst as well \citep{2002luml.conf..381B}.

\subsection{Introduction}

Pulsar radio luminosity only represents a tiny amount of their total energy losses, of the order of $\numprint{e-5}\,L_{\rm rot}$. It is therefore believed that the major part of its rotational kinetic energy is expelled through a relativistic charged particle outflow: the pulsar wind. This fact is confirmed by observations showing the interaction between this wind and its surrounding nebula. In such picture, the luminosity of the Crab nebula is explained by synchrotron radiation of ultra-relativistic electrons emanating from the central neutron star.

The problem of pulsar wind theory consists in elaborating a mechanism susceptible to convert the Poynting flux of the large amplitude low frequency strong electromagnetic wave into particle kinetic energy, as well as an acceleration process for these latter. By large amplitude we mean an electron gyro-frequency~$\nu_{\rm B}$ much greater than the wave frequency~$\nu$, in other words $\nu_{\rm B} \gg \nu$. For a pulsar, typical parameters are $\nu \approx \numprint{0.1}-\numprint{720}$~Hz and $\nu_B \gtrsim \numprint{e7}$~Hz.

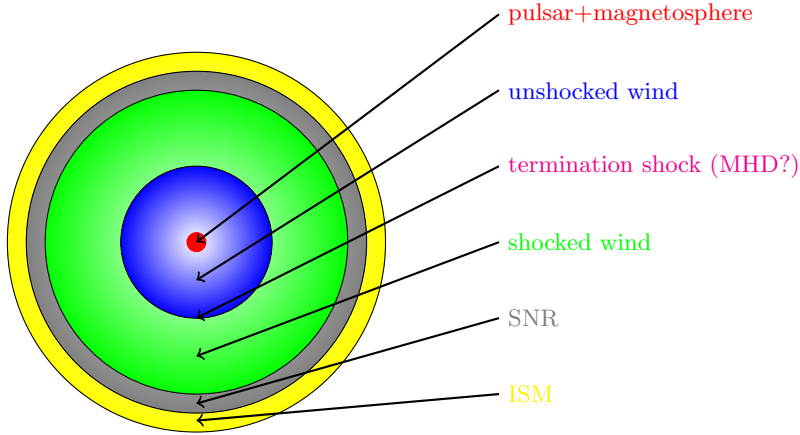
\begin{figure}
\begin{center}
 \label{fig:Nebuleuse}
 \begin{tikzpicture}
 \filldraw[inner color=white,outer color=yellow] (0,0) circle (2.5);
 \draw[black, thick, <-] (0,-2.35) -- (4,-2) node[right, text=yellow] {ISM};
 \filldraw[inner color=white,outer color=gray] (0,0) circle (2.25);
 \draw[black, thick, <-] (0,-2.125) -- (4,-1) node[right, text=gray] {SNR};
 \filldraw[inner color=white,outer color=green] (0,0) circle (2);
 \draw[black, thick, <-] (0,-1.5) -- (4,0) node[right, text=green] {shocked wind};
 \filldraw[inner color=white,outer color=blue] (0,0) circle (1);
 \draw[black, thick, <-] (0,-0.5) -- (4,2) node[right, text=blue] {unshocked wind};
 \draw[black, thick, <-] (0,-1) -- (4,1) node[right, text=magenta] {termination shock (MHD?)};
 \filldraw[red] (0,0) circle (0.125);
 \draw[black, thick, <-] (0,0) -- (4,3) node[right, text=red] {pulsar+magnetosphere};
 \end{tikzpicture}
 \caption{Link between the pulsar and its surrounding nebula. In red, the pulsar and its magnetosphere, source of $e^\pm$ pairs, in green, the wind in free almost ballistic expansion with a Lorentz factor~$\Gamma_{\rm v}$, in blue the shocked wind, in grey the supernova remnant and in yellow the interstellar medium. The termination shock is the boundary between the shocked (green) and unshocked (blue) wind.}
\end{center}  
\end{figure}

The link between the central pulsar, the supernova remnant and the nebula is well established. Let us recall the bottom line of this model, fig.~\ref{fig:Nebuleuse}. At the source, in the centre of the nebula the pulsar and its magnetosphere generates ultra-relativistic pairs~$e^\pm$. From faraway regions of the magnetosphere a cold ultra-relativistic wind forms and flows out towards the nebula, in a ballistic motion, that is a free expansion up to the termination shock (this latter being usually modelled in the ideal MHD regime) where particles are heated after crossing the shock to produce the shocked wind, in blue. This shocked wind is the main source of radiation observed in radio, optical, X~rays and gamma rays. The nebula is surrounded by the supernova remnant, in grey, itself imprisoned by the interstellar medium, in yellow. The transition between the unshocked and the shocked wind goes through the termination shock. The pre and post shock flow properties are radically different from a thermodynamic but also from a radiative point of view (particle distribution function, power law index).

\subsection{Basic theory}

It is good to remind that the exact nature of the pulsar wind remains mysterious, even basic properties such as its composition (leptonic plus a fraction of baryonic matter?) is unknown. We quickly come up against conceptual difficulties. However pulsar winds fall essentially into three kind of descriptions ordered in a decreasing plasma particle density as follows
\begin{itemize}
 \item a quasi-neutral wind of relativistic particles, usually described by the relativistic MHD formalism. This is the usual sense given to the notion of a wind. The electric current is arbitrary because generated by the relative velocity between different species of opposite charge. It requires a large particle density number.
 \item a relativistic charged wind. Here intervenes an additional complication on account of the charge separation between particles of opposite sign. The electric current is no more arbitrary but explicitly linked to the velocity of the flow and to the charge density, it is only a convective current. It implies a low particle density number.
 \item a large amplitude low frequency electromagnetic wave propagating into a low density plasma, particles surfing in a way on this wave with negligible back reaction of the plasma onto this wave. The electric current does not induce perceptible perturbations on this wave.
\end{itemize}
It is impossible to state which of this outflow prevails in pulsar wind but it is believed that the wind cannot switch from one regime to another during its propagation towards the termination shock.
The formation process of this wind in the vicinity of the neutron star, its propagation as well as its interaction with the nebula are still controversial. Theoretical investigations on pulsar winds mainly focused on propagation effects, little being known about its generation and repercussions on the nebula. The formation of the wind is the worse understood part.

\subsection{Magnetohydrodynamic models}

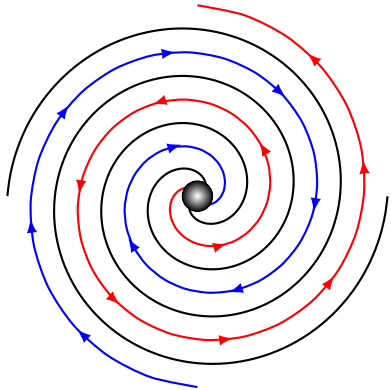
\begin{figure}
\centering
\begin{tikzpicture}
\draw [thick, domain=0:4*pi,variable=\t,smooth,samples=150]
        plot ({\t r}: {0.2*\t});
\draw [ thick, domain=-pi:3*pi,variable=\t,smooth,samples=150]
        plot ({\t r}: {0.2*(\t+pi)});
\draw [decoration={ markings,
      mark=at position 0.1 with {\arrow{latex reversed}}, 
      mark=at position 0.2 with {\arrow{latex reversed}},
      mark=at position 0.3 with {\arrow{latex reversed}},  
      mark=at position 0.4 with {\arrow{latex reversed}},  
      mark=at position 0.5 with {\arrow{latex reversed}}, 
      mark=at position 0.6 with {\arrow{latex reversed}},  
      mark=at position 0.7 with {\arrow{latex reversed}},  
      mark=at position 0.8 with {\arrow{latex reversed}},  
      mark=at position 0.9 with {\arrow{latex reversed}}},  postaction={decorate}, 
      blue, thick, domain=-0.5*pi:3.5*pi,variable=\t,smooth,samples=50]
        plot ({\t r}: {0.2*(\t+0.5*pi)});
\draw [decoration={ markings,
      mark=at position 0.1 with {\arrow{latex}}, 
      mark=at position 0.2 with {\arrow{latex}},
      mark=at position 0.3 with {\arrow{latex}},  
      mark=at position 0.4 with {\arrow{latex}},  
      mark=at position 0.5 with {\arrow{latex}}, 
      mark=at position 0.6 with {\arrow{latex}},  
      mark=at position 0.7 with {\arrow{latex}},  
      mark=at position 0.8 with {\arrow{latex}},  
      mark=at position 0.9 with {\arrow{latex}}},  postaction={decorate}, 
      red, thick, domain=-1.5*pi:2.5*pi,variable=\t,smooth,samples=50]
        plot ({\t r}: {0.2*(\t+1.5*pi)});
\filldraw[inner color=white,outer color=black] (0,0) circle (0.2);
\end{tikzpicture}
\caption{The Parker spiral structure of the solar wind. Outgoing field lines are shown in red, ingoing field lines in blue. The two black spirals correspond to places where magnetic polarity reverses. The same applies for the pulsar striped wind, see below.}
\label{fig:SpiraleParker}
\end{figure}

The modelling of pulsar winds goes back to the late 60s. Indeed, the first model of relativistic wind from a compact object was proposed by \cite{1969ApJ...158..727M} as an extension of the solar wind theory exposed by \cite{1964Natur.202..432D}, \cite{1967ApJ...148..217W} and \cite{1967JGR....72.1521M}. The solar wind is a non relativistic flow described by a fluid dominated by the pressure and not by the magnetic field. In the relativistic wind model, the magnetic field is monopolar, field lines are radial and symmetric with respect to the stellar rotation axis. On the contrary to the solar wind, pressure as well as gravity were ignored. Because of the pulsar rotation, field lines roll up forming a spiral very similar to Parker spiral \citep{1958ApJ...128..664P}. This structure will be met later again when explaining the pulsar striped wind model, fig.~\ref{fig:SpiraleParker}. Particles, required to move along these lines, are then accelerated by catapult effect. In all these computations, the dimensionless quantity, sometimes called magnetization
\begin{equation}
  \mu = \frac{e\,\Phi}{\masselec\,c^2} = \frac{e\,\Omega\,B\,R^2}{\masselec\,c^2}  
\end{equation}
introduced by \cite{1969ApJ...158..727M} plays a significant role where $\Phi$ is the electric potential drop. Physically, this parameter can be interpreted as the maximal Lorentz factor reached by particles when considering that all the Poynting flux goes into kinetic energy for the particles. From an electrostatic point of view, $\Phi$ is the maximum potential drop between the magnetic poles and the equator for an aligned rotator $\Phi=\Omega\,B\,R^2$. A probably less optimistic but better estimate of this electric potential drop is to take it along the polar cap from the centre to the rim such that $\Phi=\Omega^2\,B\,R^3/c$. However, previous studies have shown that the Lorentz factor related to this flow velocity of the wind is relatively low, the asymptotic Lorentz factor being only about~$\mu^{1/3}$. In a space-charge limited flow, acceleration can reach Lorentz factor of the order~$\mu^{1/2}$ \cite{1974ApJ...192..713M}. This upper limit reaches $\mu^{2/3}$ for a charge separated wind \citep{1984ApJ...284..384M}. A summary is presented in table~\ref{tab:FacteurLorentzAsymp}.
\begin{table}
\centering
{
\rowcolors{2}{}{green!10}
\begin{tabular}{cc}
\hline
\rowcolor{yellow} 
regime & $\gamma_{\infty}$ \\
\hline
MHD & $\mu^{1/3}$ \\
charge-separated & $\mu^{2/3}$ \\
space-charge limited & $\mu^{1/2}$ \\
\hline
\end{tabular}
}
\caption{Asymptotic Lorentz factor reached according to three plasma regimes in the wind.}
\label{tab:FacteurLorentzAsymp}
\end{table}
Test particle acceleration has also been postulated or computed by several authors. For instance \cite{1969ApJ...157..869G} claimed that the maximal Lorentz factor looks like $\gamma_{\rm max} \approx \mu^{1/3}$. On the other hand \cite{1969ApJ...157.1395O} gave $\gamma_{\rm max} \approx (\mu\,(1-r_0/r))^{2/3}$. \cite{1977MNRAS.180..125B} claimed a linear acceleration with distance such that $\gamma_{\rm max} \approx r$. \cite{1973PhRvL..31.1364K} found $\gamma_{\rm max} \approx \mu$.

Starting from the hypothesis formulated by \cite{1969ApJ...158..727M}, \cite{1970ApJ...160..971G} have added a pressure term as well as the gravitational field of the star. Solving the mass, energy and momentum conservation equations, they obtained an algebraic system. They showed that the flow passes through three critical points that are the sonic point where the velocity of the flow reaches the sound speed, the alfvenic point and the magnetosonic point. In addition \cite{1971MNRAS.152..323H} computed the relativistic breeze solution complementary to \cite{1969ApJ...158..727M}. The magnetosonic point must however lie at a finite distance according to \cite{1979MNRAS.189..397A}.

\cite{1974MNRAS.167....1R} were the first to look for a modelling of the spatial plasma distribution in the Crab nebula adopting a purely hydrodynamic point of view and assuming spherical symmetry. All the energy coming from the pulsar accumulates in the volume of the nebula which expands at a speed~$v_{\rm neb}\ll c$. At a distance~$R_{\rm s}$, the total pressure in the nebula compensates the wind dynamical pressure. In a stationary regime~$R_s/R_{\rm neb} \approx \sqrt{v_{\rm neb}/c}$, where~$R_{\rm neb}$ is the radius of the spherically symmetric nebula. Applied to the Crab nebula, the ratio is of the order~7\%. In this region a shock forms to make the transition from the ultra-relativistic wind down to a velocity of the order of~$c/\sqrt{3}$\footnote{When crossing the shock, kinetic energy of the wind has been converted into relativistic random motion and therefore becomes isotropic, thus the factor~$\sqrt{3}$.}. Farther away from~$R_{\rm s}$, the flow becomes subsonic. The pressure will approximatively be uniform in the volume comprised between the shock zone and the envelope of the nebula, $R_{\rm s} < r < R_{\rm neb}$. The wind passes therefore from a velocity~$c/\sqrt{3}$ at~$R_s$ to a velocity~$v_{\rm neb}$ at~$R_{\rm neb}$. The absence of optical radiation in the centre of the Crab nebula for $r<R_{\rm s}$ was identified with the wind zone, rather cold, underluminous and close to the pulsar.

Starting from the model of \cite{1974MNRAS.167....1R}, \cite{1984ApJ...283..694K} studied the details of the MHD shock of the nebula including the magnetic field dynamics with application to the Crab \citep{1984ApJ...283..710K}. \cite{1987ApJ...321..334E} extended the previous solution to a time-dependent moving shock solution. To satisfy the boundary conditions at the supernova remnant (velocity and pressure), the wind must terminate by a MHD shock but essentially hydrodynamic in nature, that is a flow dominated by particle pressure. They adopted a different definition of the magnetization compared to \cite{1969ApJ...158..727M} and denoted by~$\sigma$, ratio between the electromagnetic energy flux and the particle energy flux. Its dynamics is still dominated by the Poynting flux symbolised by the magnetization parameter
\begin{equation}
  \label{eq:SigmaVent}
 \sigma = \frac{\textrm{Poynting flux}} {\textrm{particle enthalpy flux}} = \frac{B^2}{\mu_0\,\Gamma_{\rm v}\,n\,\masselec\,c^2} = \frac{\mu}{\Gamma_{\rm v}}
\end{equation}
This definition is commonly used nowadays, contrary to that of Michel with the parameter $\mu$ which seems obsolete. \cite{1984ApJ...283..710K} reproduced the optical and X-ray emission of the nebula assuming a cold wind hitting the termination shock with a Lorentz factor of $\Gamma_{\rm v} \approx 10^6$. Moreover they showed that the magnetization should be of the order~$\sigma \approx 10^{-3}$, in other words, the relativistic wind emanating from the pulsar should be very dense and weakly magnetised. A copious production of $e^+e^-$ pairs in the magnetosphere could explain this high plasma density, solving simultaneously the shock problem that would only be a Poynting dominated flow. In this model, the wind magnetic field is assumed to be essentially azimuthal, only the toroidal component~$B_\varphi$ remaining non negligible. Moreover, for the aligned rotator, the field keeps an unidirectional structure, that is, field lines cross the equatorial plane always in the same sense. See also \cite{1980A&A....83....1K} for a refinement of \cite{1974MNRAS.167....1R} early model and details about geometrical, spectral and temporal features of the Crab nebula.

\cite{1994ApJ...426..269B} studied the conversion of Poynting flux into particle kinetic energy for radial and axisymmetric flow. Under such hypotheses, they showed that plasma acceleration was extremely inefficient because of magnetic pressure cancellation by magnetic tension. But if the flow could deviate from this radial motion even slightly, it would become magnetosonic and induce a significant acceleration. Unfortunately, the magnetization parameter~$\sigma$ decreases only logarithmically with the radius, which is not sufficient to explain observations of the Crab nebula. Inefficient acceleration is counterbalanced by the finite temperature of the wind as shown by \cite{1983GApFD..26..147K} but synchrotron emission cools quickly particles. Therefore, there is no simple and satisfactory explanation to the wind acceleration up to the termination shock. \cite{1998ApJ...505..835C} showed that it is impossible to transfer electromagnetic energy flux to particles in a relativistic stationary MHD flow. Only a gradual acceleration can occur and therefore $\sigma$ remains high before the termination shock which agrees with \cite{1994ApJ...426..269B} conclusions. An abrupt acceleration not far from the light cylinder should happen.

To summarize so far pulsar wind theory, a cold MHD flow in stationary regime evolving in a monopole magnetic field is always dominated by the Poynting flux if particles are injected with~$\sigma\gg 1$, the flow reaching the magneto-sonic point asymptotically. However independent estimates from the Crab nebula furnish value of the flow parameter less than~1,~$\sigma\ll1$, a required condition for sufficient confinement pressure for keeping particles inside the supernova remnant. Numerous questions remain unsettled, as for instance the precise description of the shock, the formation of a dense wind close to the pulsar surface, the nature of the large amplitude electromagnetic wave, wave in vacuum or plasma wave, the current circulation and related MHD/kinetic instabilities.

The above models are drastic simplifications of a real system because they assumed stationarity, no explicit time dependence is included. Indeed, the firsts models have been presented by \cite{1974MNRAS.167....1R} for the Crab nebula interpreted as synchrotron emission from the relativistic shocked wind in spherical geometry and more detailed by \cite{1984ApJ...283..694K, 1984ApJ...283..710K} where they introduced a thorough study of the relativistic MHD shock. The formulation relies on three hypotheses that are 
\begin{itemize}
 \item a Larmor radius smaller than the size of the nebula.
 \item negligible radiative losses, i.e. a cooling time much longer than the age of the nebula.
 \item a plasma made almost exclusively of $e^\pm$~pairs with little ions and/or heavy elements. There is therefore no time and length scales characteristics that differ because of the mass ratio.
\end{itemize}

But a pulsar and its wind are far from being stationary. The magnetic moment inclined with respect to the rotation axis generates a variable electromagnetic field that at the light cylinder gives rise to a large amplitude low frequency electromagnetic wave damped by its interaction with the surrounding plasma causing its dissipation.

\cite{1990ApJ...349..538C} was the first to recognize the importance of the time dependence of the wind structure on the energy transport mechanism. He noted that for an oblique rotator, the azimuthal component of the magnetic field in the wind change polarity alternatively in the vicinity of the rotational equatorial plan \footnote{The rotational equatorial plan is the plan perpendicular to the pulsar rotation axis and passing through its centre.}, the flux being equal in the two alternations. The wind, qualified as a striped wind, develops into a structure made of stripes that are alternating polarity from positive to negative and vice-versa, separated by a neutral surface onto which the field vanishes: the current sheet. He demonstrated that magnetic field line annihilation of opposite polarity can lead from an initial highly magnetized configuration, a flow dominated by the propagation of electromagnetic waves at $\sigma \gg 1$, to a weakly magnetized wind, dominated by particle kinetic energy at~$\sigma \ll 1$. This annihilation is also refereed to magnetic reconnection in the striped wind.

\cite{1994ApJ...431..397M} interpreted this magnetic reconnection merely in terms of inductive heating because of the plasma short-circuit necessary to maintain the current. The density of particles responsible for this electric current maintaining the striped structure decreases radially faster, like $n \propto 1/r^2$, than the amplitude of the magnetic field, like $B \propto 1/r$. However Maxwell-Amp\`ere equation imposes a radial decrease identical for both the density~$n$ and magnetic field~$B$ leading to the contradiction. The difficulty is circumvented by draining the reservoir of cold and magnetized particles making them join those that are hot and weakly magnetized. This source shrinks until exhaustion and dissipation of the field itself. More clearly, particles start lacking to maintain the current and to insure the existence of the stripes that have no other choice than to dissipate. This problem between the charge density and the current density was already noted by \cite{1975Ap&SS..32..375U}.

The striped wind shows the peculiarity of alternating polarity in the magnetic field in the equatorial plane. An oscillating current sheet emerges out of this system and separates equatorial stripes \citep{1999A&A...349.1017B}. The striped wind is considered as an entropic wave that is a wave moving with the bulk flow without entropy exchange between different parts of the fluid. Note that energy is mainly evacuated in the equatorial region. The dynamics of the striped wind is much more rich than that of a simple spherically symmetric radial wind. Indeed, \cite{2001ApJ...547..437L} have shown that the thin current sheet represents a favourable site for magnetic field annihilation in the stripes. Magnetic energy is therefore transferred to particles via reconnection. But acceleration induced by this reconnection slows down the dissipation rate estimated by a distant observer because of time dilation, rendering this mechanism inefficient to completely dissipate the magnetic field before entering the termination shock. The Lorentz factor increases faster than logarithmically but not sufficiently, only as $\Gamma_{\rm v} \propto \sqrt{r}$. The conversion could however be possible in favourable conditions with a higher than expected density of pairs through cascading \citep{2003ApJ...591..366K}. This result contradicts the general believe stipulating a domination of particles over the electromagnetic field before passage through the termination shock. Indeed, a too high magnetization at the shock would drastically increase the post-shock pressure with as a consequence an important deformation of the nebula, which is not observed. The other hypothesis meets some difficulties to explain the radio spectrum. An alternative solution consists in dissipating the magnetic field within the termination shock \citep{2005AdSpR..35.1112L, 2007A&A...473..683P, 2011ApJ...741...39S}, and would solve the problem of a flow dominated by the Poynting flux and avoid that of the radio spectrum \citep{2003MNRAS.345..153L}. Alternatively, wave dissipation in the striped wind has been studied by \cite{2003MNRAS.339..765L} who showed the decay of fast magnetosonic waves in such winds through non-linear steepening and multiple shock formation. Superluminal waves offer another interesting point of view to dissipate efficiently electromagnetic energy at the termination shock \citep{2012ApJ...745..108A}.

In all these scenarios, whatever the situation considered, after dissipation, the alternating component of the magnetic field disappears and only the DC component subsists, obtained by averaging of the magnetic field on a wavelength of the wind. In the equatorial plane, this mean value is strictly null. However, in polar regions, the same magnetic field do not change polarity, there are no stripes to annihilate. Energy is transported via magneto-sonic waves or Alfven waves. In the asymptotic region, field lines tend to the split monopole \citep{1973ApJ...186..625I, 1974ApJ...187..585M}.
\cite{1977MNRAS.180..125B} showed that any solution possesses a neutral current sheet, that the asymptotic solution resembles to a wave in vacuum and that the particle Lorentz factor increases approximatively linearly with the distance. The flow remains essentially radial after crossing the magneto-sonic point because the collimation becomes inefficient \citep{1998MNRAS.299..341B, 1998ApJ...505..835C, 1994PASJ...46..123T}. We know since the works of \cite{1978A&A....65..401A} that low frequency waves generated by the pulsar rotation are heavily damped due to the presence of a dense plasma.

The MHD model alone, as we see, cannot explain individual acceleration of particles to power law distributions but rather as simple Maxwellian in an hypothetical thermal equilibrium state. Although the wind properties are not directly accessible to observations, an indirect deduction of the magnetization, of the angular distribution of energy and of the dissipation in the equatorial plane can be gained from numerical simulations.

\subsection{Axisymmetric magnetohydrodynamical simulations}

With the advent of relativistic numerical codes noteworthy progresses have been made in the comprehension of the geometry of pulsar wind nebulae. For instance, relativistic magnetohydrodynamic simulations (RMHD) performed by \cite{2004MNRAS.349..779K} have shown that the jet+torus structure, well resolved in the Crab nebula, can be explained by a relativistic wind possessing a weak magnetization at the equator $\sigma \ll 1$ but an important Poynting flux. The transition to the nebula goes through an anisotropic termination shock braking the wind to a speed velocity about~$c/2$. Jets are formed after this shock by magnetic confinement. Moreover, synchrotron emission resulting from such a configuration reproduces faithfully X~ray observations of the Crab nebula. The rings are easily identifiable with the symmetrical jets seeming to escape from the pulsar. Since their work, numerous other simulations have reproduced similar results, see for instance \cite{2004A&A...421.1063D}, \cite{2006A&A...453..621D}, \cite{2006MNRAS.368.1717B} and \cite{2008A&A...485..337V}.

The crucial point in these relativistic MHD simulations is the presence of an anisotropy in the Poynting flux given by a prescription in luminosity according to the formal latitude dependence \citep{2002MNRAS.329L..34L}
\begin{equation}
  F(\vartheta) = \frac{F_0}{r^2} \, \left( \frac{1}{\sigma} + \alpha \, \sin^2 \vartheta \right)
\end{equation}
where $F_0$ and $\frac{1}{\sigma}$ are two constant parameters. The first term accounts for particle energy and the second for the Poynting flux reminiscent of the striped wind. The weak magnetization in the equator is accounted by magnetic dissipation in the stripes. It seems to be the most accomplished model to explain the jet+torus structure of nebulas. The plasma flow is facilitated in the equatorial plane with a magnetic compression along the axis thus forming the jet. The formation of this jet depends on the magnetization upstream the wind. For a weak value of $\sigma$, let us say $\sigma<10^{-3}$, there is no jet but for $\sigma \approx 0.1$ a jet appears with an ejection velocity of about 0.7~$c$. Stripes add to the complexity of the flow in the nebula because they can dissipate and lower the equatorial magnetization. This is necessary to explain the inner ring and the external torus. Should the opposite occur, a constant magnetization would be impossible. The intensity of the magnetic field increases by MHD compression of the wind in the shock \citep{1969Natur.222..965P, 1973ApJ...186..249P, 1974MNRAS.167....1R}. Two-dimensional axisymmetric RMHD of \cite{2004MNRAS.349..779K} have confirmed the veracity of these hypotheses. Their conclusions are the following
\begin{itemize}
 \item at high latitude, the magnetic field is still significant, inducing a jet collimation.
 \item the termination shock is closer to the neutron star at the poles than at the equator.
\end{itemize}
The a posteriori treatment of simulation data to extract synchrotron intensity maps should convince the most sceptical.

To conclude about simulation, \cite{2014MNRAS.438..278P} reported on the first full three-dimensional relativistic MHD simulations of a pulsar wind nebula. They showed that observations can be reconciled with theory even with magnetization as high as $\sigma=3$ thanks to a kink instability occurring in the polar regions, as already mentioned by \cite{1998ApJ...493..291B}.

\subsection{Wind observability}

The striped wind being a cold flow, at first sight it seems difficult to observe it even indirectly. However, some regions in the wind are detectable whereas other should remain invisible for observers on Earth. On one hand, the unshocked part of the wind, essentially cold but made of a thin hot stripe (usually also assimilated to the current sheet), generates high and very high energy photons through essentially two main channels, namely
\begin{enumerate}
 \item inverse Compton emission on target photons emanating from 
 \begin{itemize}
  \item the cosmic microwave background.
  \item synchrotron photons (then called synchrotron self-Compton emission).
  \item thermal photons from the stellar surface (heated to X~rays thus energies about  \numprint{100}~\si{\electronvolt}) \citep{2000MNRAS.313..504B}.
  \item optical/UV photons (typical energies about several~\si{\electronvolt}) from the companion in a binary system \citep{2000APh....12..335B}.
  \item infra-red dust.
  \item the surrounding nebula. 
 \end{itemize}
 \item synchrotron emission in the dense and hot stripes and incidentally in the cold and magnetized part with prediction of the associated polarisation.
\end{enumerate}
Pulses are observable from optical up to~\si{\mega\electronvolt}/\si{\giga\electronvolt} outside the magnetosphere but close to the light cylinder where emissivity is the highest. The pulsation observability condition is constrained by relativistic beaming, retardation and geometrical effects. If emission occurs at a radius~$r$ and within a range~$\Delta r$ then according to \cite{2002A&A...388L..29K} and \cite{2009A&A...503...13P, 2011MNRAS.412.1870P} pulses are restricted to regions where
\begin{equation}
 \{r,\Delta r\} \lesssim \Gamma_{\rm v}^2 \, \rlight .
\end{equation}
On the other hand, the shocked part of the wind does not contain any stripes, they have been destroyed during the passage through the termination shock. In some special regimes of the plasma flow, stripes could survive as explained by \cite{2007A&A...473..683P}. The magnetic field leaves room for very hot particles scattering photons up to~\si{\tera\electronvolt} energies. There are nor pulses at these extreme energies but orbital phase modulation is expected, see for instance the case of PSR~B1259-63 studied by \cite{1999APh....10...31K} and \cite{2011MNRAS.417..532P}.

Because radio emission must propagate through the wind to reach the observer at Earth, the wind needs to be transparent vis-\`a-vis of inverse Compton scattering of radio photons. From this condition \cite{1978MNRAS.185..297W} deduced a minimum Lorentz factor of~$\Gamma_{\rm v}>$\numprint{e4} for the Crab at a distance of $100\,\rlight$ of the pulsar. \cite{1992ApJ...395..553S} included ambient magnetic field effects altering the spectra and radio polarisation.

We therefore conclude that the essentially cold wind is observable at least partly thanks to synchrotron and inverse Compton emission from the shocked and/or unshocked wind. The luminosity is indeed sufficient to be detectable on Earth. Moreover spectral and pulsation features differ from those of the nebula, the two components are therefore distinguishable.

\section{The striped wind}
\label{sec:VentStrie}

In this paragraph, we discuss the possible observational signatures of the striped wind on the properties of high energy emission induced by synchrotron and/or inverse Compton radiation of the current sheet containing ultra-relativistic electron-positron pairs. We focus more specifically on the pulsed component of this emission, beyond the optical domain, that is $E_\gamma \gtrsim 1$~\si{\electronvolt}\footnote{In this discussion we consider optical emission as belonging to the high energy domain by comparison with radio pulses.}. The striped wind being too rarely described in details in the literature, an effort is done to give a precise and coherent description. Many of the striped wind emission features are based on the three dimensional geometry of the underlying magnetic field. It imprints on the light-curve intensity as well as on polarisation. A more detailed model about kinetic aspects of the sheet interior would allow a survey of physical conditions within it but the large span in time and spatial scales of many decades forbids a self-consistent treatment of global scales including microphysics.

Pulsar nebulas are the natural results of the shocked plasma injected by the pulsar wind. This wind although highly magnetized is assumed to be cold, relatively homogeneous and does not radiate synchrotron photons, at most high energy inverse Compton photons from scattering an exterior target field. This explains the lack of observation between the position of the neutron star and the inner parts of the nebula. Can we however hope to observe this wind even indirectly? We tempt to answer this question in the following lines.

\subsection{Structure of the striped wind}

A quantitatively accurate description of the striped wind is still not accessible. The main problems arise because of the very different time and length scales coming into play. On one hand, the global structure at large scales is dominated by the MHD regime and on the other hand the kinetic structure of the current sheet dominates at very small scales, both being difficult to reconcile from the point of view of numerical simulations. The Larmor radius is many orders of magnitude smaller than the wind wavelength. Radiation should also strongly influence the equilibrium configuration. Strong disparities in the time scale characteristics appears for instance in the three important frequencies namely the cyclotron frequency, the plasma frequency and the rotation frequency. See table~\ref{tab:Pulsarparameter} for a summary of important pulsar parameters.

Nevertheless, the magnetic field geometry at large distances in the MHD approximation is satisfactorily described by a split monopole. The use of such a structure avoids a detailed description of the closed pulsar magnetosphere. Indeed, the plasma configuration in the immediate surrounding of the neutron star, that is for distances less than the light cylinder radius remains largely unknown and ill understood. We only guess that a transition of the magnetic topology must occur approximatively at the light cylinder, switching from a confined corotating (or maybe differentially rotating) plasma to an open topology sustained by a wind of charged particles, the pulsar wind. Despite this large uncertainty, it is possible to get a simple analytical solution for the wind, independent of the precise knowledge of this magnetosphere. It is called the split monopole of which we recall some essential features.

\subsection{The split monopole}

\begin{figure}
  \centering
  \begin{tabular}{cc}
    \includegraphics[width=0.5\textwidth]{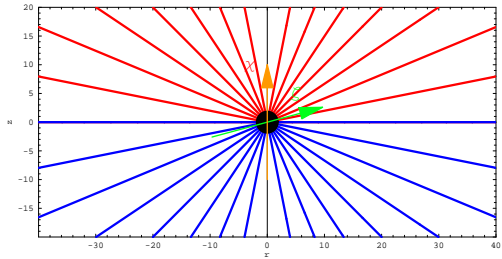}
  \end{tabular}
  \caption{Topology of the infinitely thin current sheet located in the equatorial plane. In the north hemisphere, field lines are going out from the surface, red solid lines, whereas in the south hemisphere, they go into the star, blue solid lines. The observer line of sight is shown by a green arrow.}
  \label{fig:Pulse:DoubleMonopoleMagnetiqueAligne}
\end{figure}

Exact analytical solutions of the electromagnetic field around pulsars are sufficiently rare to be of interest from a purely physical point of view, even if assuming a monopolar magnetic structure is unrealistic. The split monopole model belongs to such rare solutions and has been introduced by \cite{1973ApJ...180L.133M}. He started from the assumption that all the magnetic field lines at the surface of the star are radial and given by a monopole structure such that 
\begin{equation}
 \mathbf B = B_{\rm L} \, \frac{\rlight^2}{r^2} \, \er
\end{equation}
where $B_{\rm L}$ is the magnetic field intensity at the light-cylinder. He found an exact analytical solution of the pulsar equation~(\ref{eq:EquationPulsar}), whose solution can be summarized in the formula below for the magnetic field
\begin{equation}
 \mathbf B = B_{\rm L} \, \frac{\rlight}{r} \, \left( \frac{\rlight}{r} \, \er + \sin\vartheta \, \ephi \right) \ .
\end{equation}
The radial $B_r$ and toroidal $B_\varphi$ components have same intensity at the light cylinder radius in the equatorial plane. The electric field possesses only a latitudinal component and is fully given by the expression $E_\vartheta = c \, B_\varphi$. Particles undergo an electric drift motion combined with a movement along field lines such that the resulting velocity causes an ultra-relativistic outgoing flow (actually equal to the speed of light) with a purely radial component of the fluid such that 
\begin{eqnarray}
 \mathbf{V} = c \, \er \ .
\end{eqnarray}
The radial component of the Poynting vector is easily derived to be
\begin{equation}
\label{eq:PoyntingRadial}
 S^r = \frac{\Omega^2\,B^2\,R^4\,\sin^2\vartheta}{\mu_0\,c\,r^2} \ .
\end{equation}
We recognize the $\sin^2\vartheta$ dependence used in the RMHD simulations of section~\ref{sec:Simulations}. The Poynting vector in eq.~(\ref{eq:PoyntingRadial}) leads to a total spindown luminosity of
\begin{equation}
 L = \frac{8\,\pi}{3\,\mu_0\,c} \, \Omega^2\,B^2\,R^4
\end{equation}
and therefore a braking index equal to one, $n=1$. Thus to summarize, for a monopole we get $n=1$ and for a dipole $n=3$. In the most general case of a multipole of order~$\ell$ we can estimate the Poynting flux by very general arguments in the following way. Roughly speaking, energy is taken away by the electromagnetic wave starting at the light-cylinder~$\rlight$ that is in the wave zone. The total energy flux across the sphere of radius~$\rlight$ is $4\,\pi\,B_{\rm L}^2\,\rlight^2\,c/\mu_0$. For the multipole~$\ell$ the strength at the light-cylinder~$B_{\rm L}$ is connected to the strength at the surface~$B$ by $B_{\rm L} = B \, (R/\rlight)^{\ell+2}$. Plugging into the energy flux we get $4\,\pi\,c\,B^2\,R^{2\,\ell+4}\,\rlight^{-2\,\ell-2}/\mu_0$. Thus the Poynting flux is roughly $L\approx 4\,\pi\,B^2\,R^{2\,\ell+4}\,\Omega^{2\,\ell+2}/\mu_0\,c^{2\,\ell+1}$. We deduce the braking index to be $n=2\,\ell+1$ as noted by \cite{1991ApJ...373L..69K}. This estimate is valid irrespective of the vacuum or plasma assumption around the neutron star. A wind in the dipolar magnetosphere implies a braking index of~$n=3$ in vacuum, in FFE and in general relatively \citep{2016MNRAS.455.3779P}. Most of the Poynting flux goes away along the equatorial plane with $S^r(\vartheta) \propto \sin^2\vartheta$ explaining the setup for the RMHD simulations. The total magnetic flux through a sphere centred on the neutron star is not zero. For finding a situation that meets the overall constraint of lack of magnetic monopole, the game is to reverse the direction of the magnetic field lines when changing hemisphere. Suppose that in the northern hemisphere, field lines come out of the star. Conversely, in the southern hemisphere, field lines return to the star. By the symmetry of the problem, the total magnetic flux through a sphere centred on the star now vanish. Moreover, this magnetic topology is an exact solution of the pulsar equation with a significant peculiarity. Indeed, the junction between both magnetic monopoles of opposite magnetic charge induces a discontinuity in the equatorial plane because its polarity changes sign. In order to satisfy Maxwell-Amp\`ere equation, this discontinuity must be maintained by a surface current density (the current sheet). For an aligned rotator, this surface coincides with the rotational equatorial plane of the neutron star, fig.~\ref{fig:Pulse:DoubleMonopoleMagnetiqueAligne}. A split monopole is constituted of two magnetic monopoles with equal but opposite magnetic charge between the northern and southern hemisphere. What about an oblique rotator?

\subsection{Asymptotic MHD solution}

Near the neutron star, the split monopole approximation is certainly not verified or even justified. However, outside the light cylinder, the geometry of the pulsar wind can be assimilated to a split monopole. Indeed, to lowest order in $1/r$, only the dipole field survives as the dominant component, and combined with a radial ideal MHD flow, field lines are stretched to eventually open and resemble to a split monopole.

\begin{figure}
  \centering
  \begin{tabular}{cc}
    \includegraphics[width=0.5\textwidth]{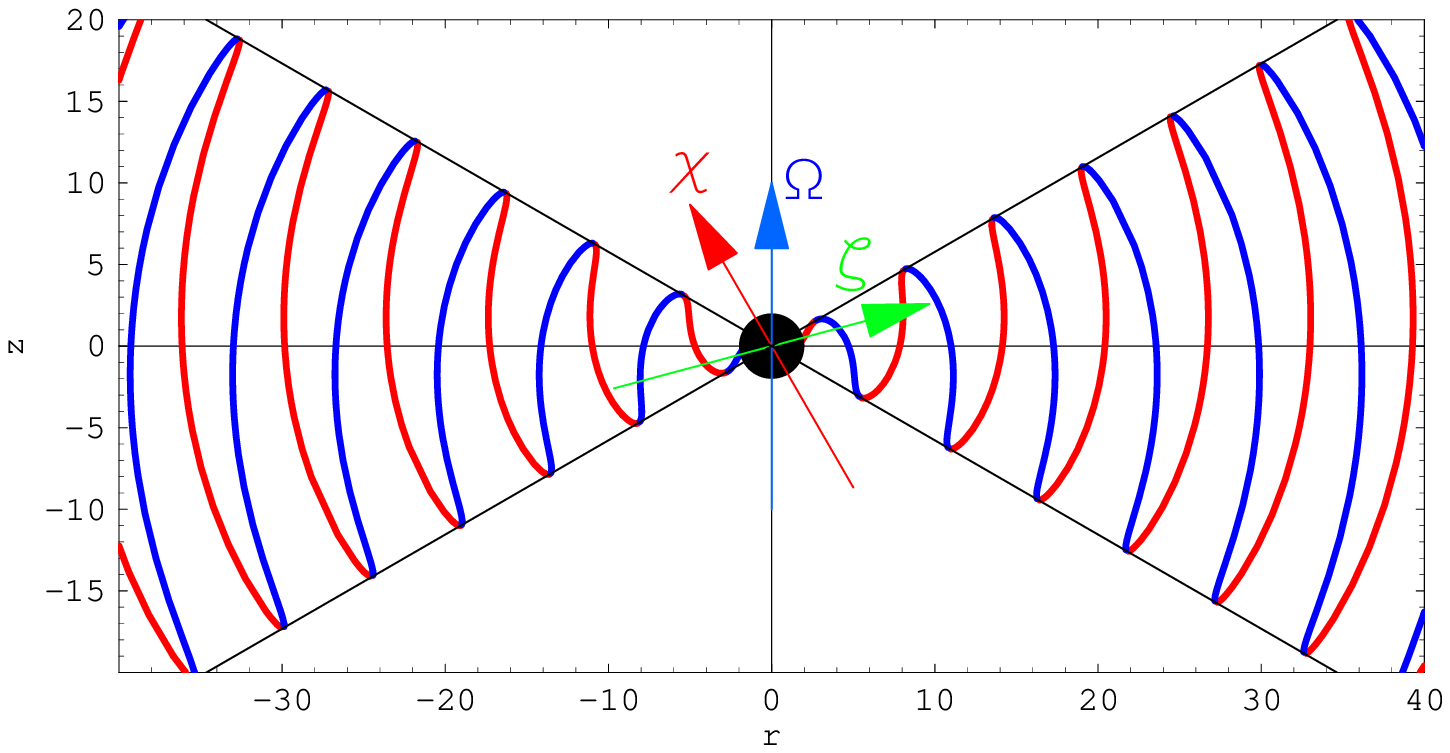} &
    \includegraphics[width=0.5\textwidth]{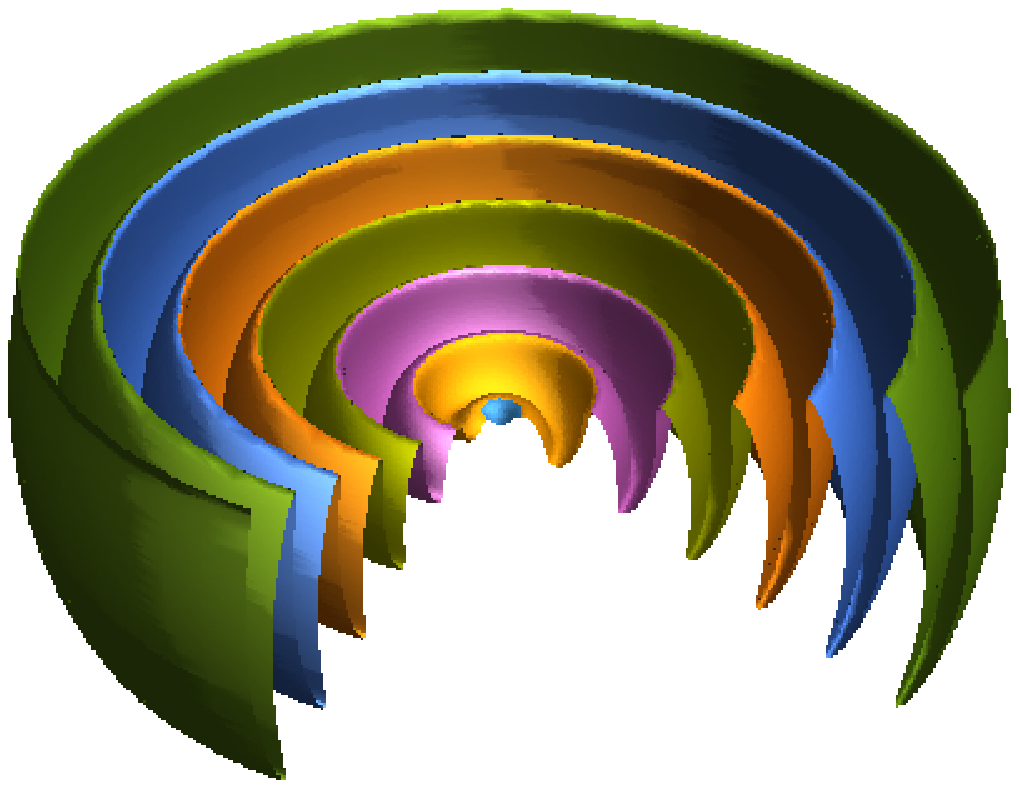}
  \end{tabular}
  \caption{Topology of the infinitely thin current sheet induced by the striped wind from the split monopole. In the north hemisphere, field lines are going out from the surface, red solid lines, whereas in the south hemisphere, they go into the star, blue solid lines. The discontinuity, or magnetic polarity reversal is depicted by this current sheet wobbling around the equatorial plane, right picture.}
  \label{fig:Pulse:DoubleMonopoleMagnetique}
\end{figure}

Remember the configuration of the system. The strongly magnetized neutron star rotates along the $(Oz)$ axis and possesses a magnetic field assumed perfectly dipolar at its surface. The expulsion of the plasma beyond the light cylinder deforms the field lines to the point that they will open, giving rise to a situation approaching the split monopole at large distances. For an oblique rotator, making an angle~$\chi$ between the magnetic moment~$\pmb{\mu}$ and the rotation axis~$\mathbf{\Omega} = \Omega \, \ez$, we have $\Omega \, \cos\chi = \pmb{\Omega} \cdot \pmb{\mu}$, the surface discontinuity oscillates and propagates at the wind velocity~$V$ in the ideal MHD approximation. The flow is only in the radial direction. This surface discontinuity is determined by finding the geometric place where the magnetic field changes sign on the stellar crust. Recall that the magnetic moment in spherical coordinates is
\begin{equation}
  \label{eq:Pulse:MomentMagnetique}
  \pmb{\mu} = \mu \, [ \sin\chi \, ( \cos (\Omega \, t) \, \ex + \sin (\Omega \, t) \, \ey ) + \cos\chi \, \ez ] \ .
\end{equation}
Let $\mathbf{n}$ be a unit vector pointing to the magnetic equator and having components 
\begin{equation}
  \label{eq:Pulse:MomentMagnetique2}
  \mathbf{n} = \sin \vartheta \, ( \cos \varphi \, \ex + \sin \varphi \, \ey ) + \cos \vartheta \, \ez \ .
\end{equation}
The magnetic equator is defined by~$\pmb{\mu} \cdot \mathbf{n} = 0$. The surface where the magnetic field changes polarity is therefore defined by
\begin{equation}
\label{eq:Pulse:SurfaceStrieEtoile}
  \Psi_s (t, r, \vartheta, \varphi) \equiv \cos \vartheta \, \cos \chi + \sin \vartheta \, \sin \chi \, \cos\left( \varphi - \Omega \, t \right) = 0 \ .
\end{equation}
This curve traced on the stellar surface is at the origin of the current sheet. Given that the plasma flow is radial and expands at a constant velocity~$V$, we replace the time dependence~$t$ by a radial propagation term of the form~$t-r/V$ to take into account this propagation effect. The current sheet will therefore be the geometric surface defined in three-dimensional space by
\begin{equation}
\label{eq:Pulse:SurfaceStrie}
  \Psi_s (t, r, \vartheta, \varphi) \equiv \cos \vartheta \, \cos \chi + \sin \vartheta \, \sin \chi \, \cos\left[ \varphi - \Omega \, \left( t - \frac{r}{V} \right) \right] = 0 \ .
\end{equation}
The equation of the surface, solved for the radial variable~$r$ is
\begin{equation}
  r_s(t, \vartheta, \varphi) = \beta_{\rm v} \, \rlight \, \left[ \pm \arccos ( - \cot\vartheta \, \cot\chi) + \frac{c\,t}{r_L} - \varphi + 2\,\ell\,\pi \right]
\end{equation}
where $\beta_{\rm v} = V/c$ and $\ell$ is an integer. It is the solution found by \cite{1999A&A...349.1017B}. The three dimensional geometry and a cross section of the current sheet are shown in fig.~\ref{fig:Pulse:DoubleMonopoleMagnetique}.
The striped wind is therefore a spiral structure rotating at the velocity of the star and moving radially at a speed close to that of light. The current sheet is infinitely thin. In the equatorial plane ($\vartheta=\pi/2$) the polar equation of the two-armed spiral is
\begin{equation}
\label{eq:SpiraleReelle}
 r = \beta \, \rlight ( \Omega \, t - \varphi + \frac{\pi}{2} + \ell \, \pi ).
\end{equation}
In reality, the striped wind possesses a certain thickness and an internal intrinsic dynamics but not described by this simple MHD approach. For more realistic models that we will consider, the wind is made of two plasma components, a strongly magnetised cold component outside the current layer and a weakly magnetized hot component inside the layer. The relativistic motion associated with the spiral structure is at the origin of the pulsed emission we now detail. \cite{2016MNRAS.457.3384T} found useful approximate analytical expressions for the current in the general oblique case by fitting full 3D force-free and MHD simulations.

\subsection{Origin of pulsed emission}

In the striped wind scenario, the dissipation of magnetic energy happens mainly in the current sheet. Charged particles are heated to relativistic temperatures and radiate synchrotron and inverse Compton emission. If moreover these particles travel at a distance~$r$ from the centre of the star such that $r \lesssim \Gamma_{\rm v}^2 \, \rlight$, the radiation is modulated at the rotation frequency of the neutron star. Explaining pulsation by such a mechanism was already suggested at the beginning era of pulsar theory as proposed by \cite{1971CoASP...3...80M} and \cite{1979SSRv...24..437A}. These ideas have been reinvestigated more recently by \cite{2002A&A...388L..29K}. These studies showed that the striped wind is a possible site for generation of incoherent high energy radiation, going from optical up to gamma rays. A priori, there is no reason to favour one emission site more than another, let it be outer gap, slot gap, polar cap or wind. However, compared to other models, the striped wind does not require knowledge about the inner magnetosphere and furnishes an analytical description of the magnetic field structure only based on fundamental geometrical hypotheses. This permits to circumvent the problems linked to arbitrariness of the magnetosphere.
\begin{figure}
\centering
\begin{tikzpicture}[scale=0.5]
\filldraw[inner color=white,outer color=gray] (0,0) circle (0.5) node [below] {pulsar};
\draw (0,0) -- (8,0);
\draw[rotate=30] (0,0) -- (8,0);
\draw[->,rotate=-20,decorate, decoration={snake}] (3,0) -- (7,0) node [above] {$\Gamma_{\rm v}$} ;
\draw[->] (2,0) arc (0:30:2) node [right] {$\vartheta\approx1/\Gamma_{\rm v}$};
\draw[thick,green] (4,0) arc (0:45:4) ;
\draw[thick,green] (4,0) arc (0:-45:4)   node [below] {shell $n+1$};;
\draw[thick,blue] (6,0) arc (0:45:6) ;
\draw[thick,blue] (6,0) arc (0:-45:6)  node [below] {shell $n$};
\begin{scope}[shift={(6,0)},rotate=-30]
\draw[red,rotate=60] (0,0) -- (2,0) ;
\draw[red] (0,0) -- (2,0) ;
\draw[red] (2,0) arc (0:60:2) ;
\draw[<->] (3,0) arc (0:60:3) node [right] {$2/\Gamma_{\rm v}$} ;
\end{scope}
\begin{scope}[rotate=30]
\begin{scope}[shift={(6,0)},rotate=-30]
\draw[red,rotate=60] (0,0) -- (2,0) ;
\draw[red] (0,0) -- (2,0) ;
\draw[red] (2,0) arc (0:60:2) ;
\draw[<->] (3,0) arc (0:60:3) node[right]  {$2/\Gamma_{\rm v}$} ;
\end{scope}
\end{scope}
\begin{scope}[rotate=-30]
\begin{scope}[shift={(6,0)},rotate=-30]
\draw[red,rotate=60] (0,0) -- (2,0) ;
\draw[red] (0,0) -- (2,0) ;
\draw[red] (2,0) arc (0:60:2) ;
\draw[<->] (3,0) arc (0:60:3) node [right] {$2/\Gamma_{\rm v}$} ;
\end{scope}
\end{scope}
\draw[->] (8,-2) -- (10,-2) node [right] {to observer} ;
\end{tikzpicture}
  \caption{Principle of pulsed emission. The spherical shells propagate radially outwards with a Lorentz factor~$\Gamma_{\rm v}$ and emit in a cone of half opening angle~$1/\Gamma_{\rm v}$ when crossing the sphere of radius~$\Rsph$, blue arc.}
  \label{fig:Pulse:PrincipeEmissionPulsee}
\end{figure}
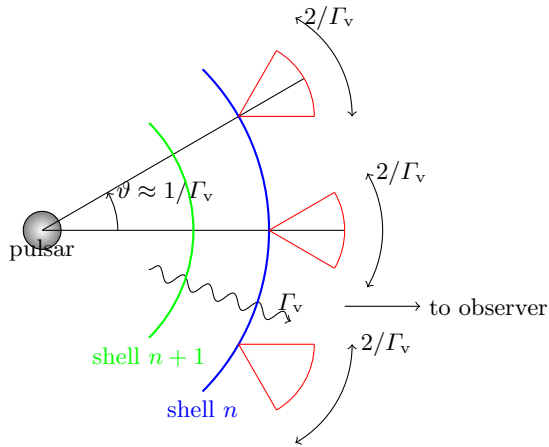

\begin{figure}
 \centering
\begin{tikzpicture}[photon/.style={decorate,decoration={snake,post length=1mm}}, scale=0.5]
\clip (0,-5) rectangle (14,5);
\draw [thick, color=red, domain=-2.*pi:0, samples=200, smooth]
  plot (xy polar cs:angle=\x r, radius={\x - 0.5*pi });
\draw [thick, color=green, domain=-2.*pi:0, samples=200, smooth]
  plot (xy polar cs:angle=\x r, radius={\x - 0.725*pi });
\draw [thick, color=blue, domain=-2.*pi:0, samples=200, smooth]
  plot (xy polar cs:angle=\x r, radius={\x - 0.95*pi });
\draw[->] (0,0) -- (8,0);
\draw[magenta,rotate=40] (0,0) -- (7,0);
\draw[magenta,rotate=-40] (0,0) -- (7,0);
\draw[black,thick] (0,0) circle (5.4) ;
\pgfmathsetmacro{\ex}{0}
\pgfmathsetmacro{\ey}{1}
\draw[orange,thick,->] (\ex,\ey) (-22.5:2) arc (-45:45:1) node [right] {$\Omega$};
\draw[red] (4,-2.25) node [below] {$t_-$} ;
\draw[green] (5,0) node [below] {$t_0$} ;
\draw[blue] (4,4.75) node [below] {$t_+$} ;
\draw[->] (7,-1) -- (9,-1) node [right] {to observer} ;
\draw[->] (1,0) arc (0:40:1) node [above] {$\varphi_{\rm rim}$};
\draw[->,thick,red,photon] (5,-3.5) -- (8,-3.5) node [right] {$\gamma_-$} ;
\draw[->,thick,red,photon] (6,0) -- (9,0) node [right] {$\gamma_0$} ;
\draw[->,thick,red,photon] (5,3.5) -- (8,3.5) node [right] {$\gamma_+$} ;
\draw[<-] (5,2) -- (7,2) node [right] {emitting shell at $\Rsph$} ;
\end{tikzpicture}
 \caption{Real shape of the current sheet not approximated by concentric spherical shells but using the true expression in the equatorial plane. Rotation is counter-clockwise. It shows the three important phases of a pulse: begin in red, middle in green and end in blue. Photons are emitted during the whole interval $t\in[t_-,t_+]$ not to be confused with the reception times $t^{\rm rec} \in[t^{\rm rec}_-,t^{\rm rec}_+]$, see text.}
 \label{fig:SpiraleReelle}
\end{figure}
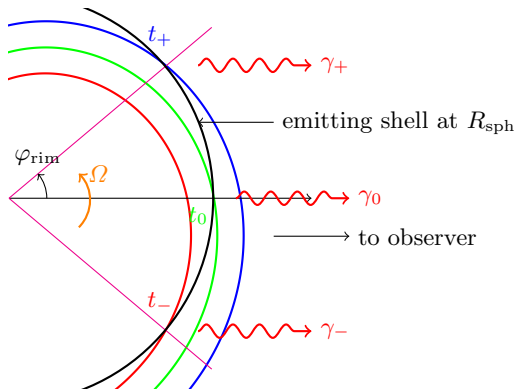

As announced earlier, pulsed emission is a direct consequence of the beaming induced by the relativistic flow. To better understand this phenomenon, let us assume for the sake of simplicity that the wind consists of thin, concentric spherical layers of hot radiating plasma. In the equatorial plane, successive sheets, marked by $n$ and $n+1$, are separated by half a wind wavelength, $\Delta l = \lambda_{\rm L}/2$ where $\lambda_{\rm L} = 2\,\pi\,\beta\,\rlight$ is the wind wavelength, fig.~\ref{fig:Pulse:PrincipeEmissionPulsee}. Moreover, assume that these sheets only radiate when crossing an imaginary sphere of radius~$\Rsph$, blue arc. The maximal time of arrival delay for photons emitted by the sheet labelled~$n$ for a distant observer is $\Delta t = \Delta R / c =  ( 1 - \cos\vartheta) \, \Rsph/c$. For an ultra-relativistic flow we simplify noting that $\vartheta \approx 1/\Gamma_{\rm v} \ll 1$. Consequently the arrival time delay is
\begin{equation}
  \label{eq:Pulse:PulseWidth} 
  \Delta t \approx \frac{\Rsph}{2\,\Gamma_{\rm v}^2\,c} 
\end{equation}
a well known result from gamma-ray burst theory. To observe pulses, the delay must be inferior to the time interval elapsed between the issuance of two consecutive layers $n$ and $n + 1$ crossing $\Rsph$ and given by $ \Delta T = \Delta l / c = \pi \, \rlight / c $. This results in a pulsed emission if \citep{1979SSRv...24..437A}
\begin{equation}
  \label{eq:Pulse:Pulsed}
  \Rsph \lesssim 2\,\pi\,\Gamma_{\rm v}^2\,\rlight \approx \Gamma_{\rm v}^2\,\lambda_{\rm L} \ .
\end{equation}
Actually this estimate is based on perfectly concentric spherical shells. In a more realistic model, care should be taken from the truly spiral structure of the wind. Thus to refine our argument, let us look at fig.~\ref{fig:SpiraleReelle}. As before the current sheet emits photons when the spiral structure crosses the sphere of radius~$\Rsph$ depicted as a solid black arc. The whole structure rotates rigidly in the direction indicated by the red arrow. The two magenta lines correspond two the region of the wind seen by the distant observer. One spiral arm crosses this sphere in the following order: beginning in red then middle in green and finally ending in blue as marked in fig.~\ref{fig:SpiraleReelle}. The related position in polar coordinates in the plane of the figure are noted $(\Rsph,-\varphi_{\rm rim})$, $(\Rsph,0)$ and $(\Rsph,\varphi_{\rm rim})$ and are measured at times respectively $t_-,t_0$ and $t_+$ meanwhile emitting photons $\gamma_-,\gamma_0$ and $\gamma_+$. From the spiral structure eq.~(\ref{eq:SpiraleReelle}) these times are related by
\begin{equation}
\Omega \, t_- + \varphi_{\rm rim} = \Omega \, t_0 = \Omega \, t_+ - \varphi_{\rm rim}
\end{equation}
which leads to the ordering $t_-<t_0<t_+$. For our purpose, in the case of a relativistic radial flow we set $\varphi_{\rm rim} \approx 1/\Gamma_{\rm v}$. Taking into account time of flight of photons from the emission site to the observer we get the reception times as
\begin{subequations}
\begin{align}
 t_-^{\rm rec} & = t_- + \frac{D-\Rsph\,\cos\varphi_{\rm rim}}{c} \\
 t_0^{\rm rec} & = t_0 + \frac{D-\Rsph}{c} \\
 t_+^{\rm rec} & = t_+ + \frac{D-\Rsph\,\cos\varphi_{\rm rim}}{c} \ .
\end{align}
\end{subequations}
Assuming the ordering as $t_-^{\rm rec}<t_0^{\rm rec}<t_+^{\rm rec}$, which must be checked a posteriori, the time observed for a pulse is of the order
\begin{equation}
 \Delta t^{\rm rec}_{+-} = t_+^{\rm rec} - t_-^{\rm rec} = \frac{2\,\varphi_{\rm rim}}{\Omega} = \frac{2}{\Gamma_{\rm v}\,\Omega}
\end{equation}
from which we deduce the duty cycle for one pulse as
\begin{equation}
\label{eq:LargeurPulse}
 \frac{\Delta t^{\rm rec}}{P} = \frac{\varphi_{\rm rim}}{\pi} = \frac{1}{\Gamma_{\rm v}\,\pi} \ll 1 \ .
\end{equation}
From this inequality (which holds only for $\Gamma_{\rm v}\gg1$) we conclude that pulsation should happen at any place in the striped wind. There should be no restriction such as eq.~(\ref{eq:Pulse:Pulsed}) where a concentric geometry was assumed. This conclusion is however incorrect. The photon $\gamma_0$ does not always succeed the photon $\gamma_-$ because 
\begin{equation}
 \Delta t^{\rm rec}_{0-} = t_0^{\rm rec} - t_-^{\rm rec} = \frac{1}{\Gamma_{\rm v}\,c} \, \left( \rlight - \frac{\Rsph}{2\,\Gamma_{\rm v}} \right)
\end{equation}
It will be the case only if $t_-^{\rm rec}<t_0^{\rm rec}$ which implies $\Rsph< 2 \, \Gamma_{\rm v} \, \rlight$. In that case, the order is preserved otherwise the photon $\gamma_0$ will be received before $\gamma_-$ even if it has been produced after the latter. The explanation lies in the additional time required by $\gamma_-$ to reach the observer. According to geometrical considerations, the second condition $t_0^{\rm rec}<t_+^{\rm rec}$ is always satisfied independently of the distance $\Rsph$. To observe pulsation, we require also $|\Delta t^{\rm rec}_{0-}| < P$ and this gives for $\Gamma_{\rm v}\gg1$ a condition similar to eq.~(\ref{eq:Pulse:Pulsed}) in the form $\Rsph\lesssim4\,\pi\,\Gamma_{\rm v}^2\,\rlight$.

In general, the striped wind model predicts two pulses per period, as observed in most light curves of gamma ray pulsars, see for instance the first and second gamma-ray pulsar catalogues described in depth in \cite{2009Sci...325..840A, 2013ApJS..208...17A}. The separation between these pulses is only function of the obliquity~$\chi$ and the inclination of the line referred as~$\zeta$. Indeed from geometric considerations about the current sheet, we derived a simple analytical relation between inclination of the line of sight~$\zeta$, obliquity of the pulsar~$\chi$ and the separation between the two pulses~$\Delta$ given by \citep{2005MmSAI..76..494K, 2011MNRAS.412.1870P}
\begin{equation}
 \label{eq:RelationZetaChi}
 | \cot\zeta \, \cot \chi | = \cos(\Delta\,\pi) \ .
\end{equation}
For pulsars observed by Fermi/LAT, the value of $\Delta$ is easily accessible. We deduce a simple relation between the two widths defining the fundamental geometry of the pulsar. Moreover with a model for radio emission, we can estimate the delay between arrival time of radio and gamma ray photons. This has been analysed in details in \cite{2011MNRAS.412.1870P}.

Moreover for an infinitely thin sheet, the width of the pulses is inversely proportional to the Lorentz factor~$\Gamma_{\rm v}$ of the flow. Related to one period~$2\,\pi$, this width is approximatively $\Delta \approx 1/\pi\,\Gamma_{\rm v}$ from eq.~(\ref{eq:LargeurPulse}). We check this with some examples of pulsed emission.
More generally, when the wind is not purely in radial expansion, the criterion for existence of pulsation changes. Consider again two spherical layers separated by half a wavelength. Suppose that the plasma expansion velocity makes an angle~$\alpha$ with respect to the radial direction. The path difference between the pulse emitted in the middle and that emitted at the edge of the sheet is
\begin{equation}
 \Delta l = \Rsph \, \left( 1-\cos\frac{1}{\Gamma_{\rm v}} \, \cos\alpha + \sin\alpha \sin\frac{1}{\Gamma_{\rm v}} \right)
\end{equation}
For high Lorentz factors it simplifies into
\begin{equation}
 \Delta l \approx \Rsph \, \left( \frac{1}{2\, \Gamma_{\rm v}^2} \, \cos\alpha + \sin\alpha \frac{1}{\Gamma_{\rm v}} \right) 
\end{equation}
Emission will be pulsed if the difference of arrival time between the two pulses~$\Delta l/c$ is less than half a period of the pulsar thus
\begin{equation}
 \Rsph < \frac{2 \, \pi \, \Gamma_{\rm v}^2 \, \rlight}{\cos\alpha + 2 \, \Gamma_{\rm v} \, \sin\alpha}
\end{equation}
Two limiting cases are worthwhile
\begin{itemize}
 \item[a)] for a strictly radial velocity with $\alpha=0$ we find again $\Rsph<2 \, \pi \, \Gamma_{\rm v}^2 \, \rlight$.
 \item[b)] for a strictly azimuthal velocity with $\alpha=\pi/2$ we have $\Rsph<\pi \, \Gamma_{\rm v} \, \rlight$.
\end{itemize}
Criterion~b) is much more constraining than~a) because it is proportional to the first power of~$\Gamma_{\rm v}$ only and not to its second power~$\Gamma_{\rm v}^2$. The angles $\alpha$ and $1/\Gamma_{\rm v}$ have to be compared to check which pulse arrives first, the middle one or the edge one: the middle pulse wins if $\alpha < 1/\Gamma_{\rm v}$.

Where do we expect to produce emission? In a realistic model the current sheet possesses a finite thickness and therefore a balance between thermal and magnetic pressure in the stripes should happen. To simplify, in the two distinct regions we have
\begin{enumerate}
 \item in the current sheet: zero magnetic field~$B=0$, constant pressure~$p$ and high particle density number~$n$ thus a hot unmagnetized plasma.
 \item between the current sheet: constant magnetic field, zero pressure, low particle density number thus a cold magnetized plasma.
\end{enumerate}
This entropy wave must be in a MHD equilibrium in the wind rest frame such that
\begin{equation}
\frac{B^2}{2\,\mu_0} + p = \frac{\textrm{constant across the wind}}{r^2}
\end{equation}
The constant has to be determined on other physical grounds.

\subsection{Emission from an infinitely thin sheet}

In the precedent paragraph, we explained the origin of pulses, provided that the emitting layer lies sufficiently close to the light-cylinder. We now study more quantitatively this pulsed emission. The mechanisms giving rise to high energy emission are diverse. We distinguish mainly between 
\begin{itemize}
 \item synchrotron emission of ultra-relativistic hot electrons/positrons pairs in the strong magnetic field of the wind.
 \item inverse Compton emission of internal or external photons, for instance those coming from a companion, the surrounding nebula, thermal photons of the surface or synchrotron photons themselves.
\end{itemize}
The intensity of emission for the synchrotron and inverse Compton radiation is proportional to the following space-time integral
\begin{multline}
  I_\nu(t) = \int_{-\infty}^{+\infty} \, \int_{R_0}^{+\infty} \, \int_{\pi/2 - \chi}^{\pi/2 + \chi} \, \int_0^{2\pi} j_\nu(\mathbf{r}, t') \, \delta( r - r_s(\vartheta,\varphi,t') ) \nonumber \times \\ 
  \times \delta\left(t' - ( t - \frac{||\mathbf{R}_{\rm obs}-\mathbf{r}||}{c} )\right) \, r^2 \, \sin\vartheta \, dt' \, dr \, d\vartheta \, d\varphi \ .
\end{multline}
Integration must be performed on the current sheet making sure to include the retardation effects due to propagation at finite speed of the photons. The observer is located at the point~$\mathbf{R}_{\rm obs}$ where the unit vector is $\mathbf{n}_{\rm obs} = \mathbf{R}_{\rm obs} / R_{\rm obs}$. Emission starts at an arbitrary radius~$R_0\lesssim\Gamma_{\rm v}^2\,\rlight$ and  $t'=t-\mathbf{r} \cdot \mathbf{n}_{\rm obs}/c$ corresponds to retarded time associated to emission at point~$\mathbf r$ in the sheet. The Dirac distributions insure emission only when being on the current sheet, thus the $\delta( r - r_s(\vartheta,\varphi,t') )$ term, and when observation time is related to retarded time of emission~$t'$ of a photon emanating from the point $\mathbf r$ of the sheet, thus the $\delta\left(t' - ( t - \frac{||\mathbf{R}_{\rm obs}-\mathbf{r}||}{c} )\right)$ term. 

\begin{figure}
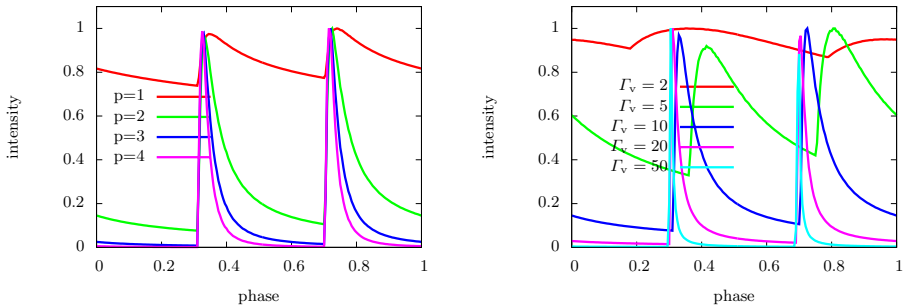

  \centering
  \begin{tabular}{cc}
  \resizebox{.45\textwidth}{!}{\input{fig6a.tex}} &
  \resizebox{.45\textwidth}{!}{\input{fig6b.tex}}
  \end{tabular}
  \caption{Sample of synchrotron emission light curves for different power law indices $p=\{1,2,3,4\}$ with $\Gamma_{\rm v} = 10$ on the left and for different Lorentz factors $\Gamma_{\rm v} = \{2,5,10,20,50\}$ with $p=2$ on the right. Intensities are normalized to $I_{\rm max}=1$.}
  \label{fig:Pulse:Synchrotron}
\end{figure}

\begin{figure}
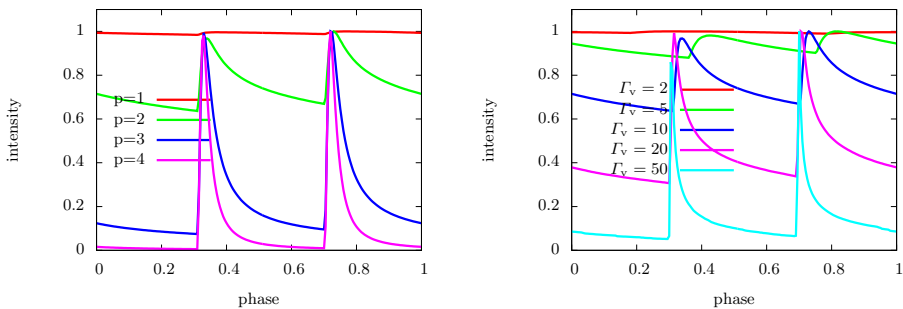

  \centering
  \begin{tabular}{cc}
  \resizebox{.45\textwidth}{!}{\input{fig7a.tex}} &
  \resizebox{.45\textwidth}{!}{\input{fig7b.tex}}
  \end{tabular}
  \caption{Sample of inverse Compton emission light curves for different power law indices $p=\{1,2,3,4\}$ with $\Gamma_{\rm v} = 10$ on the left and for different Lorentz factors $\Gamma_{\rm v} = \{2,5,10,20,50\}$ with $p=2$ on the right. Intensities are normalized to $I_{\rm max}=1$.}
  \label{fig:Pulse:InverseCompton}
\end{figure}

Synchrotron and inverse Compton emissivities, far from the low and high frequency cut off, are given respectively by
\begin{subequations}
\begin{align}
  j_\nu^{\rm sync}(\mathbf{r}, t) & = K_e(\mathbf{r}, t) \, \nu^{-(p-1)/2} \, \mathcal{D}^{(p+3)/2} \, B^{(p+1)/2} \\
  j_\nu^{\rm IC}(\mathbf{r}, t)   & = K_e(\mathbf{r}, t) \, \nu^{-(p-1)/2} \, \mathcal{D}^{p+2} \, n_\gamma(\varepsilon)
\end{align}
\end{subequations}
Relativistic beaming effects are symbolised by the usual Doppler factor
\begin{equation}
 \label{eq:Doppler}
 \mathcal{D} = \frac{1}{\Gamma_{\rm v} \, ( 1 - \pmb{\beta_{\rm v}} \cdot \mathbf{n}_{\rm obs})} \ .
\end{equation}
The power law dependence on $\mathcal{D}$ is different for $j_\nu^{\rm sync}$ and $j_\nu^{\rm IC}$, thus affecting the pulse shape depending on the distribution of particles but also following the emission process considered. The light curves exhibit peaks that are more or less pronounced. For pedagogical purposes, we show a sample of light curves for a prescribed volume emissivity. The impact of different Lorentz factors and spectral indices are shown in fig.~\ref{fig:Pulse:Synchrotron} for synchrotron emission and in fig.~\ref{fig:Pulse:InverseCompton} for inverse Compton emission. Synchrotron profiles differ from inverse Compton profiles but the general trend is the same: a decrease in the full width half maximum when the power law index increase and/or when the Lorentz factor is augmented.


Knowing the shape of pulses for a given frequency, now we are interested in the spectral power density from this radiation. We have seen that pulsed emission comes from relativistic Doppler beaming. The exact function of this dependence in the Doppler factor is determined by the same power law spectral density. It is therefore essential to know this spectral power density for estimating the shape of the pulses at a given frequency. We study the synchrotron and inverse Compton emission. Emissivity is chosen so as to include the power law dependence of the frequency of observation considered. The shape of the pulses depend not only on the Lorentz factor of the wind and the emission process itself, IC or synchrotron, but also on the observation frequency before the low cut off, between the two cut off frequencies and after the high frequency cut off. Light curves are shown in the case of inverse Compton emission of the gamma ray binary PSR~B1259-63 in \cite{2011MNRAS.417..532P}. In the striped wind model for high-energy emission, we observe a natural narrowing of the pulse width at highest energies due to the increased sensitivity on relativistic beaming.

\subsection{Emission of a more realistic model}

In reality, the current sheet in the striped wind is not infinitely thin but possesses a finite thickness defined by its internal dynamics therefore a spatial extension induced by pressure in the gas heated to relativistic temperatures. To account for this finite thickness of the current sheet , it is necessary to integrate emissivity in the whole three dimensional volume of the wind and not only on the 2D current sheet. To extract meaningful light-curves we need to set the parameters of a realistic wind model. These can be divided into three groups
\begin{enumerate}
\item geometrical properties:
\begin{itemize}
    \item the obliquity~$\chi$ of the pulsar.
    \item the inclination~$\zeta$ of the line of sight with respect to rotation axis.
    \end{itemize}
    \item magnetic field configuration:
    \begin{itemize}
    \item no radial component~$B_r=0$ but toroidal components given by $B_\theta,B_\varphi \propto 1/r$.
    \item the current sheet represented as a discontinuous $B_\varphi$, replaced by a transition layer of thickness~$(\Delta_\varphi)$ inducing a smooth $B_\varphi$ polarity reversal.
    \item accompanied by a significant~$B_\theta$ component in the current sheet.
  \end{itemize}
  \item dynamical properties (emitting particles):
    \begin{itemize}
    \item the Lorentz factor $\Gamma_{\rm v}$ of the wind.
    \item the power law index~$p$ of the particle distribution.
    \item the electron/positron number density~$K(\vec{r},t)$ such that the distribution function, isotropic in momentum space~$\vec{\mathcal{P}}$, is
      $$
      N(E,\vec{\mathcal{P}},\vec{r},t) = K(\vec{r},t) \, E^{-p} .
      $$
      Pressure balance implies that a strong magnetic field is associated with low density plasma and conversely.
    \end{itemize}
\end{enumerate}

We keep the structure of the split monopole but consider only the toroidal component $B_\varphi$, the two other components being negligible. So the wind velocity is perpendicular to the magnetic field which simplifies Lorentz transformation of the electromagnetic field between wind frame and observer frame. Pulsed emission arises in the striped wind via inverse Compton radiation from synchrotron photons from the nebula or cosmic microwave background. Particle distribution is mono-energetic. This method is applied to Geminga, see the phase resolved spectra in \cite{2009A&A...503...13P}. Moreover \cite{2012MNRAS.424.2023P} showed that the gamma ray luminosity of Fermi/LAT pulsars can be interpreted as synchrotron emission from the striped wind current sheet as already mentioned by \cite{1996A&A...311..172L}. In a stationary state, the radiative losses are compensated by magnetically reheated particles through magnetic reconnection. The Lorentz factor of the wind is then estimated as well as the reconnection rate in the relativistic plasma. \cite{2013A&A...550A.101A} investigated the properties of synchrotron radiation in the current sheet assuming a thermal population of particles and found spectra that peak around the GeV with gamma-ray efficiency in agreement with Fermi/LAT observations. However, due to magnetic reconnection in the stripes, \cite{2015MNRAS.449L..51M} identified two regimes of particle acceleration, the first limited by radiation reaction and the second by the size of the accelerating region that strongly impacts on the pulsed inverse Compton spectra in the sub-TeV band. Reconnection in the current sheet has also been investigated by \cite{2014ApJ...780....3U}.

For binary systems with two neutron stars of which at least one is a pulsar, geodetic precession causes a secular variation in the inclination of the line of sight. We deduce a variation in the light curve not only in radio but also at higher energies, including X-rays and gamma-rays. We therefore undertook using the striped wind model to compute these phase-resolved light curves. Some systems will maybe permit a detection of this precession in the decades ahead as was shown in \cite{2015A&A...574A..51P}.

But the striped wind could also be responsible for a non-pulsed emission causing giant gamma-ray flares around~400~MeV lasting for hours to days like deferred by \cite{2011ApJ...741L...5S, 2012ApJ...749...26B} and \cite{2013ApJ...765...52S}. \cite{2013MNRAS.436L..20B} have interpreted this phenomenon as a signature of relativistic magnetic reconnection operating explosively in the striped wind due to instability caused by the presence of several neighbouring current sheets. Already two alternations of the field are sufficient to get violent reconnection. This is known as the double tearing mode. \cite{2013MNRAS.436L..20B} work was followed by some numerical improvements \citep{2015PPCF...57a4034P} and extraction of synchrotron radiation signature in a post-processing procedure \citep{2015MNRAS.454.2972T}.

\begin{figure}
\centering
\input{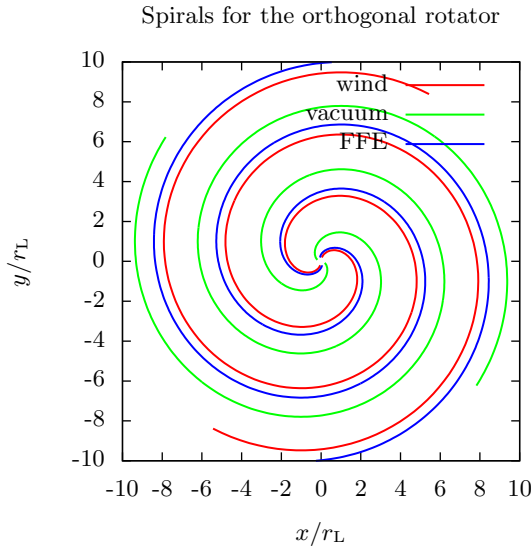}
\caption{Comparison of the location of the spiral structure for the striped wind (red), the vacuum (green) and the force-free (blue) solution for the orthogonal rotator. In current sheet models, dissipation and radiation outside the light cylinder essentially occurs within a small thickness around this spiral region.}
\label{fig:ComparaisonSpirales}
\end{figure}

Force-free simulations have shown that the wind outside the light-cylinder resembles to the split monopole solution with a dominant toroidal magnetic field component. The split monopole is a simple and good analytical solution at large distances but inadequate to represent the closed magnetosphere. It is therefore illuminating to compared the phase shift in the two-armed spirals found in striped wind, the vacuum and the force-free solution for the orthogonal rotator. The results of the comparisons are made in fig.~\ref{fig:ComparaisonSpirales}. The shift is evident and the ordering is, vacuum first, force-free second and split monopole third. So we conclude that using the split monopole to compute simultaneously polar cap radio emission and current sheet high-energy emission leads to a time lag between both pulses which is not the same as for the more realistic force-free solution. If the dipole geometry inside the light cylinder is taken into account, we expect the delay between radio and gamma-ray to be less than the lags reported in \cite{2011MNRAS.412.1870P}. This can explain the 0.1 phase excess noted in this earlier work.

\subsection{Kinetic aspects related to the current sheet}

The physical conditions inside the current sheet are badly constraints. A detailed description would require the modelling of the overall electromagnetic circuitry in the magnetosphere and even beyond the light cylinder. Unfortunately, such studies are still not available to a sufficient level of refinement. Also, we have to restrict ourselves by formulating hypothesis that one can judge more or less reasonable only based on observational consequences that will be drawn. It seems that several pulsars emitting mainly in gamma-rays show light curves possessing not two but three or four distinct peaks. It is already clear that Vela exhibits a third pulse. Moreover, this third pulse drifts with frequency \citep{2009ApJ...696.1084A}. This observation tends therefore to eliminate the striped wind because it only predicts two pulses per rotation, one for each polarity of the current sheet, sometimes linked to the north pole, sometimes to the south pole. The nature of the stripes combined to drift motion of $e^\pm$~pairs in different directions leads to important modification of the light curve shape. We could expect up to four pulses more or less well separated, depending on the internal dynamic in the current sheet. 

From a more fundamental point of view, \cite{2007PPCF...49..297P, 2007PPCF...49.1885P} have studied in details the kinetic aspects related to tearing instability for a relativistic current sheet and for the Weibel instability via a linear analysis of Vlasov-Maxwell equations. In the longer term  simulations of reconnection will benefit from a relativistic approach including kinetic works previously cited. It could also be supplemented with radiation effects that are dominant in the stripes.

\subsection{Polarization of the synchrotron pulsed emission}

To probe the structure of any magnetic field, measurement of synchrotron emission polarization is often invoked. Synchrotron emission indeed shows a high degree of linear polarization when the magnetic field in which bathes the leptons is ordered. The striped wind magnetic field possesses such ordered topology. We therefore expect to observe a specific pulsed polarisation signature in the emission of the wind. Fortunately such observations exist for at least one pulsar. Indeed \cite{2009MNRAS.397..103S} have reported with high precision the optical phase-resolved polarization properties of the Crab pulsar. A study of the polarization of the synchrotron emission reveals strongly discriminating for later comparison between models and observations.

The polarisation of total emission of the wind is performed by summation on the distribution function of emitting particles, here the electrons, that is integrated in whole three dimensional space and for all time. Electrons are assumed to possess a stationary distribution independent of time, isotropic in momentum and following a power law in energy. The asymptotic structure of the field such that given by the ideal MHD cannot give information about the field interior to the current sheet. It is seen as a singularity. Actually, it has a finite thickness and an inner structure, but the distribution of particles and magnetic field are still inaccessible to simulations. We only have no choice but to prescribe a priori its characteristics. Comparisons between our model and observations of pulsed emission in optical by \cite{2009MNRAS.397..103S} are shown in \cite{2005ApJ...627L..37P}. Agreement between our model and these are satisfactory. However, a $B_\vartheta$ component was necessary to fit properly the data. A complete catalogue of polarization properties from pulsed synchrotron emission has been compiled for all possible geometries. The results have been synthesized in \cite{2013MNRAS.434.2636P} using a slightly different prescription for radiation in the cold and hot part of the wind. Comparing the expectations about phase-resolved polarisation in optical from several models as presented by \cite{2004ApJ...606.1125D} and our wind model in \cite{2005ApJ...627L..37P}, it is obvious that firm conclusions could be drawn from such observations. Unfortunately, we still await such an instrument able to detect linear polarisation of photon above several tenth of eV with also a sufficient temporal resolution. Attempts to use current technology such as INTEGRAL to search for linearly polarized emission above 200 keV have been conducted by \cite{2008ApJ...688L..29F} who found in a survey of the Crab Nebula a strong signal polarized along the pulsar rotation axis with some improvements from \cite{2013ApJ...769..137C}. INTEGRAL was also used by \cite{2008Sci...321.1183D} to detected gamma-ray polarized emission from the Crab. Recently \cite{2016MNRAS.456L..84C} reported hard X-ray polarization measurements of the Crab pulsar with a balloon-borne polarimeter called PoGOLite \citep{2008APh....30...72K}.

There is still a strong debate about the precise location of pulsed high-energy emission from pulsars. However, the general consensus now is that it must come from regions close or outside the light-cylinder.

\section{Summary}
\label{sec:Summary}

Neutron stars and their most frequent observational manifestations as pulsars are exquisite space laboratories to explore physics under extreme gravitational and electromagnetic fields. However, in order to correctly interpret the multi-wavelength signal detected on Earth, we need a better understanding of the interrelations between the electromagnetic field, the gravitational field, the plasma and the radiation field. We are still far from a complete and exhaustive answer to the physics undergoing in pulsar magnetosphere. Our sparse ideas need a more thorough and systematic investigations. The wealth of observational data at all wavelengths acquired from systems containing a neutron star (isolated pulsars, isolated neutron stars, magnetars, binary pulsars) complicates our tentative to synthesize in a clear manner the subject. Most importantly it renders our task difficult to extract useful informations.

However, efforts have been made towards a more self-consistent treatment of the electromagnetic, gravitation, plasma and radiation interactions leading to more quantitative results. Numerical simulations promise to unveil some aspects of pulsar physics but the huge gaps between the microphysics and the global structure of this system requires even more clever ideas to perform realistic meaningful simulations. Improvement in numerical algorithms are certainly welcome but it is hopeless to expect to get the final answer simply by running such codes claimed to contain all the physics starting from first principles. Simulations must be taken with caution and we should never forget that numerical simulations are useful to encourage critical thinking rather than being a substitute to brainwork.

\section{For further information}

For a thorough discussion on several important aspects of pulsar dynamics to more than an introductory level, the reader is referred to some excellent books. For instance \cite{1991tnsm.book.....M} although outmoded from an observational point of view, contains a still topical and interesting theoretical description of neutron stars magnetosphere. \cite{1993ppm..book.....B} discuss in details the kinetic aspects of magnetospheric plasma with pair creation and radiation mechanisms. Accretion powered pulsars offer some other insight into neutron star magnetospheric physics \citep{2007WSSAA..10.....G}. The two last chapters of \cite{1999stma.book.....M} are devoted to pulsar electrodynamics. \cite{1999PhR...318..227M} proposed a pedagogical and convincing discussion of the usefulness of non neutral plasmas in the context of pulsars. The general theory of these non neutral plasmas is presented in depth by \cite{Davidson1990}.


%

\end{document}